\def\thisversion{4 November 2018}
\documentclass[aps,prc,superscriptaddress,twoside,twocolumn,%
                nofootinbib,showpacs]{revtex4-1}

\usepackage{amsmath,amssymb}
\usepackage{graphicx}
\usepackage{braket}
\usepackage{slashed}
\usepackage{newtxtext,newtxmath}
\usepackage{mathrsfs}
\usepackage{csquotes}
\usepackage{dcolumn}
\newcolumntype{d}[1]{D{.}{.}{#1}}

\graphicspath{{figs/}}

\newcommand*{\fs}[1]{\slashed{#1}}
\newcommand*{\IC}{{\text{int}}}
\newcommand*{\LD}[2][N]{#1L$_#2$}
\newcommand*{\hQ}{\hat{Q}}
\newcommand*{\bsf}[1]{{\textbf{\textsf{#1}}}}
\newcommand*{\Gzero}{\bsf{G}_{\bsf{0}}}
\newcommand*{\Mpp}{M_{\pi\pi}}

\newcommand*{\qqtext}[1]{\qquad\text{#1}\qquad}
\newcommand*{\qtext}[1]{\quad\text{#1}\quad}
\newcommand*{\GD}[1]{\left\{#1\right\}^\mu}
\newcommand*{\Fpipi}{\mathbb{F}}
\newcommand*{\Cpipi}{\mathbb{C}}
\newcommand*{\Mpipi}{\mathbb{M}}
\newcommand*{\Rpipi}{\mathbb{R}}
\newcommand*{\Tpipi}{\mathbb{T}}
\newcommand*{\Spipi}{\mathbb{S}}
\newcommand*{\Kpipi}{\mathbb{K}}

\newcommand*{\BGamma}{\boldsymbol{\Gamma}}

\allowdisplaybreaks

\begin{document}

\title{Theory of two-pion photo- and electroproduction off the nucleon}

\author{Helmut~Haberzettl}
 \email{helmut.haberzettl@gwu.edu}
 \affiliation{\mbox{Institute for Nuclear Studies and Department of Physics, The George Washington University, Washington, DC 20052, USA}}
\author{Kanzo~Nakayama}
 \email{nakayama@uga.edu}
 \affiliation{Department of Physics and Astronomy, University of Georgia, Athens, GA 30602, USA}
\author{Yongseok Oh}
 \email{yohphy@knu.ac.kr}
 \affiliation{Department of Physics, Kyungpook National University, Daegu 41566, Korea}
 \affiliation{Asia Pacific Center for Theoretical Physics, Pohang, Gyeongbuk 37673, Korea}

\date{\thisversion}

\begin{abstract}
A field-theoretical description of the electromagnetic production of two
pions off the nucleon is derived and applied to photo- and electroproduction
processes, assuming only one-photon exchange for the latter. The developed
Lorentz-covariant theory is complete in the sense that all explicit
three-body mechanisms of the interacting $\pi\pi N$ system are considered
based on three-hadron vertices. The modifications necessary for incorporating
$n$-meson vertices for $n\ge 4$ are discussed. The resulting reaction
scenario subsumes and surpasses all existing approaches to the problem based
on hadronic degrees of freedom. The full three-body dynamics of the
interacting $\pi\pi N$ system is accounted for by the Faddeev-type ordering
structure of the Alt-Grassberger-Sandhas equations. The formulation is valid
for hadronic two-point and three-point functions dressed by arbitrary
internal mechanisms
--- even those of the self-consistent nonlinear Dyson-Schwinger type (subject
to the three-body truncation) --- provided all associated electromagnetic
currents are constructed to satisfy their respective (generalized)
Ward-Takahashi identities. It is shown that coupling the photon to the
Faddeev structure of the underlying hadronic two-pion production mechanisms
results in a natural expansion of the full two-pion photoproduction current
$\Mpp^\mu$ in terms of multiple  dressed loops involving two-body subsystem
scattering amplitudes of the $\pi\pi N$ system that preserves gauge
invariance as a matter of course order by order in the number of (dressed)
loops. A closed-form expression is presented for the entire gauge-invariant
current $\Mpp^\mu$ with complete three-body dynamics. Individually
gauge-invariant truncations of the full dynamics most relevant for practical
applications at the no-loop, one-loop, and two-loop levels are discussed in
detail. An approximation scheme to the full two-pion amplitude for
calculational purposes is also presented. It approximates, systematically,
the full amplitude to any desired order of expansion in the underlying
hadronic two-body amplitude. Moreover, it allows for the approximate
incorporation of all neglected higher-order mechanisms in terms of a
phenomenological remainder current. The effect and phenomenological
usefulness of this remainder current is assessed in a tree-level calculation
of the $\gamma N \to K K \Xi$ reaction.
\end{abstract}

\pacs{25.20.Lj, 13.75.Gx, 13.75.Lb, 25.30.Rw}


\maketitle

\section{Introduction}  \label{sec:introduction}

The experimental study of double-pion production off the nucleon has a fairly
long history, with some of the earliest experiments going back to more than
half a century~\cite{PH54,FC57,FM67,Hauser67,ABBHHM68,ABBHHM69,BCGG72}. In the
last two decades, with the availability of sophisticated experimental
facilities at MAMI in Mainz, GRAAL in Grenoble, ELSA in Bonn, and the CLAS
detector at Jefferson Lab (JLab), the emphasis of experiments with both real
and virtual photons is clearly on using this reaction as a tool to study and
extract the properties of excited baryonic states that form at intermediate
stages of the
reaction~\cite{BMAA95,HABK97,ZABB97,ZABB99,KAAB00,WABH00,LABH01,%
KAAB04,ABBB03,CLAS-02,GDH-A2-03,CLAS-05,%
GDH-A2-05,GDH-A2-07,AABB07,CLAS-09b,CB-A2-09,CLAS-09,%
SAPHIR05,AABB08,CB-ELSA-07c,CBELSA/TAPS-08,CBELSA/TAPS-08c,CBELSA-08,CBELSA-07, %
CLAS07b,CBELSA/TAPS-09, %
CBELSA/TAPS-14,CBELSA/TAPS-15,CBELSA/TAPS-15b,CBELSA/TAPS-15c,CLAS-17}. For
comprehensive accounts on the pre-2013 activities in double-meson photo- and
electro-production processes in particular, and on baryon spectroscopy in
general, we refer to Refs.~\cite{KR09,CR13}.

Baryon spectroscopy has long been plagued by the so-called \emph{missing
resonance\/} problem~\cite{IK77,KI80}, which refers to resonances predicted by
nonrelativistic quark models but not found in $\pi N$ scattering experiments.
One of the possible explanations for this problem is that those resonances may
dominantly undergo sequential decays rather than direct decays into $\pi N$. An
integral part of a comprehensive baryon-spectroscopy program, therefore, is the
determination of sequential decay modes of baryons, in addition to direct
one-step decays. To understand sequential decays, it is essential to
investigate the production of two (or more) mesons. Indeed, analyses of some
experiments in two-pion and $\pi\eta$ photoproduction processes provide
evidence for sequential decays of $N$ and $\Delta$
resonances~\cite{WABH00,CBELSA-07,CBELSA/TAPS-15c,CBELSA-07b,AABB08,ABKN11}.

As the database for two-meson photo- and electroproduction increases, the need
for more complete theoretical descriptions of such processes will increase as
well to help in understanding their reaction dynamics. Theoretically, the study
of double-pion electroproduction off the nucleon is a challenging problem
because, unlike single-pion production, its correct description needs to
combine baryon and meson degrees of freedom on an equal footing because the two
pions in the final state can come off a decaying intermediate meson state, and
not just off intermediate baryons as a sequence of two single-pion productions.
This therefore requires accounting for all competing internal
photo-subprocesses like, for example, the baryonic $\gamma N \to \pi N$ and the
purely mesonic $\gamma \rho \to \pi\pi$ in a consistent manner.

Several groups have theoretically studied two-meson photo- and
electro-productions employing a variety of approaches. The Bonn-Gatchina group
has performed a multi-channel partial-wave analysis of the existing two-pion
and $\pi\eta$ photoproduction data~\cite{CBELSA-07b,ABKN11} by extending its
single-channel photoproduction partial-wave analyses. Double-pion
photoproduction near threshold is described by chiral perturbation
theory~\cite{BKMS94a,BKM95b,BKM96} and the 2004 data from MAMI on $\pi^0\pi^0$
photoproduction off the proton~\cite{KAAB04} seem to be consistent with its
predictions. Unitary chiral perturbation theory has been applied in the
analyses of $\pi\eta$ and $K\pi\Sigma$
photoproduction~\cite{DOS05,DOS06,DOM10}. In $\pi\eta$ photoproduction, cross
sections as well as spin-observables $I^s$ and $I^c$ were computed. The results
are in good agreement with the existing data of
Refs.~\cite{AABB08,CBELSA/TAPS-08c,CBELSA/TAPS-09}.

At present, the most detailed model calculation of two-pion photoproduction is
that of the EBAC/ANL-Osaka group~\cite{KJLMS09}. It is an extension of their
dynamical coupled-channels approach for single pseudoscalar-meson production
developed over recent years~\cite{MSL06} by describing the basic two-meson
production mechanisms as isobar-type approximations obtained by attaching the
vertices for $ \Delta \to \pi N$, $\rho \to \pi\pi$, and $\sigma \to \pi\pi$
transitions to the corresponding single-meson production amplitudes,
\textit{viz.}, $\gamma N \to \pi \Delta$, $\gamma N \to \rho N$, and $\gamma N
\to \sigma N$ amplitudes, respectively, obtained in the dynamical
coupled-channels approach~\cite{MSL06}. This model includes the hadronic $\pi N
\to \pi \pi N$ channel~\cite{KJLMS08}, and the $S_{11}(1535)$, $S_{31}(1620)$,
and $D_{13}(1520)$ resonances are found to be relevant to two-pion
photoproduction up to $W=1.7~\mbox{GeV}$.

The majority of existing model calculations of two-meson photo- and
electro-photoproduction processes are based on straightforward tree-level
effective Lagrangian approaches. In photoproduction, these models have been
applied to
two-pion~\cite{GO94,GO95,NOVR00,NO01,Roca04,OHT97,HOT98,HKT02,FA05,MRAB01,MRAB03},
and $\pi\eta$~\cite{JOH01} productions. Two-pion photoproduction in nuclear
medium has been also studied within tree-level
approximations~\cite{ROV02,HOT97}. In the strangeness sector, the $K\bar{K}$
photoproduction has been investigated within the tree-level effective
Lagrangian approach~\cite{ONL04} as well as the $KK\Xi$
photoproduction~\cite{NOH06,MON11}. The latter calculation includes a
generalized four-point contact current to keep the resulting amplitude gauge
invariant. The $\pi\eta$ and $\pi\pi$ electroproduction reactions were studied
in a similar framework~\cite{APSTW05,KCL07}. A variation of the tree-level
approximation in the analyses of two-pion electroproduction is adopted in
Refs.~\cite{MBLEFI08,CLAS-12,MBCE15}. For the theoretical description of
$\pi\pi$-production observables, we refer to Ref.~\cite{RO04} and references
therein. Because of their simplicity, tree-level approximations are widely used
in the analyses of two-meson photo- and electro-production processes. Despite
their simplicity, they often provide insights into dominant aspects of the
reaction mechanism in a more transparent way than more involved approaches.

The purpose of the present work is to  derive two-meson photoproduction
amplitudes which include the full miscroscopic details contained in the tree-
and four-point hadronic vertices and thus offer the theoretical framework for
exploiting the underlying reaction dynamics in a detailed and systematic manner
beyond simple tree-level models. The derivation proceeds analogous to
single-meson photoproduction, based on the field-theoretical approach of
Haberzettl~\cite{Haberzettl97}, where the photoproduction amplitude is obtained
by attaching a photon to the full $N \to \pi N$ three-point hadronic vertex
using the Lehmann-Symanzik-Zimmermann (LSZ) reduction~\cite{LSZ55} which allows
to express the full photoproduction amplitude in term of the gauge-derivative
procedure proposed in Ref.~\cite{Haberzettl97}. For the two-meson case, we
attach the photon to the full $N \to \pi\pi N$ four-point hadronic vertex,
whose microscopic structure is described in a nonlinear three-body Faddeev-type
approach. The gauge-derivative device provides a very convenient tool to
identify and link all relevant microscopic reaction mechanisms in a consistent
manner. Similar to the single-meson photoproduction amplitude, the resulting
two-meson photoproduction amplitude is analytic, covariant, and (locally)
gauge-invariant as demanded by the generalized Ward-Takahashi
identity~\cite{Ward50,Takahashi57}. Local gauge invariance, in particular, is
important in electromagnetic processes because it requires consistency of all
contributing mechanisms. Its violation may thus point to missing mechanisms, as
was demonstrated for the $NN$ bremsstrahlung reaction which is one of the most
basic hadron-induced processes. In Refs.~\cite{NH09,HN10},  it was shown how to
solve the long-standing problem of describing the high-precision KVI data by
including in the model a properly constructed interaction current that obeys
the generalized Ward-Takahashi identity required by local gauge invariance.

Since particle number is not conserved in meson dynamics, the full two-meson
photoproduction amplitude as described here is highly nonlinear, thus making
truncations unavoidable  in practical calculations. However, to help with the
incorporation of higher-order contributions, we present a scheme that expands
the amplitude in powers of the underlying two-body hadronic $T$-matrix elements
and, in addition, provides a procedure for accounting for  neglected
higher-order contributions in a phenomenological manner. In principle, at
least, the approximation can be refined to any desired accuracy. Local gauge
invariance is maintained at each level of the approximation.

%
\begin{figure*}[!t]\centering
\includegraphics[width=.8\textwidth]{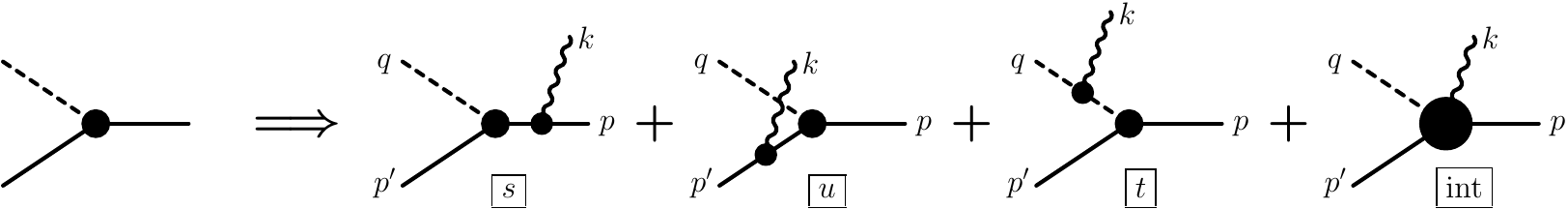}
\caption{\label{fig:Msutc}%
Generic topological structure of the single-pion production current $M^\mu$ for
$\gamma N\to \pi N$ of Eq.~(\ref{eq:Msuti}). Here and throughout this paper, time proceeds from right
to left in all diagrams. Attaching the incoming photon (wavy line) to the three
external legs of the $\pi NN$ vertex on the left (where the solid lines are
nucleons and the dashed line is the pion) produces the first three diagrams on
the right labeled $s$, $u$, and $t$, after the corresponding Mandelstam
variables of the intermediate off-shell particle. The three resulting currents
are denoted by $M^\mu_s$, $M^\mu_u$, and $M^\mu_t$, respectively. Coupling the
photon to interior of the vertex produces the interaction current $M^\mu_\IC$
depicted by the last contact-type four-point vertex. This generic structure is
independent of how the vertex is dressed in detail, or even if it is dressed at
all. For bare vertices, the interaction current of the last diagram is the
Kroll-Ruderman term~\cite{KR54}. The labels at the external lines indicate the
four-momenta of the respective particles satisfying four-momentum conservation,
$p+k=p'+q$.}
\end{figure*}
%

A preliminary account of a part of the main results of the present work can be
found in the conference proceedings of Ref.~\cite{HKO12}. The present paper is
organized as follows. In the subsequent Sec.~\ref{sec:1pion}, we recapitulate
some features of the theory of single-pion production off the nucleon of
Ref.~\cite{Haberzettl97} so that we can establish the relevant techniques and
tools to tackle the double-pion production problem. Then in
Sec.~\ref{sec:topology}, using the basic topological properties of the process,
we derive a formulation of the \emph{hadronic\/} two-pion production process
$N\to \pi\pi N$ that incorporates all relevant degrees of freedom and all
possible final-state mechanisms of the dressed $\pi \pi N$ system. We do this
by employing the Faddeev-type~\cite{Faddeev60,Faddeev} three-body
Alt-Grassberger-Sandhas equations~\cite{AGS67} to sum up the corresponding
multiple-scattering series. The actual electromagnetic production current
$\Mpp^\mu$ is then constructed in Sec.~\ref{sec:photon}, by applying the gauge
derivative~\cite{Haberzettl97} to couple the (real or virtual) photon to the
hadronic process found in Sec.~\ref{sec:topology}. (This procedure is sometimes
referred to as `gauging' of the underlying hadronic mechanisms.) We show that
the resulting closed-form expression for the complete current satisfies the
generalized Ward-Takahashi identity and thus is \textit{locally} gauge
invariant. We also show that the full current can be decomposed in a systematic
manner into a sum of contributions that are directly related to topologically
distinct hadronic two-pion production mechanisms of increasing complexity and
that each of these partial currents is gauge invariant separately. This finding
is important from a practical point of view because it allows one, to a certain
extent, to separate the technical issue of maintaining gauge invariance from
the question of how complex the reaction mechanisms must be to describe the
physics at hand. In Sec.~\ref{sec:approx}, an approximation scheme to the full
two-meson photoproduction amplitude is presented based on the expansion in
powers of the underlying two-body hadronic interactions. Section~\ref{sec:appl}
contains the application of the approximation scheme in the lowest order to
describe the $\gamma N \to K K \Xi$ reaction. Finally, we present a summarizing
assessment and discussion in the concluding Sec.~\ref{sec:summary}. Two
appendices then provide additional material. Appendix~\ref{app:4meson}
discusses the incorporation of four-meson vertices like $\omega\to\pi\pi\pi$
and Appendix~\ref{app:GenContactCurrent} provides a derivation of generic
phenomenological contact currents for arbitrary hadronic transition that
satisfy local gauge-invariance constraints, with specific applications to four-
and five-point contact currents needed for the present formalism.

\section{\boldmath Foundation: The $\gamma N \to \pi N$ Problem}\label{sec:1pion}

A necessary prerequisite to understanding the photoproduction of two pions is
to understand the photoproduction of a single pion off the nucleon. To this
end, we recapitulate here some features of the theoretical formulation of that
process following the field-theoretical treatment of Ref.~\cite{Haberzettl97}.
This will also help us establish some of the necessary tools for the
description of two-pion production.

The basic topological structure of the single-pion production current $M^\mu$
was given a long time ago~\cite{GG54} by observing how the photon can couple to
the underlying hadronic single-pion production process $N\to \pi N$. As shown
in Fig.~\ref{fig:Msutc}, there are two distinct types of contributions,
respectively called class A and class B in Ref.~\cite{GG54}. Class A contains
the three contributions $M^\mu_s$, $M^\mu_u$, and $M^\mu_t$ coming from the
external legs of the $\pi NN$ vertex that have poles in the Mandelstam
variables $s$, $u$, and $t$, and class B is the non-polar contact-type current
$M^\mu_\IC$ originating from the interaction of the photon with the interior of
the vertex. The full current $M^\mu$, therefore, can be written as
\begin{equation}
  M^\mu = M^\mu_s+M^\mu_u+M^\mu_t +M^\mu_\IC~,
  \label{eq:Msuti}
\end{equation}
as indicated in Fig.~\ref{fig:Msutc}.
This structure is based on topology alone and therefore independent of the details of
the individual current contributions.

These details matter, of course, if one wishes to derive the currents for
practical applications. In general, an electromagnetic current for a hadronic
process is defined by first employing minimal substitution for the connected
part of the hadronic Green's functions and then taking the functional
derivative with respect to the electromagnetic four-potential $A^\mu$, in the
limit of vanishing $A^\mu$. The current is then obtained by removing the
propagators of the external hadron legs from this derivative in an LSZ
reduction procedure~\cite{LSZ55}. The gauge-derivative procedure of
Ref.~\cite{Haberzettl97} provides a formally equivalent method that is much
simpler to handle in practice because it essentially amounts to the simple
recipe of attaching a photon line to any topologically distinct feature of a
hadronic process expressed in terms of Feynman diagrams and summing up the
corresponding contributions to obtain the full current.

\begin{figure}[!t]
\centering
  \includegraphics[width=.8\columnwidth]{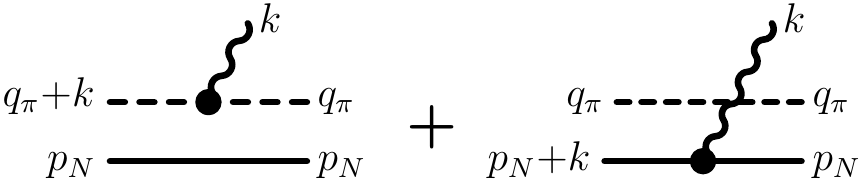}
\caption{\label{fig:dmu}%
Graphical representation of the dual-current contribution $G_0d^\mu G_0$ of
Eq.~(\ref{eq:G0dmuG0}) for the photon being coupled to the free $\pi N$
propagator $G_0=S(p_N)\circ \Delta(q_\pi)$ of Eq.~(\ref{eq:G0=SoDelta}).}
\end{figure}

For the single-pion photoproduction process at hand, the connected part of the
free $\pi NN$ Green's function is given by $G_0 F S$, where $F$ is the $\pi NN$
vertex shown on the left-hand side of Fig.~\ref{fig:Msutc}, $S$ is the
propagator of the incoming nucleon leg and
\begin{equation}
  G_0 = S(p_N^{}) \circ \Delta(q_\pi^{})
  \label{eq:G0=SoDelta}
\end{equation}
is the product of the outgoing nucleon and pion propagators $S$ and $\Delta$,
respectively, written here with generic momenta for the nucleon and the pion,
where the sum of the respective four-momenta $p_N^{} + q_\pi^{}$ is the fixed available
total momentum. Within a loop integration, this free $\pi N$ propagator would
correspond to a convolution integration of these momenta, as indicated by
``$\circ$'' here. The LSZ expression for the photoproduction current may now be
written as~\cite{Haberzettl97}
\begin{equation}
  M^\mu = -G_0^{-1}\left\{G_0 F S\right\}^\mu S^{-1}~,
  \label{eq:Mcurrentdefined}
\end{equation}
where $\{\cdots\}^\mu$ is the short-hand notation for the gauge derivative
introduced in Ref.~\cite{Haberzettl97}, with $\mu$ indicating the Lorentz index
of the incoming photon. Being a derivative, the product rule applies and we
obtain
\begin{align}
  M^\mu   &=-G_0^{-1}\left\{G_0\right\}^\mu F
  -\left\{F \right\}^\mu -  F\left\{ S\right\}^\mu S^{-1}
  \nonumber\\
  &= d^\mu G_0 F +  M^\mu_\IC + FSJ_N^\mu~,
  \label{eq:MmuFT}
\end{align}
where in the last step
\begin{subequations}\label{eq:GDdetail}
\begin{align}
-\left\{ S\right\}^\mu &= S J_N^\mu S~,
\\
  -\left\{F\right\}^\mu &= M^\mu_\IC~, \label{eq:MintGDeriv}
\\
  -\left\{G_0\right\}^\mu &=G_0 d^\mu G_0~,
  \label{eq:dmu}
\end{align}
\end{subequations}
were used~\cite{Haberzettl97}, which relate the corresponding gauge derivatives
to the nucleon current operator $J^\mu_N$ of the incoming nucleon, the
interaction current $M^\mu_\IC$ for the interior of the vertex $F$, and the
dual-current contribution of the free $\pi N$ system,
\begin{align}
  G_0d^\mu G_0 &= S \circ (\Delta J^\mu_\pi\Delta)+(SJ^\mu_NS) \circ \Delta   ~,
   \label{eq:G0dmuG0}
\end{align}
as depicted in Fig.~\ref{fig:dmu}, which sum up attaching the photon to $G_0$
in terms of the corresponding nucleon current $J^\mu_N$ and the pion current $J_\pi^\mu$.
The three polar currents in Eq.~(\ref{eq:Msuti}) obviously are given here by
\begin{subequations} \label{eq:Msut}
\begin{align}
  d^\mu G_0 F &= M^\mu_t + M^\mu_u~,
\\
  FSJ_N^\mu &= M^\mu_s~,
\end{align}
\end{subequations}
which completes matching the field-theoretical result of Eq.~(\ref{eq:MmuFT})
with the topological one in Eq.~(\ref{eq:Msuti}).

%
%

Note here that with the external momenta of the photo-process given as in
Fig.~\ref{fig:Msutc}, namely,
\begin{equation}
  \gamma(k) + N(p) \to \pi(q)+N(p')~,
\end{equation}
it was not necessary to write out the momentum dependence of any of the
elements of the preceding equations because it can easily be found explicitly
by knowing that the photon carries a momentum $k$ into the element to which it
is attached.

\subsection{Gauge invariance}

Gauge invariance as the manifestation of $U(1)$ symmetry is of fundamental
importance for any photo-process because it provides a conserved (on-shell)
current and thus implies charge conservation. The requirement of \emph{local\/}
gauge invariance~\cite{PS}, in particular, implies the very existence of the
electromagnetic field and thus is of fundamental importance for the formulation
of \emph{consistent\/} reaction dynamics of photo-processes, which goes beyond
the mere on-shell constraint of charge conservation.

For single-pion photoproduction, local gauge invariance is formulated in terms
of the generalized Ward-Takahashi identity (WTI)~\cite{Haberzettl97,Kazes59}
\begin{eqnarray}
  k_\mu M^\mu &=& S^{-1}(p') Q_{N_f} S(p'-k) F_u
  \nonumber\\   &&\mbox{}
   +\Delta^{-1}(q) Q_\pi \Delta (q-k) F_t
     \nonumber\\   &&\mbox{}
    -F_s S(p+k)Q_{N_i} S^{-1}(p)~,
   \label{eq:gWTIpiN}
\end{eqnarray}
where the four-momenta are those shown in Fig.~\ref{fig:Msutc} and the vertices
$F_x$ are the $\pi NN$ vertex functions in the specific kinematic situations
described by the Mandelstam variables $x=s,u,t$ in the figure. The charge
operators for the initial and final nucleons are represented by $Q_{N_i}$ and
$Q_{N_f}$, respectively, and $Q_\pi$ is the charge operator for the outgoing
pion. The inverse propagators here ensure that this four-divergence vanishes
for matrix elements with all hadron legs on-shell and thus provides a conserved
current. The generalized  WTI as such, however, is an \emph{off-shell\/}
constraint, thus providing a continuous dynamical link between the transverse
and longitudinal regimes. This is analogous to the usual single-particle
Ward-Takahashi identities~\cite{Ward50,Takahashi57} for the nucleon current,
\begin{equation}
  k_\mu J^\mu_N (p_N^{} + k, p_N^{}) =  S^{-1}(p_N^{} + k) Q_N - Q_N S^{-1}(p_N^{})~,
  \label{eq:WTIN}
\end{equation}
and for the pion current,
\begin{equation}
k_\mu J^\mu_\pi(q_\pi^{} + k, q_\pi^{})
    =  \Delta^{-1}(q_\pi^{} + k) Q_\pi - Q_\pi \Delta^{-1}(q_\pi^{})~,
  \label{eq:WTIpi}
\end{equation}
which are also off-shell relations. Note that the validity of these two
equations, which apply to the currents associated with the external legs in
Fig.~\ref{fig:Msutc}, and the generalized  WTI of Eq.~(\ref{eq:gWTIpiN})
immediately imply that the four-divergence of the interaction current is given
by
\begin{equation}
  k_\mu M^\mu_\IC =Q_{N_f}F_u + Q_\pi F_t -  F_s \, Q_{N_i}~.
  \label{eq:WTIint}
\end{equation}
In fact, given the usual single-particle WTIs of Eqs.~(\ref{eq:WTIN}) and
(\ref{eq:WTIpi}), Eqs.~(\ref{eq:gWTIpiN}) and (\ref{eq:WTIint}) are equivalent
formulations of gauge invariance of the photoproduction amplitude, with one
condition implying the  respective other. However, for practical purposes, in
particular, in a semi-phenomenological approach, the interaction-current
condition (\ref{eq:WTIint}) is actually a more versatile tool because it lends
itself very easily to phenomenological recipes that help ensure gauge
invariance~\cite{Haberzettl97,HBMF98a,HNK06,HHN11,HWH15}. The fact that all of
these four-divergences are off-shell relations and therefore remain valid
within whatever context the corresponding currents appear will be of immediate
and direct relevance for two-pion production-current considerations in
Sec.~\ref{sec:photon}.

To facilitate the investigation of gauge invariance for the two-pion production
case later on, we will now expand the meaning of the charge operators $Q_i$ of
particle $i$. We first note that the charge operators appearing in all of the
preceding relations only act on the isospin dependence within the $\pi NN$
vertices $F_x$, i.e., their placements before or after a vertex cannot be
changed, but otherwise they can appear anywhere in an equation. In all of the
preceding equations, however, the charge operators $Q_i$ have always been
placed at the locations where the momentum of the particular particle line
\emph{increases\/} by the momentum $k$ of the incoming photon. Therefore,
following Ref.~\cite{Haberzettl97}, we define the operator $\hQ_i$ which
injects the photon momentum $k$ into the equation where it is placed as well as
having the role of the charge operator $Q_i$.
We can then omit \emph{all\/} explicit momenta in the equations because they
can be recovered unambiguously from knowing the given external momenta of the
process at hand. We can even go further to introduce~\cite{Haberzettl97}
\begin{equation}
  \hQ = \sum_i \hQ_i~,
\end{equation}
where the summation is taken to be context-dependent, i.e., wherever $\hQ$ is
placed in an equation, the sum extends over all particles that appear in that
place in the equation. We may then write the generalized  WTI of
Eq.~(\ref{eq:gWTIpiN}) equivalently and very succinctly as
\begin{equation}
k_\mu (G_0 M^\mu S) =  \hQ (G_0 F S) - (G_0 F S) \hQ ~,
\label{eq:gWTIGreenNew}
\end{equation}
i.e., as a commutator of $\hQ$ and the connected $\pi NN$ Green's function $G_0
F S$. Here, $\hQ$ appearing on the left of $G_0FS$ subsumes the outgoing pion
and nucleon, and $\hQ$ on the right only comprises the incoming nucleon. The
physical current $M^\mu$ on the left is amended with the propagators $S$ and
$G_0$ of the incoming and outgoing particles, respectively, similar to the
external propagators in the Green's function $G_0 F S$. For the interaction
current, the formulation equivalent to Eq.~(\ref{eq:WTIint}) is
\begin{equation}
  k_\mu M^\mu_\IC = \hQ F - F \hQ~,
  \label{eq:WTIintQ}
\end{equation}
and the single-particle WTIs of Eqs.~(\ref{eq:WTIN}) and (\ref{eq:WTIpi}) may
be written as
\begin{subequations}\label{eq:WTIJQ}
\begin{align}
  k_\mu (S J^\mu_N S) &=  \hQ S - S\hQ ~,
  \\[1ex]
  k_\mu (\Delta J^\mu_\pi \Delta) &=  \hQ \Delta - \Delta \hQ~,
\end{align}
\end{subequations}
where the propagators $S$ and $\Delta$ are single-particle Green's functions
for the nucleon and the pion, respectively, in complete analogy to
Eq.~(\ref{eq:gWTIGreenNew}).

The structures of all equations here are similar: For a physical current, the
four-divergence of the current, with propagators attached to its external legs,
is expressed as a commutator of $\hQ$ with the corresponding (connected)
Green's function. For an interaction current describing only the interaction
with the interior of a hadronic process, the four-divergence is given by the
commutator of $\hQ$ with the underlying hadronic process. This finding is
generic and holds true irrespective of how complicated the photo-process at
hand actually is. The $\hQ$ device will prove to be invaluable for
investigating the gauge invariance of the two-pion production process.

\subsection{Dressing propagators and vertices}\label{sec:II.B}

In the preceding discussion, we have not touched upon the question if, and if
yes, to what extent, the propagators of the nucleon and pion and the $\pi NN$
vertex need to be dressed. As far as gauge invariance is concerned, the answer
is very simple: for gauge invariance to hold true \emph{any\/} degree of
dressing that ensures the validity of Eqs.~(\ref{eq:WTIJQ}) for the propagators
and of Eq.~(\ref{eq:WTIintQ}) for the interaction current is sufficient. Local
gauge invariance, therefore, only requires that the single-particle and the
interaction currents be constructed consistently with each other by keeping the
overall structure of the production current depicted in Fig.~\ref{fig:Msutc}.
Besides that, it does not demand or imply any particular degree of dressing.

\begin{figure*}[t]\centering
  \includegraphics[width=.9\textwidth,clip=]{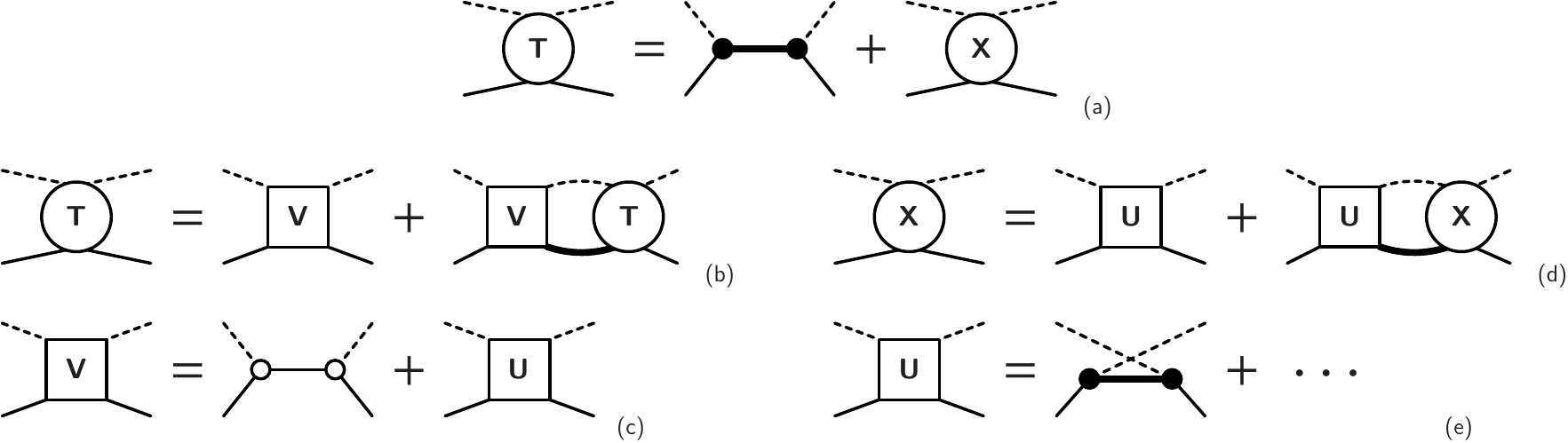}
\caption{\label{fig:piNScattering}%
Generic structure of the pion-nucleon $T$ matrix employing pions and nucleons
as the only hadronic degrees of freedom~\cite{Haberzettl97}. (a) Splitting of
$T$ into $s$-channel pole part and non-pole $X$. (b) Bethe-Salpeter integral
equation for $T$, with (c) the driving term $V$ that contains an
\emph{undressed\/} $s$-channel exchange. (d) Bethe-Salpeter integral equation
for non-pole $X$, with (e) \emph{fully dressed\/} non-pole driving term $U$.
Dressed vertices are represented by solid circles, while undressed ones are
denoted by open circles. Dressed (internal) nucleons are shown as thick lines;
undressed ones as thin lines; pions are shown as dashed lines. Note that the
$s$-channel pole term in the driving term $V$ is bare [because it gets dressed
by the equation (b) itself] whereas, in the full theory, all mechanisms in the
non-pole $U$ are fully dressed via the Dyson-Schwinger-type mechanisms as shown
in Fig.~\ref{fig:Fdressed}.}
\end{figure*}

Even the simplest example, where the nucleon and pion propagators and their
currents as well as the $\pi NN$ vertex are essentially bare, satisfies the
generalized  WTI of Eq.~(\ref{eq:gWTIpiN}), as long as the masses are physical
and the interaction current is the well-known Kroll-Ruderman
current~\cite{KR54}. The key to maintaining gauge invariance, therefore, is
\emph{consistency\/} among all ingredients of a particular formulation of the
reaction dynamics. Exploiting this consistency requirement in cases where gauge
invariance does not follow --- which nearly always is the case as soon as one
introduces any kind of dressing mechanisms --- is found to be indeed a powerful
tool for constraining the interaction current by ensuring the validity of
Eq.~(\ref{eq:WTIint})~\cite{Haberzettl97,HBMF98a,HNK06,HHN11}.

\begin{figure}[!t]\centering
\includegraphics[width=.7\columnwidth,clip=]{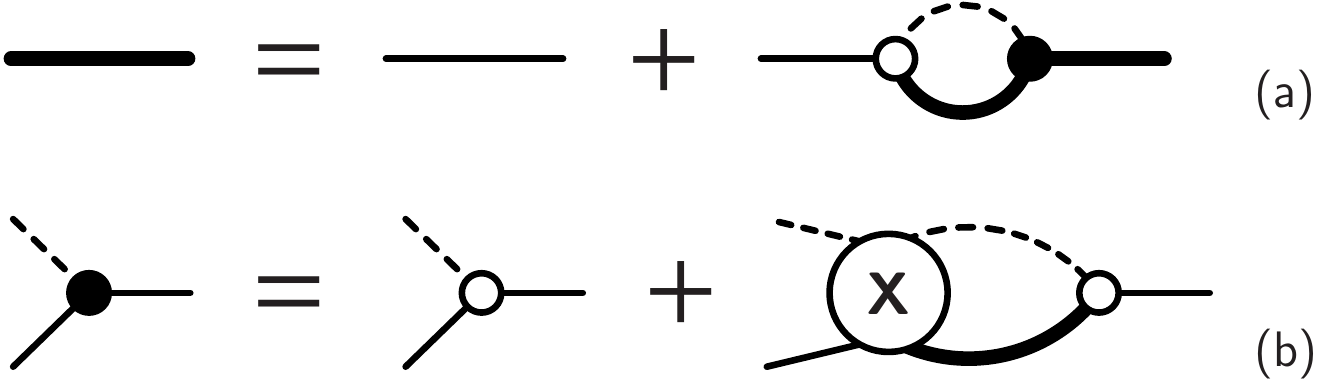}
  \caption{\label{fig:Fdressed}%
Dressing mechanisms for (a) the nucleon propagator $S$ and (b) the $\pi NN$
vertex $F$ according to Eq.~(\ref{eq:Fdressed}) that appears in the nucleon's
self-energy contribution $\Sigma$ shown in (a) as a loop. The notation is the
same as in Fig.~\ref{fig:piNScattering}.}
\end{figure}

%
\begin{figure}[!t]\centering
  \includegraphics[width=.9\columnwidth,clip=]{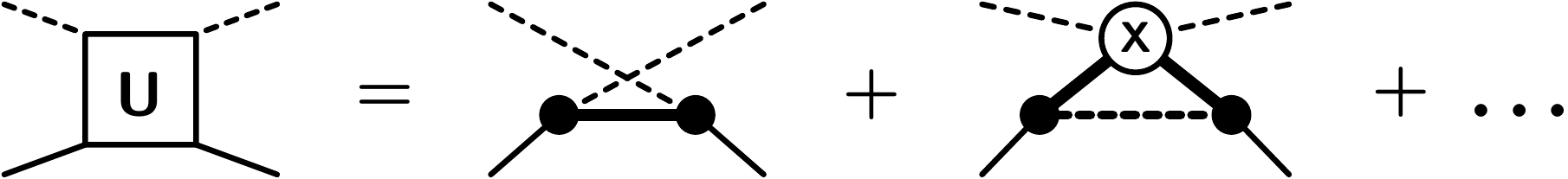}
  \caption{\label{fig:uXchange}%
More detailed description of the driving term $U$ in
Fig.~\ref{fig:piNScattering}(e). In addition to the basic $u$-channel exchange,
$U$ also contains \emph{nonlinear\/} contributions where the full amplitude $X$
given in Fig.~\ref{fig:piNScattering}(d) is dressed by hadron
loops~\cite{Haberzettl97}. (The lowest-order contribution from the nonlinear
dressing mechanism of the second diagram on the right-hand side here appears in
the fourth graph of Fig.~\ref{fig:hadronsbare}.) }
\end{figure}
%

We will find, in Sec.~\ref{sec:photon}, similar consistency constraints for the
present problem of two-pion production. However, to understand better the
structure of the problem, we need to look in more detail at some of the
features of the dressing mechanisms resulting from the theoretical treatment of
single-pion photoproduction of Ref.~\cite{Haberzettl97}, since the underlying
field theory for both single-pion and two-pion production is the same. The full
dressing mechanisms of single-pion production originate from the
Dyson-Schwinger-type structure that governs the pion-nucleon scattering problem
whose equations are summarized diagrammatically in
Figs.~\ref{fig:piNScattering} and \ref{fig:Fdressed}. There is no need here to
recapitulate all features of the treatment of Ref.~\cite{Haberzettl97}
providing these structures. Relevant for the problem at hand is only the fact
that the bare $\pi NN$ vertex $f$ from the underlying interaction Lagrangian is
dressed by the \emph{non-polar\/} part $X$ of the full $\pi N$ $T$ matrix,
i.e.,
\begin{equation}
  F = f +XG_0 f
  \label{eq:Fdressed}
\end{equation}
depicted in Fig.~\ref{fig:Fdressed}(b).
Here, $X$ solves the Bethe-Salpeter-type equation,
\begin{equation}
X = U + UG_0X
\label{eq:BSX}
\end{equation}
shown in Fig.~\ref{fig:piNScattering}(d), whose non-polar driving term $U$ is
given in the lowest order by the $u$-channel exchange of
Fig.~\ref{fig:piNScattering}(e). At higher orders, $U$ also contains nonlinear
contributions where the full $X$ itself is dressed by loops, as shown in the
example of Fig.~\ref{fig:uXchange}. (See also Ref.~\cite{Haberzettl97}.) In
principle, therefore, everything in Eq.~(\ref{eq:BSX}) is dressed fully by the
nonlinear Dyson-Schwinger mechanisms.

\begin{figure}[t!]\centering
  \includegraphics[width=.95\columnwidth,clip=]{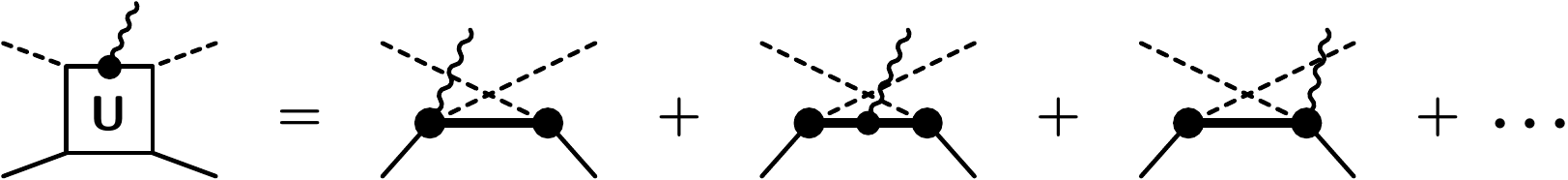}
  \caption{\label{fig:Ucurr}%
Contribution to the five-point interaction current $U^\mu$ based on coupling
the photon to the interior of the lowest-order $u$-channel exchange in the
non-polar driving term $U$ in Fig.~\ref{fig:uXchange}. The currents arising
from the higher-order loops are discussed in Ref.~\cite{Haberzettl97}.}
\end{figure}

According to Eq.~(\ref{eq:MintGDeriv}), the four-point interaction current
$M^\mu_\IC$ is obtained by applying the gauge derivative to the dressed vertex
$F$. Using the explicit dressing equation~(\ref{eq:Fdressed}), this
reads~\cite{Haberzettl97}
\begin{equation}
  M^\mu_\IC = \left(1+XG_0\right)f^\mu + X^\mu G_0 f + XG_0 d^\mu G_0 f~,
  \label{eq:MintDetails}
\end{equation}
where $f^\mu$ is the (bare) Kroll-Ruderman current and $X^\mu$ is the
five-point interaction current resulting from applying the gauge derivative to
Eq.~(\ref{eq:BSX}), i.e.,
\begin{equation}
  X^\mu = \left(1+XG_0\right)U^\mu\left(G_0X +1\right) +X G_0d^\mu G_0 X~.
  \label{eq:Xmu}
\end{equation}
Here, $U^\mu$ is the five-point interaction current (whose lowest order is
shown in Fig.~\ref{fig:Ucurr}) obtained by coupling the photon to all elements
of the driving term $U$. We see here that the internal dressing structure of
the interaction current $M^\mu_\IC$ in Eq.~(\ref{eq:MintDetails}) is quite
complex; it contains, in particular, the full hadronic final-state interaction
in terms of the non-polar $\pi N$ scattering matrix $X$. One can use
Eq.~(\ref{eq:Xmu}) to bring Eq.~(\ref{eq:MintDetails}) into a form better
suited for practical applications, but there is no need to pursue this here
(for more details; see Ref.~\cite{Haberzettl97} for formal derivations and
Refs.~\cite{HNK06,HHN11} for practical aspects).

What we do need for the present purpose, however, is the proof that $X^\mu$
satisfies the usual gauge-invariant constraint of an interaction current. This
proof was given already in Eq.~(72) of Ref.~\cite{Haberzettl97}, but we repeat
it here because it will introduce the general techniques of handling such
four-divergences that we will need later on. For this purpose, let us restrict
$U$ to be given only by the $u$-channel exchange shown in
Fig.~\ref{fig:uXchange}. We emphasize that neglecting higher orders is done
here only to simplify the derivation. In general, the proof will go through for
any possible mechanism at any order~\cite{Haberzettl97}. For a simple
$u$-channel exchange, we may write $U$ as
\begin{equation}
   U = F_i S F_f~, \label{eq:Udef}
\end{equation}
where the indices $i$ and $f$ on the dressed vertices $F$ indicate whether the
corresponding pion leg is an initial or a final particle for the $\pi N\to \pi
N$ process. The current $U^\mu\equiv-\{U\}^\mu$ resulting from coupling the
photon to $U$ is then given by the three diagrams shown in
Fig.~\ref{fig:Ucurr}, i.e.,
\begin{equation}
  U^\mu = M^\mu_i S F_f + F_i SJ_N^\mu S F_f + F_i S M^\mu_f~.
\end{equation}%
%
%
We note here that, because the photon couples into the fully dressed $\pi NN$
vertices of the $u$-channel exchange~(\ref{eq:Udef}), the currents $M^\mu_i$
and $M^\mu_f$ are the full four-point interaction currents of
Eq.~(\ref{eq:MintDetails}), with $i$ and $f$ indicating the direction of the
pion leg. This type of nonlinearity is a natural and unavoidable consequence of
the fact that particle number is not conserved in any process involving mesons.
Using the four-divergences of Eqs.~(\ref{eq:WTIintQ}) and (\ref{eq:WTIJQ}), we
obtain
\begin{align}
k_\mu U^\mu &= (\hQ F_i - F_i \hQ)SF_f+F_i (\hQ S - S \hQ)F_f
  \nonumber\\
  &\quad\mbox{}+ F_i S (\hQ F_f -F_f \hQ)
  \nonumber\\
  & =\hQ U - U\hQ~,
  \label{eq:kU}
\end{align}%
%
\begin{figure*}[!t]\centering
  \includegraphics[width=.8\textwidth,clip=]{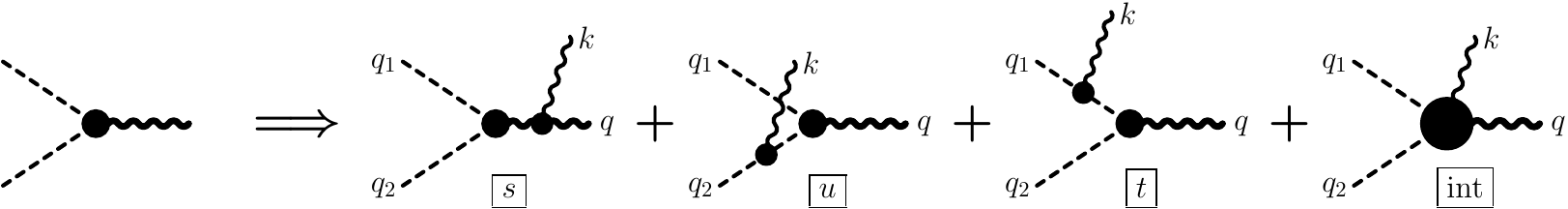}
  \caption{\label{fig:Mrho}%
Generic topological structure of the two-pion production current off another
meson (depicted here as a heavy wavy line), with $\gamma \rho\to \pi\pi$ shown
as an example. Attaching the photon to the hadronic $\pi\pi\rho$ vertex on the
left produces a structure exactly analogous to the $\pi NN$ case, with three
contributions arising from the external legs, and one from the interior
interaction region. }
\end{figure*}
%
and thus
\begin{align}
k_\mu X^\mu &= (1+XG_0)(k_\mu U^\mu)(G_0X +1)
+X G_0 (k_\mu d^\mu) G_0 X
\nonumber\\
&=(1+XG_0)(\hQ U- U \hQ)(G_0X +1)
\nonumber\\
&\qquad\mbox{}
 +X (\hQ G_0 -G_0 \hQ) X
\nonumber\\
&= \hQ X- X\hQ~,
  \label{eq:kX}
\end{align}
where
\begin{equation}
G_0(k_\mu d^\mu)G_0=\hQ G_0 -G_0 \hQ
\label{eq:kd}
\end{equation}
was used, which follows from the definition of $d^\mu$ and the  WTIs of
Eq.~(\ref{eq:WTIJQ}). Both four-divergences of $U^\mu$ and $X^\mu$, therefore,
produce the generic structure associated with interaction currents discussed at
the end of the preceding subsection. For this generic result to hold, it is
irrelevant whether we are dealing with four-point currents like $M^\mu_\IC$ or
five-point currents like $U^\mu$ or $X^\mu$. The result (\ref{eq:kX}), in
particular, will be relevant for the gauge-invariance proof of the two-pion
photoproduction current given in Sec.~\ref{sec:photon} in the context of
Eq.~(\ref{eq:kK}).

\subsection{\boldmath Topologically analogous problem: $\gamma\rho \to \pi\pi$}\label{sec:gammarho}

The underlying field theory of single pion photoproduction just discussed
above~\cite{Haberzettl97} contains  pions, nucleons, and photons as explicit
degrees of freedom. The resulting topological structure is complete in the
sense that even if in actual practical applications one needs to expand the
meaning of ``pion'' and ``nucleon'' to generically stand for all possible
mesons and baryons, respectively, this structure does not change. The situation
is different for two-pion production processes because, as we will discuss in
more detail in Sec.~\ref{sec:topology}, two pions can be produced not only
sequentially off baryons but also directly through the decay of mesons, and
this will add topological features to the problem that cannot be expressed in
the generic picture of pions and nucleons alone with their interaction being
described by the $\pi NN$ vertex. In the following, therefore, we need to
introduce the $\rho$ meson as an additional generic meson degree of freedom
that can decay into two pions, i.e.,
\begin{equation}
  \rho \to \pi \pi~.
\end{equation}
As with pions and nucleons, in an actual application, ``rho'' can then be
expanded to subsume all mesons that can decay into two pions.

As far as the interaction with photons is concerned, we now also need to
consider the photon-induced process,
\begin{equation}
  \gamma \rho \to \pi\pi
\end{equation}
as being on par with the $\gamma N \to \pi N$ reaction. Topologically, the
production current for this reaction has the structure depicted in
Fig.~\ref{fig:Mrho}, which is in complete analogy to the pion production of the
nucleon shown in Fig.~\ref{fig:Msutc} because both types of processes are based
on the interaction of the photon with a hadronic \emph{three\/}-point vertex.

\begin{figure}[t!]\centering
  \includegraphics[width=.7\columnwidth,clip=]{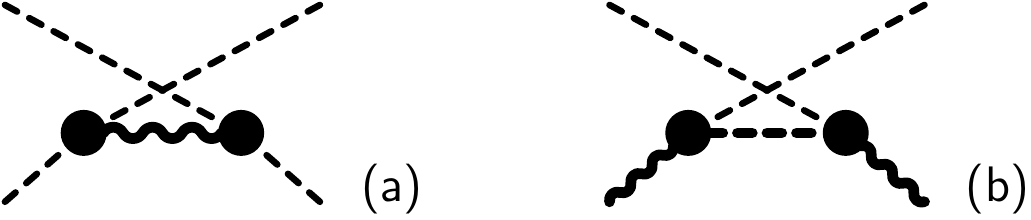}
  \caption{\label{fig:Upipi}%
Lowest-order non-polar $u$-channel exchanges for (a) $\pi\pi\to\pi\pi$ and (b) $\pi\rho\to\pi\rho$ scattering.}
\end{figure}

The hadronic final-state interaction of the $\pi\pi$ system for this process
can be depicted in a structure similar to Fig.~\ref{fig:piNScattering}, with
all external lines being pions and the primary interaction being given by the
$\pi\pi\rho$ vertex. Relevant for the following, in particular, is the fact
that one can also split the full $T$ matrix into a pole part and a non-pole
part $X$ whose lowest-order driving term is a $u$-channel exchange as depicted
in Fig.~\ref{fig:Upipi}(a). The same is true for any meson-meson scattering
problem whose basic interaction is described in terms of a bare three-meson
vertex. Figure~\ref{fig:Upipi}(b) shows the corresponding non-polar driving
terms for $\pi\rho\to\pi\rho$.

As we shall see, the details of the underlying meson-meson scattering problem
does not matter for the following. What matters is only the generic topological
structure of the production current shown in Fig.~\ref{fig:Mrho} and the fact
that non-polar contributions $X$ to the scattering amplitude satisfy a
Bethe-Salpeter-type equation of the generic structure given in
Eq.~(\ref{eq:BSX}) that is driven at lowest order by non-polar $u$-channel
exchanges, like the ones shown in Fig.~\ref{fig:Upipi}. All other details can
be left to be worked out in an actual application.

\section{Hadronic two-pion production}\label{sec:topology}

We now turn to the problem of the production of two pions off a nucleon. Before
looking at the photon-induced process, we first consider all possible
\emph{hadronic\/} transitions
\begin{equation}
  N \to \pi \pi N~,
  \label{eq:pipiN}
\end{equation}
including all possible dressing mechanisms. We will then derive the associated
photoproduction current by attaching the photon in all possible ways to the
dressed hadronic process. This is done in complete analogy to how the
single-pion-production current is obtained from the fully dressed $\pi NN$
vertex as visualized in Fig.~\ref{fig:Msutc}.

\begin{figure}[t!]\centering
  \includegraphics[width=.95\columnwidth,clip=]{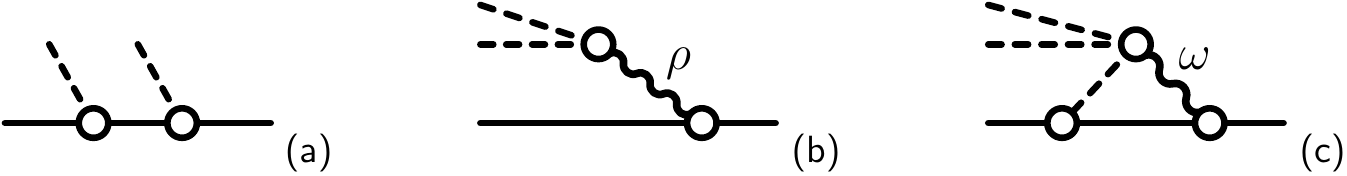}
  \caption{\label{fig:basicprocess}%
Basic two-pion production processes: (a) sequential production along the
nucleon line, and (b) intermediate production of a $\rho$ meson decaying into
two pions. Part (c) provides an example of another mechanism based on
intermediate multi-pion vertices. In this example, an $\omega$ meson produced
off the nucleon decays into three pions, with one pion subsequently being
absorbed by the nucleon. The mesons $\rho$ and $\omega$ here subsume any meson
having two-pion and three-pion decay modes, respectively.}
\end{figure}

Describing the process $N \to \pi \pi N$ within the generic field-theory
framework of pions, rho mesons, and nucleons, there are three basic interaction
vertices that are relatively easy to deal with, namely the
\emph{three\/}-hadron vertices $\pi NN$, $\pi\pi\rho$, and $\rho NN$. These
interactions provide the basic sequential production mechanisms shown in
Figs.~\ref{fig:basicprocess}(a) and \ref{fig:basicprocess}(b). However, there
exist also multi-pion processes where a meson decays into three or more pions
that cannot be resolved experimentally as being due to a sequence of
three-meson interactions. For the $\omega$ meson, for example, the dominant
decay mode is $\omega\to \pi^+\pi^0\pi^-$. Hence, one of the simplest examples
of two-meson production due to a four-meson interaction is depicted in
Fig.~\ref{fig:basicprocess}(c) showing an intermediate $\omega\pi\pi\pi$ vertex
where one of the pions is subsequently absorbed by the nucleon.

It should be clear that the full dynamical treatment of processes initiated by
the three-pion vertex requires at least a four-body treatment of the
intermediate $\pi\pi\pi N$ system. In general, any process initiated by an
$n$-pion meson vertex would require employing the dynamics of at least an
$(n+1)$-body system. Such treatments clearly are beyond the scope of what is at
present practically possible, and we will deal with this complication by, at
first, ignoring multi-pion vertices like the one depicted in
Fig.~\ref{fig:basicprocess}(c). We will restrict, therefore, the present
formulation to the three-body dynamics of the $\pi\pi N$ system that is
initiated by the two types of processes depicted in
Figs.~\ref{fig:basicprocess}(a) and \ref{fig:basicprocess}(b) based solely on
three-hadron interactions. As we shall see, this does \emph{not\/} exclude
incorporating processes initiated by multi-pion vertices like the one in
Fig.~\ref{fig:basicprocess}(c) at some later stage because all photo-processes
that can be related to independent hadronic production mechanisms can be
treated independently. Hence, we may safely ignore such multi-pion processes
now, and we will revisit the problem later, in Appendix~\ref{app:4meson}.

\begin{figure}[t!]\centering
  \includegraphics[width=1.0\columnwidth,clip=]{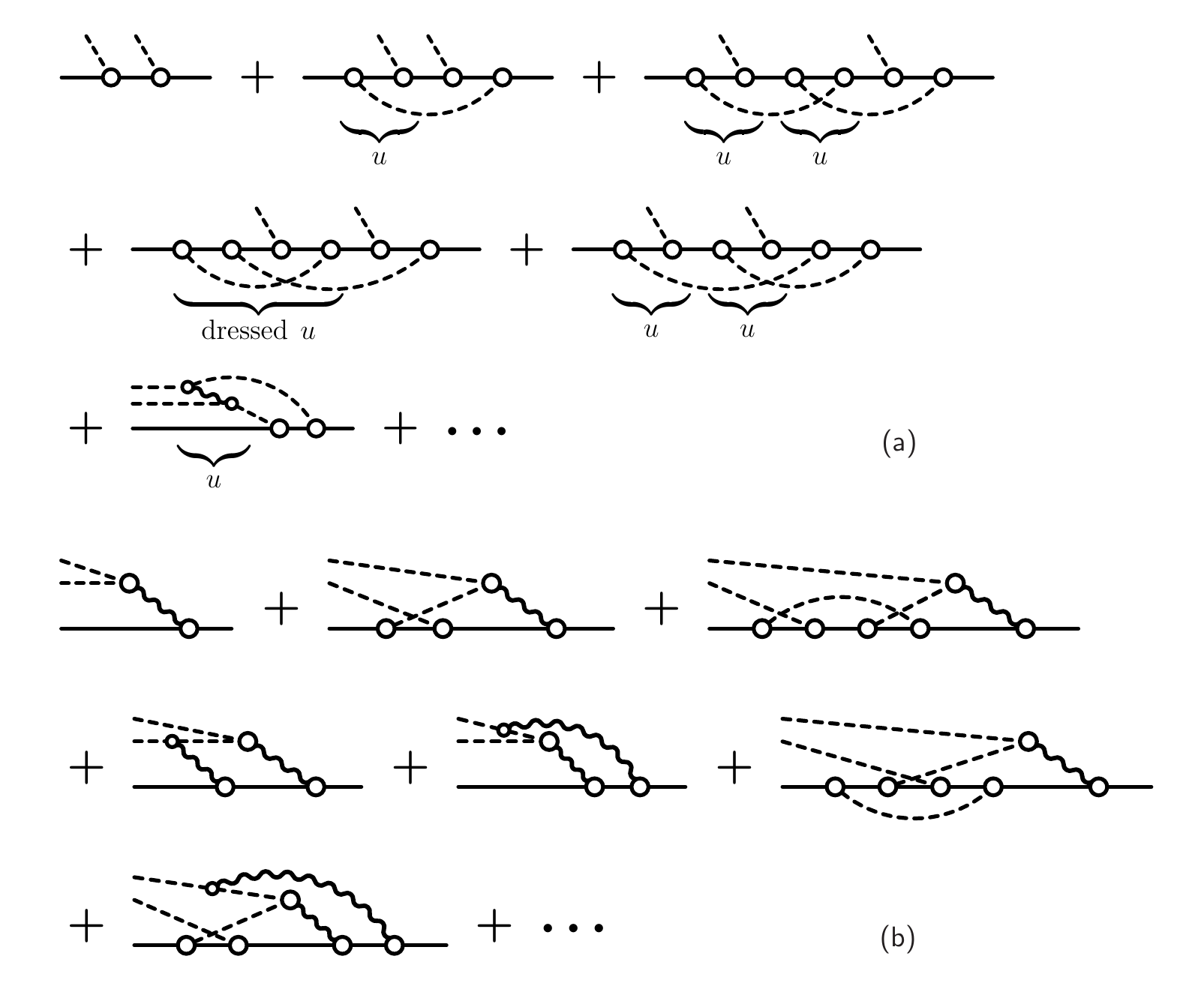}
  \caption{\label{fig:hadronsbare}%
Hadronic two-pion production processes (a) along a nucleon line and (b) via an
intermediate meson (wavy line) that can decay into two pions. Shown here for
both cases are only those bare graphs up to the two-pion-loop level that do not
contribute to the dressing of individual vertices, i.e., the loops shown here
always straddle at least two vertices. The braces under the diagrams for (a)
indicate basic $u$-channel-type exchanges. The $u$-channel exchange in the
fourth diagram is dressed by a pion loop, corresponding to the nonlinear loop
mechanism shown in the second diagram on the right-hand side of
Fig.~\ref{fig:uXchange}. The intermediate wavy line in the $u$-channel-type
exchange of the last diagram in (a) indicates a meson that can couple to two
pions. (Note that we could equally well interpret this as a $t$-channel
exchange. In fact, when symmetrizing the indistinguishable physical pions, both
types of exchanges are incorporated on an equal footing as a matter of course.)
}
\end{figure}

For the time being, therefore, we only consider the two basic  $N\to\pi\pi N$
processes initiated by the two bare transitions depicted in
Figs.~\ref{fig:basicprocess}(a) and \ref{fig:basicprocess}(b).
Figure~\ref{fig:hadronsbare} shows the first few terms of higher-order loop
corrections of the basic processes. In the figure, we have omitted all
contributions that can be subsumed in the dressing mechanisms of individual
three-point vertices. In other words, the diagrams shown in
Fig.~\ref{fig:hadronsbare} depict the first few contributions of the
multiple-scattering series describing the three-body final-state interaction
(FSI) within the $\pi\pi N$ system.

\begin{figure}[t!]\centering
  \includegraphics[width=1.0\columnwidth,clip=]{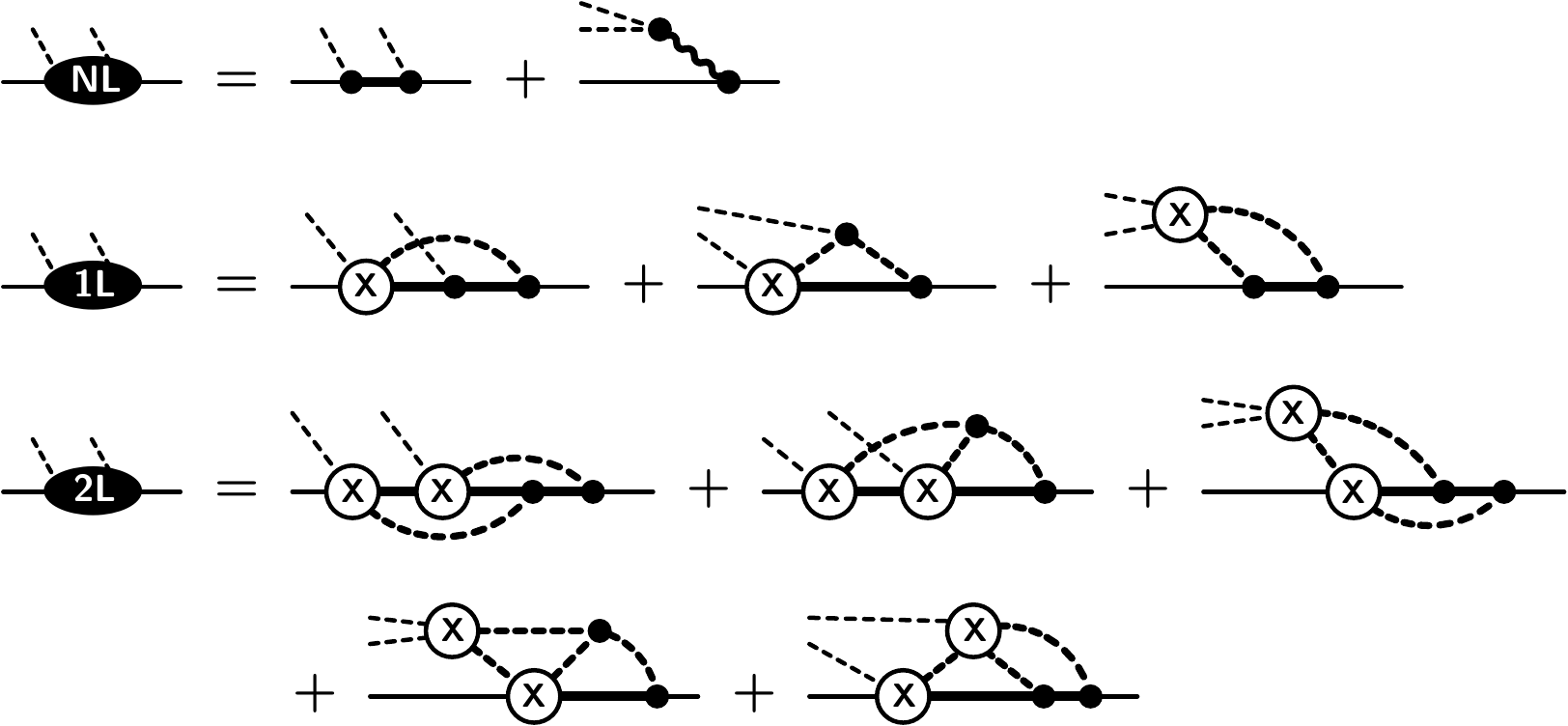}
  \caption{\label{fig:All2Pi}%
Grouping of hadronic two-pion production mechanisms off the nucleon involving
no loop (NL), one loop (1L), and two loops (2L). (Anticipating the outcome of
taking into account the symmetry of the indistinguishable pions, we do not
differentiate between diagrams that differ only by labeling the two pions.) The
thick interior lines subsume all particles permitted by the process with the
solid lines indicating baryons and the dashed lines mesons. The thick wavy line
stands for those mesons (like $\rho$, $\omega$, etc.) that can decay into two
pions (for intermediate mesons, such mesons are subsumed under the heavy dashed
line). Summations over all permitted internal particles are implied. All
vertices are fully dressed and various meson-baryon or meson-meson scattering
processes indicated by \textsf{X} are \emph{non-polar\/}, i.e., they \emph{do
not\/} contain $s$-channel driving terms because their contributions are
already subsumed in the full dressing of the vertices.}
\end{figure}

Inspecting the diagrams in Fig.~\ref{fig:hadronsbare} and noting that the
$u$-channel exchanges appearing there are the beginnings of the two-body
multiple-scattering series,
\begin{equation}
  X=U + UG_0 U +\cdots~,
\end{equation}
it is a simple exercise to sum up all contributions up to the level of two
\emph{dressed\/} loops, i.e., the internal particle propagators and vertices in
the resulting diagrams shown in Fig.~\ref{fig:All2Pi} are fully dressed, and
all meson-baryon and meson-meson FSI scattering processes are described by
\emph{non-polar\/} scattering matrices $X$ because all $s$-channel pole
contributions are accounted for in fully dressed sequential two-meson vertices.
In drawing Fig.~\ref{fig:All2Pi}, we have relaxed the restriction to nucleons,
pions, and rho mesons, and allowed the graphs to subsume all possible meson and
baryon states that may contribute to the process of $N\to \pi\pi N$. The
diagrams are grouped into no-loop (NL), one-loop (1L) and two-loop (2L)
contributions in increasing complexity of the hadronic final-state interactions
mediated by non-polar $X$ amplitudes.

%

We could now attach the photon to the hadronic diagrams in
Fig.~\ref{fig:All2Pi} and derive the corresponding production currents. The
explicit results given in Sec.~\ref{sec:expand} presumably will be sufficient
for most, if not all, practical purposes. For the fundamental theoretical
understanding of the process, however, it would be interesting to  derive a
closed-form expression for the entire two-pion photo-process similar to what is
possible for the single-pion production. And one would like to do so in a
manner that maintains gauge invariance. To this end, we note that after the
first interaction in the 1L graphs of Fig.~\ref{fig:All2Pi}, the $\pi\pi N$
system loses its memory about which of the two NL graphs of
Fig.~\ref{fig:All2Pi} was responsible for its initial creation and only retains
the memory about the last two-body interaction, i.e., whether it was a $\pi N$
or a $\pi\pi$ reaction. Ignoring for the moment nonlinear effects that allow
the creation of an arbitrary number of pions, all subsequent interactions,
therefore, are governed by the dynamics of a three-body system. [We add here
parenthetically that apart from the generic implications of a
Dyson-Schwinger-type framework which is tantamount to having infinitely many
mesons, the multi-pion aspect will also enter the picture through the
driving-term's nonlinearities discussed in the context of
Eq.~(\ref{eq:AGSdrive}); see also Fig.~\ref{fig:NLexample}.]

%

\subsection{Alt-Grassberger-Sandhas equations}\label{sec:Faddeev}

The solution of the non-relativistic quantum-mechanical three-body scattering
problem was given by Faddeev~\cite{Faddeev60,Faddeev}. One of the most decisive
aspects of the Faddeev approach is the manner in which the information about
the sequence of interactions percolates through the system such that all
interactions at all orders are possible, but double-counting of sequential
interactions within the same two-body subsystem of the three particles is
precluded, thus making the solutions unique. This basically is just an
``accounting'' problem and as such also valid in a relativistic context.%
\footnote{For relativistic Faddeev-type treatments of three-quark systems see,
e.g., Ref.~\cite{Eichmann2016} and references therein.} We may therefore
translate the structure of the Faddeev equations to the present problem by (1)
simply assuming covariant relativistic kinematics, (2) realizing that the
proper counterparts of the non-relativistic two-body $T$-matrices are the
corresponding non-polar scattering matrices $X$ because non-relativistic
potentials correspond to non-polar driving terms, and (3) allowing for
non-trivial nonlinearities of the type analogous to those for the $\pi N$
problem depicted in Fig.~\ref{fig:uXchange}.

The particular variant of the Faddeev approach we will use in the present work
are the Alt-Grassberger-Sandhas (AGS) equations~\cite{AGS67,Sandhas72} because
they are given in terms of transition operators that are symmetric in their
initial and final cluster configuration and thus can be applied to the present
problem requiring only minor modifications related to relativistic kinematics
and the fact that the particle number is not conserved.%
\footnote{The original Faddeev equations~\cite{Faddeev60,Faddeev}, by contrast,
  correspond to a Green's function description of the scattering process that
  contains unwanted disconnected contributions~\cite{Sandhas72} that need to be
  removed to be useful for the present context.}

First, to organize the information, we assume that the pions are
distinguishable and label them as $\pi_1$ and $\pi_2$. (The
indistinguishability of pions can easily be taken care of when calculating
observables by appropriately symmetrizing the amplitudes.) Accordingly, we
introduce three two-cluster channels $\alpha,\beta,\gamma=1,2,3$ by grouping
the three particles as
\begin{align}
  \text{``1''} &= (\pi_1^{} N,\pi_2^{})~,
   \nonumber\\
  \text{``2''} &= (\pi_2^{} N,\pi_1^{})~,
  \label{eq:FindexDef}
  \\
  \text{``3''} &= (\pi_1^{} \pi_2^{},N)~.
  \nonumber
\end{align}
Each ($2+1$) three-body configuration, therefore, consists of a two-body
subsystem and a single-particle spectator. The indices $\alpha$, $\beta$,
$\gamma$, etc.\ may also refer to the two-body subsystem of these two-cluster
configurations.

\begin{figure}[t!]\centering
  \includegraphics[width=\columnwidth,clip=]{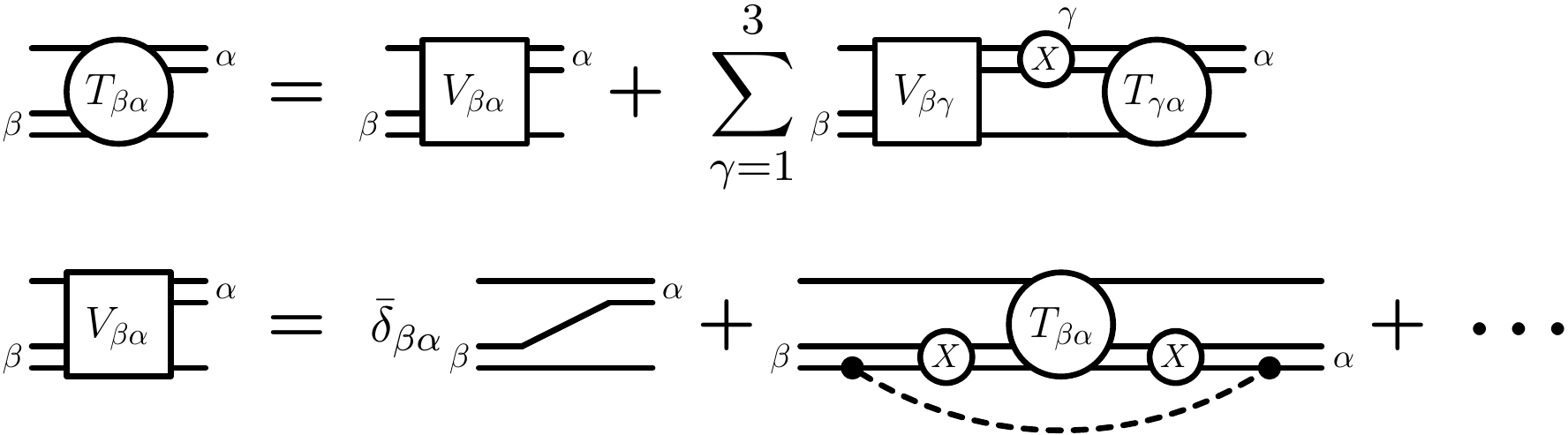}
  \caption{\label{fig:AGSeqs}%
Generic structure of the Faddeev-type AGS three-body equations (\ref{eq:AGS})
and its driving terms (\ref{eq:AGSdrive}). Depending on the cluster indices
$\alpha$, $\beta$, and $\gamma$ defined in Eq.~(\ref{eq:FindexDef}), two of the
horizontal lines depict pions and one the nucleon. The (dashed) meson loop
around $G_0 X_\beta G_0 T_{\beta\alpha} G_0X_\alpha G_0$ in the last diagram of
the bottom line provides one (of many) nonlinear contributions to the solution
(see also Fig.~\ref{fig:NLexample}). The nature of the meson depends on which
particles are connected by the loop.}
\end{figure}

The AGS equations~\cite{AGS67,Sandhas72} can be written within the present
context as
\begin{equation}
  T_{\beta\alpha} = V_{\beta\alpha} +\sum_{\gamma=1}^3 V_{\beta\gamma} G_0 X_\gamma G_0 T_{\gamma\alpha}~,
  \label{eq:AGS}
\end{equation}
with $\alpha,\beta,\gamma =1,2,3,$ where $T_{\beta\alpha}$ describes the
transition from an initial two-cluster configuration $\alpha$ to the final
configuration $\beta$. The equation is depicted in Fig.~\ref{fig:AGSeqs}. For
each two-body subsystem within the intermediate configurations $\gamma$, the
full interaction is given by the corresponding non-polar scattering matrix $X$
of the two-body subsystem of $\gamma$ that has to be extended into the
three-body space such that the propagation of the single spectator particle
within $\gamma$ is unaffected. Hence, we may write in a generic manner,
\begin{equation}
  X_\gamma = [X]_\gamma\circ t_{s,\gamma}^{-1}~,
  \label{eq:Xextend}
\end{equation}
where $[\cdots]_\gamma$ denotes the restriction to the two-body subspace within
the $\gamma$ cluster, and $t_{s,\gamma}$ is a generic notation for the
single-particle \emph{spectator} propagator within the $\gamma$ cluster. We
thus have
\begin{equation}
  G_0 X_\gamma G_0 = \left[G_0 X G_0\right]_{\gamma} \circ t_{s,\gamma}~,
\end{equation}
where $G_0$ on the left-hand side describes the free intermediate propagation
of the three particles within the $\pi\pi N$ system, i.e.,
\begin{equation}
  G_0 = \Delta_1(q_{\pi_1}^{}) \circ \Delta_2(q^{}_{\pi_2}) \circ S(p_N^{})~,
\end{equation}
which is the straightforward three-body extension of the two-body $G_0$ of
Eq.~(\ref{eq:G0=SoDelta}), whereas $G_0$ within the $[\cdots]_\gamma$ brackets
denotes the two-body restriction as given in Eq.~(\ref{eq:G0=SoDelta}). The
meaning of $G_0 X_\gamma G_0$ within the present three-body context, therefore,
is simply $[G_0 X G_0]_\gamma$ as a \emph{two\/}-body expression convoluted
with the spectator propagator $t_{s,\gamma}$ of the free third particle that is
unaffected by the two-body interaction $X$.

\begin{figure}[t!]\centering
  \includegraphics[width=\columnwidth,clip=]{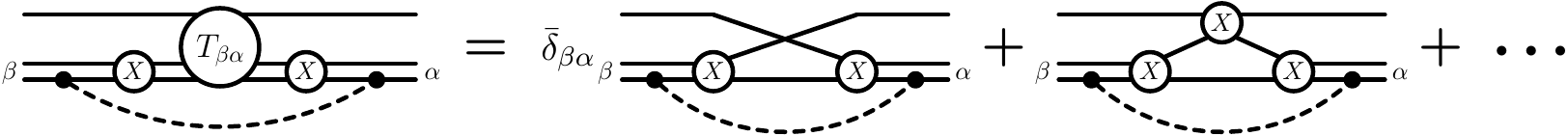}
  \caption{\label{fig:NLexample}%
First two lowest-order contributions of the nonlinear dressing
$N_{\beta\alpha}$ of the AGS driving term (\ref{eq:AGSdrive}). The loop is seen
here to straddle at least two $X$ amplitudes. Higher-order iterates of
$T_{\beta\alpha}$ produce loops around any number of $X$ amplitudes.}
\end{figure}

The driving terms $V_{\beta\alpha}$ of Eq.~(\ref{eq:AGS}) are given as
\begin{equation}
  V_{\beta\alpha} =  \bar{\delta}_{\beta\alpha} G^{-1}_0 +  N_{\beta\alpha}~,
  \label{eq:AGSdrive}
\end{equation}
where
\begin{equation}
  \bar{\delta}_{\beta\alpha} = 1-\delta_{\beta\alpha}
\end{equation}
is the anti-Kronecker symbol that vanishes if the initial and final two-body
groupings of the $\pi\pi N$ system are the same. The elements $N_{\beta\alpha}$
describe the \emph{nonlinear\/} dressing of $T_{\beta\alpha}$ in the manner
depicted in Fig.~\ref{fig:NLexample}, in analogy to the nonlinear $\pi N$
dressing mechanisms shown in Fig.~\ref{fig:uXchange}. It is crucial here that
this dressing happens around $G_0 X_\beta G_0 T_{\beta\alpha} G_0X_\alpha G_0$,
i.e., the loop particle must connect particles of the initial and final
two-body systems to avoid double-counting with the mechanisms described by
Fig.~\ref{fig:uXchange} or with higher-order iterations of $X_\gamma$
contributions. Nonlinearities, like $N_{\beta\alpha}$, are absent from the
original AGS equations~\cite{AGS67,Sandhas72} because they assume the particle
number to be conserved. For three-body processes involving pions, however,
terms like this one are necessary in principle (even if it is very difficult to
calculate in practice) because internally infinitely many pions may contribute.

We emphasize that there are limits to the three-body treatment of the $\pi\pi
N$ system even if one takes into account nonlinear dressings of the driving
terms of the kind shown in Fig.~\ref{fig:AGSeqs}. For example, if the loop
particle for the last graph in Fig.~\ref{fig:AGSeqs} is the nucleon, the AGS
amplitude enclosed by the loop is a three-\emph{meson\/} amplitude and thus
outside the scope of the three-body treatment of the $\pi \pi N$ system.
Moreover, in general, depending on how many mesons one considers to be created
at intermediate stages, much more complicated $N$-body-type nonlinearities will
result. It is possible in this way to  recover some of the complexities of the
problem associated with multi-pion vertices discussed in connection with the
mechanism of Fig.~\ref{fig:basicprocess}(c), for example. We will consider
additional three-body-force-type mechanisms associated with such processes in
more detail in Appendix~\ref{app:4meson}. In general, of course, the actual
calculation of such higher-order contributions in practical applications is
quite challenging, to say the least, and we will, therefore, limit the detailed
derivations in the following to the ``pure'' Faddeev contribution
$\bar{\delta}_{\beta\alpha}G_0^{-1}$, and only mention in passing the
ramifications of including nonlinearities in the driving term. Suffice it to
say that the present formulation is consistent and correct for the system of
two explicit pions and one nucleon where each of the particles may be fully
dressed by any mechanism compatible with three-body dynamics.

Before we implement the AGS approach for the present problem, it is convenient
to introduce a short-hand notation by defining operator-valued $3 \times 3$
matrices according to
\begin{subequations}
\begin{align}
  \bsf{T}_{\beta\alpha} &= T_{\beta\alpha}~,
  \\
  \bsf{V}_{\beta\alpha} &= V_{\beta\alpha}~,
  \\
  (\bsf{G}_{\bsf{0}})_{\beta\alpha}^{} &= \delta_{\beta\alpha} G_0 X_\alpha G_0~.
  \label{eq:G0BoldDefined}
\end{align}
\end{subequations}
This permits us to write the AGS equation~(\ref{eq:AGS}) as a matrix equation
in the form of
\begin{equation}
  \bsf{T} =\bsf{V}+\bsf{V}\bsf{G}_{\bsf{0}}\bsf{T}~,
  \label{eq:TmatLS}
\end{equation}
which has the familiar Lippmann-Schwinger (LS) form of all scattering problems.
Note in this context that the three-body dressing mechanism depicted in
Fig.~\ref{fig:NLexample} corresponds to the dressing of
$\bsf{G}_{\bsf{0}}\bsf{T}\bsf{G}_{\bsf{0}}$, i.e., exactly analogous to the
dressing of $G_0 XG_0$ depicted in the right-most diagram of
Fig.~\ref{fig:uXchange} for the two-body $\pi N$ problem.

\subsection{Three-body Faddeev treatment of hadronic two-pion production}

\begin{figure}[t!]\centering
  \includegraphics[width=.95\columnwidth,clip=]{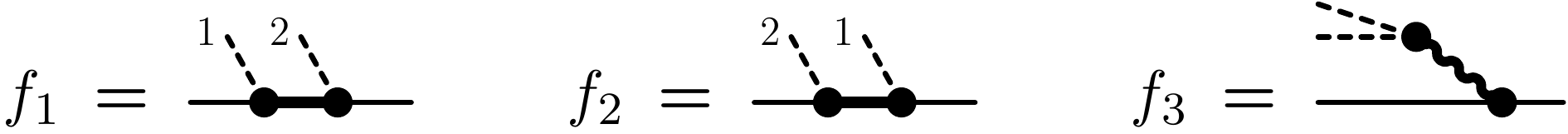}
  \caption{\label{fig:Fvertices}%
Definition of the basic $\pi\pi N$ vertices $f_\alpha$ assuming distinguishable
pions. The pion lines of the first two diagrams are labeled accordingly. The
cluster index $\alpha=1,2,3$ defined in Eq.~(\ref{eq:FindexDef}) describes the
hadron pair of the final three-point vertex in $f_\alpha$.}
\end{figure}

\begin{figure*}[!t]\centering
  \includegraphics[width=.8\textwidth,clip=]{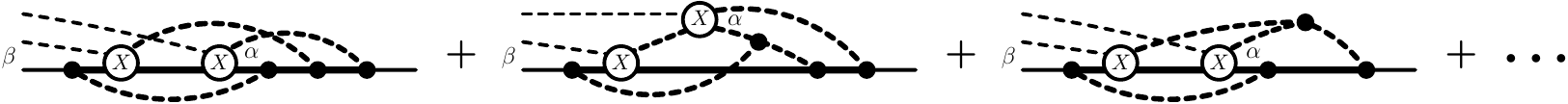}
  \caption{\label{fig:NLloops}%
Lowest-order nonlinear contributions $N_{\beta\alpha}G_0 f_\alpha$ employing
the mechanism of Fig.~\ref{fig:NLexample}. The internal meson lines (thick
dashes) depicts any meson compatible with the process. The loops may connect
any two particle respectively from the $\alpha$ and $\beta$ two-body systems,
i.e., each graph here represents only one example of four possible
contributions.}
\end{figure*}

Following the reasoning that the primary dynamics of the $\pi\pi N$ system
beyond the one-loop level is given by three-body dynamics, the
multiple-scattering series providing the final-state interactions within the
$\pi\pi N$ system can be summed up in terms of the three-body transition
operators $T_{\beta\alpha}$ of the AGS approach, and we immediately find that
the hadronic two-pion production can be described by three components $M_\beta$
($\beta=1,2,3$) given by
\begin{align}
M_\beta&=\sum_{\alpha}\left(\delta_{\beta\alpha}
+\sum_\gamma T_{\beta\gamma} G_0  X_\gamma G_0 \bar{\delta}_{\gamma\alpha}\right) f_\alpha
\nonumber\\&\qquad\mbox{}
+~\sum_\gamma\bigg(\delta_{\beta\gamma}
+ T_{\beta\gamma} G_0  X_\gamma G_0\bigg) \sum_{\alpha} N_{\gamma\alpha}G_0f_\alpha~,
\label{eq:MAGS}
\end{align}
where $f_\alpha$ describes the three basic production mechanisms shown in
Fig.~\ref{fig:Fvertices}. The second term here is only present because of
nonlinearities like those depicted in Fig.~\ref{fig:NLexample}; it is absent in
a standard three-body treatment. Expanding the right-hand side to second order
in $X_\gamma$ produces
\begin{align}
  M_\beta&=  f_\beta
+ \sum_{\gamma,\alpha} \bar{\delta}_{\beta\gamma} \bar{\delta}_{\gamma\alpha}   X_\gamma G_0 f_\alpha
\nonumber\\
&\qquad\mbox{}
+ \sum_{\gamma,\kappa,\alpha} \bar{\delta}_{\beta\gamma}\bar{\delta}_{\gamma\kappa} \bar{\delta}_{\kappa\alpha}
X_\gamma G_0 X_\kappa G_0 f_\alpha
\nonumber\\
&\qquad\mbox{}
+ \sum_{\alpha} N_{\beta\alpha}G_0f_\alpha
\cdots
\label{eq:Mexpand}
\end{align}
where the first three terms correspond precisely to the structure up to two
loops shown in Fig.~\ref{fig:All2Pi}, with the terms here corresponding to the
NL, 1L, and 2L graph groups of that figure. The lowest-order nonlinear effects
contained in the last explicit term here are of second order in $X_\gamma$,
like the preceding term, but they are of \emph{third\/} order in the (dressed)
loop structure, as shown in Fig.~\ref{fig:NLloops}.

Defining formal three-component vectors with elements
\begin{subequations}
\begin{align}
  \bsf{f}_\alpha &= f_\alpha~,
\\
  \bsf{F}_\beta &= \sum_\alpha \bar{\delta}_{\beta\alpha }f_\alpha~,
\\
  \tilde{\bsf{F}}_\beta &= \sum_\alpha N_{\beta\alpha } G_0 f_\alpha~,
\\
    \bsf{M}_\beta &= M_\beta~,
\end{align}
\end{subequations}
we may rewrite Eq.~(\ref{eq:MAGS}) as
\begin{equation}
  \bsf{M} = \left(\bsf{I} + \bsf{T} \bsf{G}_{\bsf{0}}\right) \bsf{F}
  +\left(\bsf{1} + \bsf{T} \bsf{G}_{\bsf{0}}\right) \tilde{\bsf{F}}
~,
  \label{eq:Mmat}
\end{equation}
where the matrix $\bsf{I}$ provides
\begin{equation}
  \bsf{I}\bsf{F}= \bsf{f}
  \qqtext{with}
    \bsf{I}_{\beta\alpha} = \tfrac{1}{2}-\delta_{\beta\alpha}~.
    \label{eq:Mmat-aux}
\end{equation}
One easily verifies that $\bsf{I}$ is indeed the inverse of the matrix whose
elements are the anti-Kronecker symbols. Here, $\bsf{1}$ is the unit matrix of
the three-body system with elements $\delta_{\beta\alpha}$ and
$\bsf{F}+\tilde{\bsf{F}}=\bsf{V}G_0 \bsf{f}$

In summary, the present description of the $N\to \pi\pi N$ process is given by
\begin{equation}
  \Fpipi = \sum_\beta M_\beta = \sum_\beta \bsf{M}_\beta~.
  \label{eq:Fpipi}
\end{equation}
The $\pi\pi NN$ ``vertex'' $\Fpipi$ constructed in this manner provides a
complete description of the reaction dynamics at the three-body level of the
dressed $\pi\pi N$ system (subject to the general limitations of three-body
dynamics discussed earlier).

\section{Attaching the Photon}\label{sec:photon}

Using the LSZ reduction, the full double-pion production current is given in
terms of the gauge derivative by
\begin{equation}
  \Mpp^\mu =-G_0^{-1} \left\{G_0 \Fpipi S\right\}^\mu S^{-1}~,
  \label{eq:2piCurrent-0}
\end{equation}
where $S$ describes the incoming nucleon propagator and
$G_0=\Delta_1\circ\Delta_2\circ S$ is the outgoing propagation of the free
$\pi\pi N$ system. Hence, we have
\begin{equation}
\Mpp^\mu =D^\mu G_0 \Fpipi + \Fpipi^\mu + \Fpipi S J_N^\mu~,
\label{eq:2piCurrent}
\end{equation}
where  $J_N^\mu$ describes the current of the incoming nucleon.
Here, $D^\mu$ is the three-body generalization of $d^\mu$ of Eq.~(\ref{eq:G0dmuG0}), \emph{viz.\/}
\begin{align}
  G_0D^\mu G_0
  &\equiv -\left\{G_0\right\}^\mu
  \nonumber\\[.5ex]
  &= (SJ^\mu_N S) \circ \Delta_1\circ \Delta_2
   +   S \circ (\Delta_1 J^\mu_{\pi_1}\Delta_1)\circ \Delta_2
  \nonumber\\[.5ex]
  &\quad\mbox{}
   +   S \circ \Delta_1 \circ (\Delta_2 J^\mu_{\pi_2}\Delta_2)~,
   \label{eq:G0DmuG0}
\end{align}
i.e., it subsumes the three currents of the outgoing legs analogous to what is
depicted in Fig.~\ref{fig:dmu} for the two-body case. The five-point
\emph{interaction current\/} $\Fpipi^\mu\equiv -\{\mathbb{F}\}^\mu$ contains
all mechanisms where the photon is attached to the \emph{interior\/} of the
hadronic two-pion production mechanisms given by Eq.~(\ref{eq:Fpipi}).

Then with
\begin{equation}
  k_\mu D^\mu = G_0^{-1} \hQ -\hQ G_0^{-1}
\end{equation}
and the  WTI of Eq.~(\ref{eq:WTIN}) for the nucleon current, the
four-divergence of $\Mpp^\mu$ reads
\begin{equation}
  k_\mu \Mpp^\mu = G_0^{-1} \hQ G_0 \Fpipi -\hQ \Fpipi +k_\mu \Fpipi^\mu + \Fpipi\hQ - \Fpipi S\hQ S^{-1}~,
\end{equation}
which shows that the four-divergence of the interaction current $\Fpipi^\mu$,
in analogy to Eq.~(\ref{eq:WTIintQ}),  must be given by
\begin{equation}
k_\mu \Fpipi^\mu =\hQ \Fpipi - \Fpipi\hQ
\label{eq:gWTIintM}
\end{equation}
to produce the generalized WTI,
\begin{equation}
  k_\mu \Mpp^\mu = G_0^{-1} \hQ G_0 \Fpipi  - \Fpipi S\hQ S^{-1}~.
  \label{eq:gWTIpipi}
\end{equation}
This provides a conserved current in the usual manner when all external hadrons
are on-shell. More explicit form of this result will be given later in
Eq.~(\ref{eq:gWTIL}).

\begin{figure}[t!]
  \includegraphics[width=.85\columnwidth,clip=]{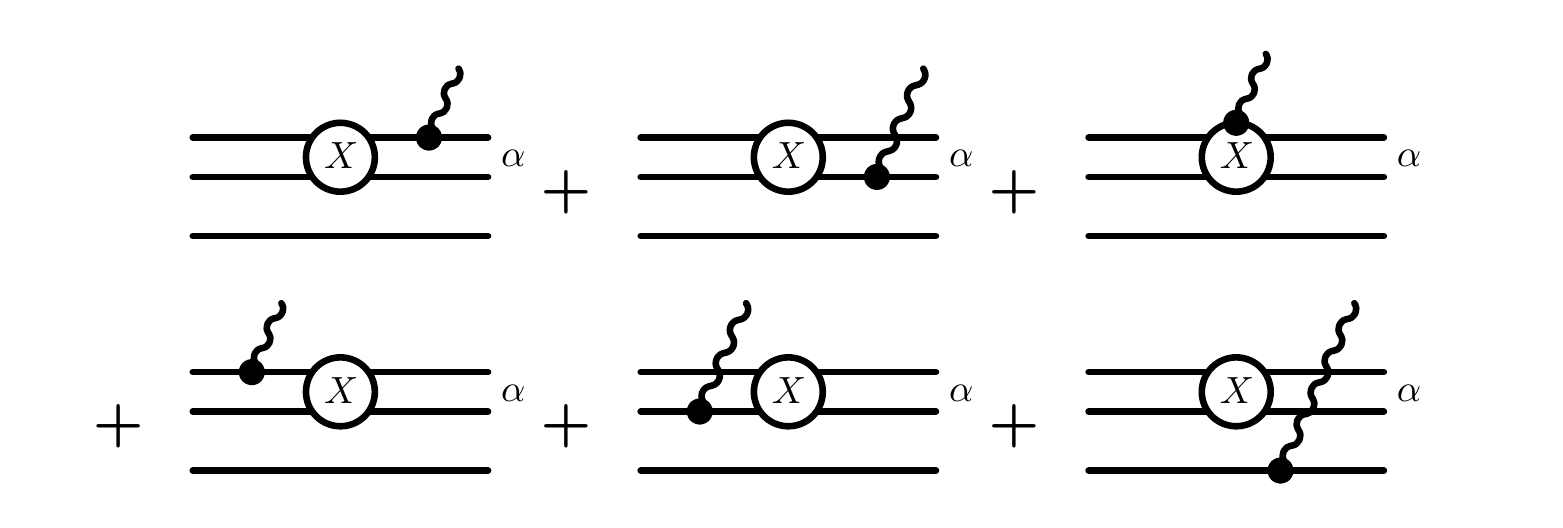}
  \caption{\label{fig:G0mu}%
Representation of the current $(\Gzero^\mu)_{\beta\alpha}$ of
Eq.~(\ref{eq:G0mu}). As in Fig.~\ref{fig:AGSeqs}, two of the horizontal lines
depict pions and one the nucleon. Note that the negative contribution of the
spectator current in Eq.~(\ref{eq:Xmu3}) cancels one  of the spectator
contributions of the two $D^\mu$ currents in Eq.~(\ref{eq:G0mu}), leaving only
one spectator current given as the last diagram here.}
\end{figure}

\subsection{Proof of gauge invariance}

To verify Eq.~(\ref{eq:gWTIintM}), let us define
\begin{equation}
  \bsf{M}^\mu \equiv -\left\{\bsf{M}\right\}^\mu
\end{equation}
as the vector whose components provide $\Fpipi^\mu$ according to
Eq.~(\ref{eq:Fpipi}) as
\begin{equation}
  \Fpipi^\mu = \sum_\beta \bsf{M}^\mu_\beta ~.
  \label{eq:FpipiMuSum}
\end{equation}
Taking the gauge derivative of the matrix relation (\ref{eq:Mmat}), the
interaction-current component vector is given as
\begin{align}
\bsf{M}^\mu &= \left(\bsf{I} + \bsf{T} \bsf{G}_{\bsf{0}}\right) \tilde{\bsf{F}}^\mu
  +\bsf{T}^\mu \bsf{G}_{\bsf{0}} \tilde{\bsf{F}}   +\bsf{T} \bsf{G}_{\bsf{0}}^\mu  \tilde{\bsf{F}}~,
\end{align}
where
\begin{align}
  \bsf{T}^\mu = \left(\bsf{1}+\bsf{T}\Gzero\right)\bsf{V}^\mu \left(\Gzero\bsf{T}+\bsf{1}\right)
  +\bsf{T}\Gzero^\mu\bsf{T}
\end{align}
is a straightforward consequence of applying the gauge derivative to the LS
equation~(\ref{eq:TmatLS}), in complete analogy to $X^\mu$ of
Eq.~(\ref{eq:Xmu}). Hence,
\begin{equation}
\bsf{M}^\mu = \left(\bsf{I} + \bsf{T} \bsf{G}_{\bsf{0}}\right) \tilde{\bsf{F}}^\mu
  +  \left(\bsf{1}+\bsf{T}\Gzero\right)\bsf{K}^\mu \left(\bsf{1}+\bsf{T}\Gzero\right)\tilde{\bsf{F}}~,
  \label{eq:MmuBold}
\end{equation}
where the elements of $\tilde{\bsf{F}}^\mu$,
\begin{equation}
  \tilde{\bsf{F}}^\mu_\beta  = \sum_\alpha \bar{\delta}_{\beta\alpha} \bsf{F}^\mu_\alpha~,
  \qtext{with}
  \bsf{F}^\mu_\alpha = -\left\{\bsf{F}_\alpha\right\}^\mu~,
\end{equation}
are the interaction currents associated with the elementary processes depicted
in Fig.~\ref{fig:Fvertices}, and
\begin{equation}
  \bsf{K}^\mu \equiv-\left\{\bsf{V}\Gzero\right\}^\mu = \bsf{V}^\mu\Gzero+\bsf{V}\Gzero^\mu
\end{equation}
is the current associated with the kernel of the LS equation (\ref{eq:TmatLS}).
The current matrix $\Gzero^\mu$ reads
\begin{equation}
(\Gzero^\mu)_{\beta\alpha} = \delta_{\beta\alpha} G_0 \big( D^\mu G_0 X_\alpha +  X^\mu_\alpha  +  X_\alpha G_0 D^\mu \big)G_0~,
\label{eq:G0mu}
\end{equation}
where, using Eq.~(\ref{eq:Xextend}), we obtain
\begin{align}
  X^\mu_\alpha
  &\equiv -\left\{X_\alpha\right\}^\mu
  \nonumber\\
  &=  [X^\mu]_\alpha\circ t_{s,\alpha}^{-1}
     -[X]_\alpha\circ J^\mu_{s,\alpha}~,
     \label{eq:Xmu3}
\end{align}
which is the three-body extension of the two-body interaction current
$[X^\mu]_\alpha$. The current of the spectator particle within the three-body
cluster $\alpha$ is represented by $J^\mu_{s,\alpha}\equiv
\{t_{s,\alpha}^{-1}\}^\mu$. The negative sign of this term is crucial for
avoiding double-counting of spectator contributions. In Eq.~(\ref{eq:G0mu}),
for example, it cancels out one of the spectator $D^\mu$ currents in
(\ref{eq:G0mu}), as shown in Fig.~\ref{fig:G0mu}.

\begin{figure}[t!]\centering
  \includegraphics[width=.95\columnwidth,clip=]{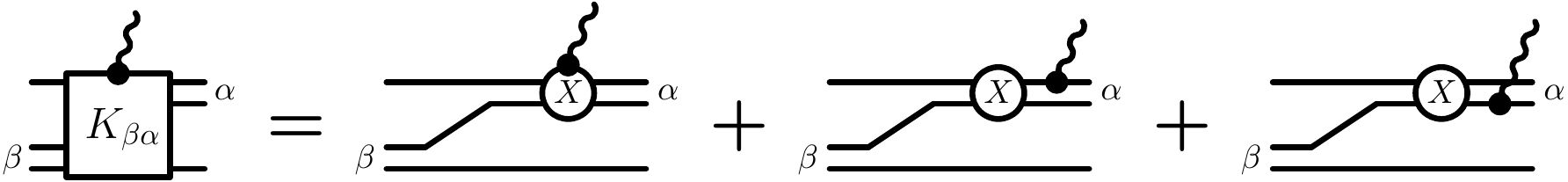}
  \caption{\label{fig:Kcurrent}%
Interaction-current matrix element $\bsf{K}^\mu_{\beta\alpha}$ of the kernel of
the AGS equation with explicit terms shown in the approximation of
Eq.~(\ref{eq:lowkernelC}), i.e., without nonlinear terms $N_{\beta\alpha}$. As
in Fig.~\ref{fig:AGSeqs}, two of the horizontal lines depict pions and one the
nucleon. In this approximation, due to the cancelation mechanism explained in
the caption of Fig.~\ref{fig:G0mu}, there is no current associated with the
spectator particle, i.e., only the first three diagrams of Fig.~\ref{fig:G0mu}
contribute.}
\end{figure}

In detail, the AGS-kernel matrix is given by
\begin{equation}
(\bsf{V}\Gzero)_{\beta\alpha} = \bar{\delta} _{\beta\alpha} X_\alpha G_0 + N_{\beta\alpha} G_0 X_\alpha G_0~.
\label{eq:VG0_0}
\end{equation}
If we neglect the nonlinearities $N_{\beta\alpha}$, we have
\begin{equation}
 \bsf{K}^\mu_{\beta\alpha} \to  \bsf{K}^\mu_{\beta\alpha} = \bsf{V}_{\beta\alpha} G_0 \big(X_\alpha^\mu G_0  + X_\alpha  G_0 D^\mu G_0\big)~.
 \label{eq:lowkernelCa}
\end{equation}
Using Eqs.~(\ref{eq:G0DmuG0}) and (\ref{eq:Xmu3}), we may write this as
\begin{equation}
 \bsf{K}^\mu_{\beta\alpha} \to  \bsf{K}^\mu_{\beta\alpha} = \bsf{V}_{\beta\alpha} G_0 \big[X^\mu G_0 + X  G_0d^\mu G_0\big]_\alpha~,
 \label{eq:lowkernelC}
\end{equation}
as shown in Fig.~\ref{fig:Kcurrent}. In this approximation, therefore, using
the known four-divergences of $X^\mu$ and $d^\mu$ given in Eqs.~(\ref{eq:kX})
and (\ref{eq:kd}), one immediately obtains
\begin{equation}
k_\mu \bsf{K}^\mu = \hQ \bsf{V}\Gzero-\bsf{V} \Gzero \hQ~.
\label{eq:kK}
\end{equation}
One may use here $\hQ$ for the entire three-body system, even though
Eqs.~(\ref{eq:kX}) and (\ref{eq:kd}) only contain the corresponding operator
for the two-body subsystem, because the spectator contribution of $\hQ$, along
the lower line on the right-hand side of Fig.~\ref{fig:Kcurrent}, cancels
between the two terms on the right-hand side of Eq.~(\ref{eq:kK}) since no
interaction takes place along that line. One can show that the
result~(\ref{eq:kK}) remains true even if the nonlinearities $N_{\beta\alpha}$
are taken into account. The proof requires tedious calculations and is not very
illuminating; it will be omitted here.

To evaluate the four-divergence of $\bsf{M}^\mu$, we use the
four-divergence~(\ref{eq:kK}) and the fact that the current
$\tilde{\bsf{F}}^\mu$ satisfies the generic relations of any interaction-type
current, i.e.,%
\footnote{This will be proved explicitly in the next section in the context of
Fig.~\ref{fig:MNL}. }
\begin{equation}
k_\mu\tilde{\bsf{F}}^\mu = \hQ\tilde{\bsf{F}}-\tilde{\bsf{F}}\hQ ,
 \label{eq:ktildeF}
\end{equation}
and then we easily find
\begin{equation}
k_\mu \bsf{M}^\mu
  =
  \hQ\bsf{M} -\bsf{M}\hQ~,
\end{equation}
which upon using Eq.~(\ref{eq:FpipiMuSum}) immediately verifies the validity of
Eq.~(\ref{eq:gWTIintM}) as stipulated. Hence, the current $\Mpp^\mu$ of
Eq.~(\ref{eq:2piCurrent}) constructed with the help of the hadronic mechanisms
(\ref{eq:Mmat}) is indeed (locally) gauge invariant, and its generalized WTI is
given by Eq.~(\ref{eq:gWTIpipi}).

We can now write down the closed-form equation,
\begin{equation}
\Mpp^\mu =\sum_\beta\left(D^\mu G_0 M_\beta + M_\beta^\mu + M_\beta S J_N^\mu\right)
\label{eq:2piCurrenta}
\end{equation}
for the full two-pion photoproduction current $\Mpp^\mu$, where
\begin{equation}
  M^\mu_\beta \equiv \bsf{M}^\mu_\beta
\end{equation}
is the two-body component of $\bsf{M}^\mu$ in Eq.~(\ref{eq:MmuBold}) that
contains the full three-body final-state interactions of the problem. For
practical applications, this presumes that the full two-pion production
mechanisms $M_\beta$ of Eq.~(\ref{eq:MAGS}) can be calculated. In view of their
complexity, this cannot be done easily in practice. One can show, however, that
one can expand the full current in contributions of increasing complexity,
similar to the NL, 1L, and 2L contributions in Fig.~\ref{fig:All2Pi}, which
satisfy \emph{independent\/}  WTIs of their own. Maintaining local gauge
invariance, therefore, is not predicated on being able to calculate the full
current $\Mpp^\mu$.

\subsection{Expanding the two-pion production current}\label{sec:expand}

To see how one may expand the full current, we define
\begin{equation}
  \bsf{M}_0 =  \bsf{F}
  \qtext{and} \bsf{M}_i = \left(\bsf{V}\Gzero\right)^i\tilde{\bsf{F}}
  \qtext{for} i=1,2,3,\ldots~,
\end{equation}
which implies, formally, that $\bsf{M}$ of Eq.~(\ref{eq:Mmat}) can be written as
\begin{equation}
  \bsf{M} =\sum_{i=0}^\infty \bsf{M}_i~.
  \label{eq:Mexpand1}
\end{equation}
Note that, without the nonlinearities $N_{\beta\alpha}$, the matrix elements of
the AGS kernel $\bsf{V}\Gzero$ are just given by $\bar{\delta}_{\beta\alpha}
X_\alpha G_0$, as seen from Eq.~(\ref{eq:VG0_0}). The expansion
(\ref{eq:Mexpand1}), therefore, provides the three-body multiple-scattering
series of the final-state interactions within the $\pi\pi N$ system as a
sequence of two-body interactions $X_\alpha$. One can show very easily, by the
same techniques used in verifying the gauge invariance of the full current
$\Mpp^\mu$ that the same is true order by order by coupling the photon to
$\bsf{M}_i$.

\begin{figure}[t!]\centering
  \includegraphics[width=\columnwidth,clip=]{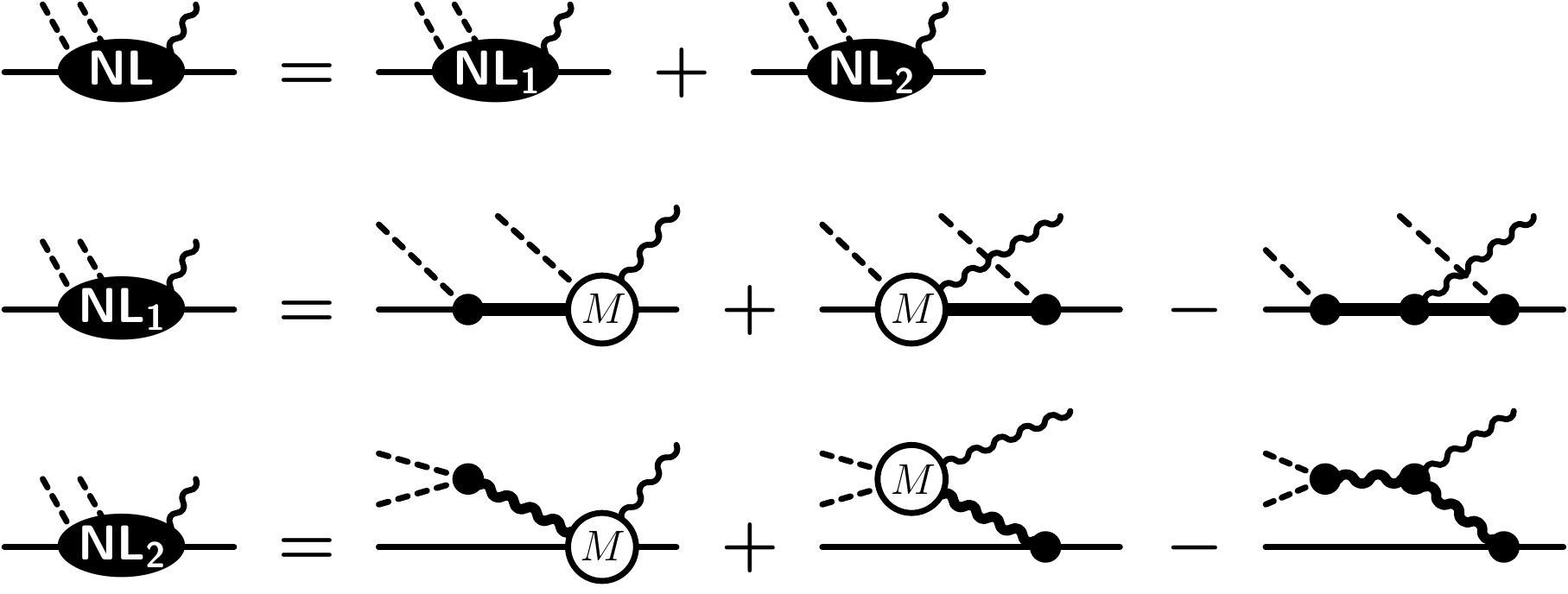}
  \caption{\label{fig:MNL}%
Two-pion photoproduction at the no-loop level where the photon is attached to
the NL diagrams of Fig.~\ref{fig:All2Pi}. The two contributions \LD{1} and
\LD{2} correspond to the two NL diagrams in  Fig.~\ref{fig:All2Pi} in the given
order. The photoproduction amplitudes labeled $M$ are comprised of four generic
terms each, similar to the pion-production case shown in Fig.~\ref{fig:Msutc}
or the $\pi\pi$-production of Fig.~\ref{fig:Mrho}. The subtractions correct the
double counting resulting from the photon being attached to the respective
intermediate particle in both preceding diagrams, i.e., when expanding all
amplitudes $M$, each group consists of seven topologically distinc diagrams
(see also Fig.~\ref{fig:NoLoopExplicit}). Each group of \LD{1} and \LD{2}
diagrams satisfies an \emph{independent\/} gauge-invariance constraint.}
\end{figure}

For the NL graphs of $\bsf{M}_0$, whose components are shown in
Fig.~\ref{fig:Fvertices}, the two-pion currents depicted in Fig.~\ref{fig:MNL}
are gauge invariant as a matter of course because the corresponding
\emph{gauge-invariant\/} subprocess currents indicated by $M$ in the diagrams
trivially add up to make each of the NL$_1$ and NL$_2$ currents in
Fig.~\ref{fig:MNL} gauge invariant \emph{separately\/}. This can be found
immediately by taking the four-divergence of each current. These are simple
examples for something which is generally true: Coupling the photon to
topologically \emph{independent\/} hadronic processes (like the two distinct
processes summed up in the NL contributions of Fig.~\ref{fig:All2Pi}) will
produce naturally \emph{independent\/} gauge-invariance constraints.
This means that \emph{each component} of $\bsf{M}_0=\bsf{F}$ is gauge invariant separately.%
\footnote{Note that Fig.~\ref{fig:MNL} only shows topologically different
currents, i.e., no distinction is made for graphs  that differ only by
numbering the pions.} Since the components of $\tilde{\bsf{F}}^\mu$ are given
by sums of NL currents, this also implies an explicit proof for the
gauge-invariance relation~(\ref{eq:ktildeF}).

To investigate the gauge invariance of higher-order currents, we only need to
look at the properties of the interaction-type currents
$\bsf{M}^\mu_i\equiv-\left\{\bsf{M}_i\right\}^\mu$ because the contributions
resulting from the four external legs of any $\gamma N\to \pi\pi N$ current are
trivial. We must show, therefore, that each of such currents satisfies a
constraint similar to Eq.~(\ref{eq:gWTIintM}). We write
\begin{equation}
  \bsf{W}\equiv (\bsf{V}\Gzero)^i
\end{equation}
and find for the current $\bsf{W}^\mu\equiv -\left\{(\bsf{V}\Gzero)^i\right\}^\mu$,
\begin{align}
  \bsf{W}^\mu&=\sum_{k=0}^{i-1}\left(\bsf{V}\Gzero\right)^k\bsf{K}^\mu \left(\bsf{V}\Gzero\right)^{i-1-k}~,
\end{align}
which gives its four-divergence as
\begin{align}
k_\mu\bsf{W}^\mu&=
\sum_{k=0}^{i-1}\left(\bsf{V}\Gzero\right)^k(\hQ\bsf{V}\Gzero-\bsf{V}\Gzero\hQ)
 \left(\bsf{V}\Gzero\right)^{i-1-k}
\nonumber\\
&=\hQ \bsf{W}-\bsf{W}\hQ~.
\end{align}
This indeed is the generic result for an interaction-type current.
With
\begin{equation}
\bsf{M}^\mu_i = \bsf{W}^\mu\tilde{\bsf{F}}+ \bsf{W}\tilde{\bsf{F}}^\mu~,
\end{equation}
we thus find
\begin{align}
  k_\mu\bsf{M}^\mu_i &= (\hQ\bsf{W}-\bsf{W}\hQ)\tilde{\bsf{F}}+\bsf{W}(\hQ\tilde{\bsf{F}}-\tilde{\bsf{F}}\hQ)
  \nonumber\\
  &= \hQ\bsf{M}_i-\bsf{M}_i\hQ ,
  \label{eq:QM-MQ}
\end{align}
which, once again, provides the generic gauge-invariance constraint for
interaction currents. In other words, in view of the trivial gauge-invariance
contributions from external legs, the current
\begin{equation}
\bsf{M}_{i,\pi\pi}^\mu =D^\mu G_0 \bsf{M}_i + \bsf{M}^\mu_i + \bsf{M}_iS J_N^\mu
\end{equation}
is also gauge invariant for each two-body component $\beta$ of this equation.

\begin{figure}[t!]\centering
  \includegraphics[width=\columnwidth,clip=]{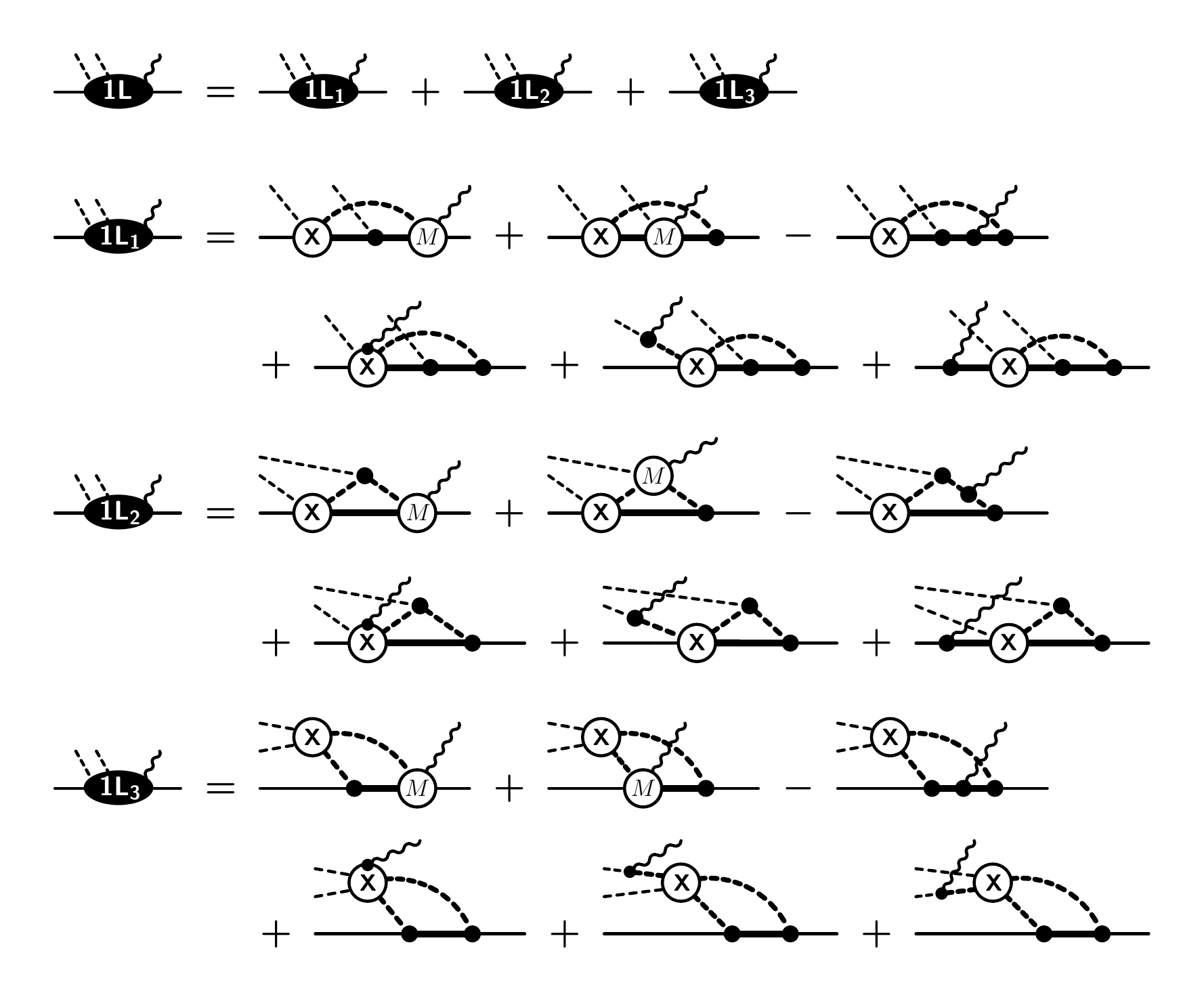}
  \caption{\label{fig:Loop1}%
Two-pion-production currents resulting from coupling the photon to the 1L
diagrams in Fig.~\ref{fig:All2Pi}. The subtractions correct double counting of
the corresponding mechanisms. Expanding each four-point current labeled $M$
into its generic four terms, all current groups comprise ten diagrams each.
Attaching the photon to the interior of $X$, as indicated in the respective
fourth diagram of each group, produces the five-point interaction current
$X^\mu$ detailed in Eq.~(\ref{eq:Xmu}) (see also Fig. 6 in
Ref.~\cite{Haberzettl97}). Each group \LD[1]{i} ($i=1,2,3$) obeys an
\emph{independent\/} gauge-invariance constraint. }
\end{figure}

\begin{figure}[t!]\centering
  \includegraphics[width=\columnwidth,clip=]{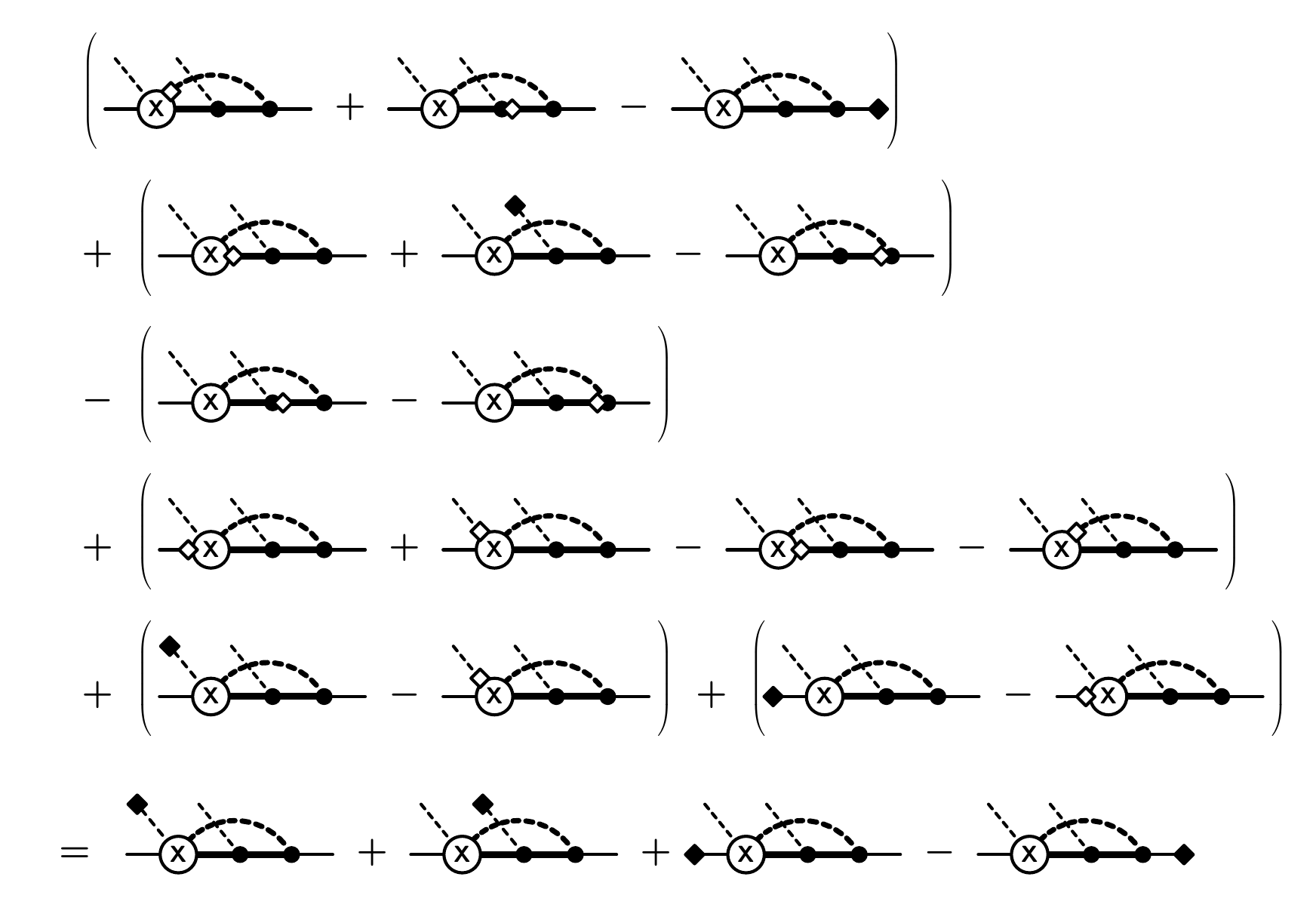}
  \caption{\label{fig:gaugeproof}%
Graphical proof of gauge invariance for the \LD[1]{1} current of
Fig.~\ref{fig:Loop1}. Each bracketed group above the equal sign corresponds to
the four-divergence of one of the six graphs of \LD[1]{1} in the same order as
they appear in Fig.~\ref{fig:Loop1}. In other words, the contents of each group
is the result of applying the appropriate WTI to the corresponding current. The
open diamond symbols indicate the action of the $\hQ$ charge operators and show
where the photon four-momentum $k$ needs to be injected into the hadronic
graphs so that the four external hadron momenta are exactly the same as for the
photo-process. If the diamond sits right next to a vertex or the amplitude $X$,
there is no propagator between the diamond and the vertex or amplitude along
that leg because it was canceled by the inverse propagators in the
corresponding Ward-Takahashi identities~(\ref{eq:WTIJQ}); see text. The solid
diamond symbols  at the ends of some external legs indicates that in addition
to a momentum $k$ being injected, there are residual \emph{inverse\/}
propagators with the four-momenta of the respective external particles that
vanish when taken on-shell. Hence, the resulting expression in the last line,
given explicitly in Eq.~(\ref{eq:gWTIL}), vanishes if all four external hadrons
are on-shell, thus providing a conserved current.}
\end{figure}

This consideration shows that attaching the photon in all possible ways to
\emph{any\/} topologically independent hadronic production process will provide
an \emph{independent\/} current that is constrained by its own
Ward-Takahashi-type identity. The two topologically independent NL processes
depicted in Fig.~\ref{fig:MNL} are among the simplest examples for this fact.
Figure~\ref{fig:Loop1} provides the currents resulting from attaching the
photon to the three 1L diagrams of Fig.~\ref{fig:All2Pi}. The three
\emph{independent\/} currents labeled 1L$_{i}$ ($i=1,2,3$) in
Fig.~\ref{fig:Loop1} must be gauge invariant separately. The corresponding
proofs are implied by the result found in Eq.~(\ref{eq:QM-MQ}). Nevertheless,
we shall prove gauge invariance for the example of the current 1L$_{1}$ in
Fig.~\ref{fig:Loop1} because it comprises contributions from single-particle
currents, single-meson production currents, and the five-point interaction
currents $X^\mu$ given in Eq.~(\ref{eq:Xmu}), and thus provides a non-trivial
explicit example of how the \emph{consistency\/} among all contributing current
mechanisms ensures gauge invariance of the entire process. The procedure is
most transparent in the graphical manner as shown in Fig.~\ref{fig:gaugeproof}.
Writing the underlying hadronic process, i.e., the first of the three 1L
diagrams in Fig.~\ref{fig:All2Pi}, as $H_{\pi\pi}$ and the corresponding
current as $H_{\pi\pi}^\mu$, its four-divergence can now simply be read from
the final line in Fig.~\ref{fig:gaugeproof} as
\begin{align}
  k_\mu H_{\pi\pi}^\mu &= \Delta_1^{-1}(q_1) Q_{\pi_1} \Delta_1 (q_1-k)\, H_{\pi\pi}^{(\pi_1)}
              \nonumber\\
              &\quad\mbox{}
              + \Delta_2^{-1}(q_2) Q_{\pi_2}\Delta_2 (q_2-k)\,  H_{\pi\pi}^{(\pi_2)}
              \nonumber\\
              &\quad\mbox{}
              +S^{-1}(p') Q_{N_f}S(p'-k) \, H_{\pi\pi}^{(N_f)}
              \nonumber\\
              &\quad\mbox{}
              - H_{\pi\pi}^{(N_i)}\,S(p+k) Q_{N_i}S^{-1}(p)~,
              \label{eq:gWTIL}
\end{align}
where the explicit four-momenta are those of the photo-process
\begin{equation}
  \gamma(k)+N_i(p) \to \pi_1^{}(q_1^{}) + \pi_2^{} (q_2^{}) + N_f(p')
  \label{eq:photomomenta}
\end{equation}
in a self-explanatory symbolic notation. The functions $H^{(h)}_{\pi\pi}$
describe the hadronic process $H_{\pi\pi}$ in the respective dynamical
situations of the four diagrams of the final line in Fig.~\ref{fig:gaugeproof},
i.e., $h=\pi_1$, $\pi_2$, $N_f$ indicate that, as compared to the momentum
dependence of the photo-process, the corresponding outgoing pion or nucleon leg
is \emph{reduced\/} by the photon momentum $k$, and for $h=N_i$, the initial
nucleon leg is \emph{increased\/} by $k$. The momenta at the four external
hadron legs of each $H^{(h)}_{\pi\pi}$, therefore, add up to conserve the
overall four-momentum, similar to the photo-process (\ref{eq:photomomenta}).
Equation~(\ref{eq:gWTIL}) is completely analogous to the generalized  WTI for
the single-pion photoproduction given in Eq.~(\ref{eq:gWTIpiN}).

The \emph{off-shell\/} result (\ref{eq:gWTIL}) is the appropriate generalized WTI
for \emph{any\/} two-pion production current resulting from a topologically
distinct hadronic two-pion production process $H_{\pi\pi}$. It is true for any
one of the hadronic processes depicted in Fig.~\ref{fig:All2Pi}, and it will
remain true for any one of the higher-order loop contributions.

Similar to the one-loop currents of Fig.~\ref{fig:Loop1}, one can now easily
derive as well the currents for the two-loop graphs in Fig.~\ref{fig:All2Pi}.
Each group of 13 current diagrams resulting from each of the two-loop hadron
graphs in Fig.~\ref{fig:All2Pi} is then gauge invariant separately. Moreover,
higher-order-loop contributions can be constructed by expanding the hadron
equation (\ref{eq:MAGS}) beyond what is given in Eq.~(\ref{eq:Mexpand}). In
general, each gauge-invariant group of graphs with $n$ loops consists of $7+3n$
members of which $5+n$ graphs contain a three-point current along a hadron line
and $2+2n$ contain a four-point current resulting from the photon interaction
with the interior of a three-hadron vertex. Each of these $n$-loop extensions
is straightforward and may be easily derived following the examples given here
explicitly. However, we expect that for most, if not all, practical purposes,
the NL and 1L currents of Figs.~\ref{fig:MNL} and \ref{fig:Loop1} may be
sufficient, and so we see no immediate need to go into more details here.

Before closing this section, we reiterate that in the formalism presented here,
nucleons, pions and rho mesons are to be understood as generic placeholders for
any and all baryonic or mesonic states compatible with the reaction in
question. In particular, all intermediate states must subsume all baryons and
mesons allowed for a particular reaction. This means that the nucleon lines in
the intermediate states in the diagrams in Figs.~\ref{fig:MNL} and
\ref{fig:Loop1} represent not only the nucleons but also any baryons that may
contribute to the process at hand, i.e., the baryon resonances. Also, the pion
as well as the $\rho$ meson lines appearing in the intermediate states in those
diagrams represent any meson that may contribute. In two-pion photoproduction,
for example, one of the relevant baryons in the intermediate states is the
$\Delta$ which couples strongly to $\pi N$. For the same process, the $\sigma$
meson should also be taken into account wherever the $\rho$ meson appears since
both mesons couple strongly to $\pi\pi$. Moreover, pure transverse
transition-current contributions such as due to the Wess-Zumino anomalous
couplings $\gamma\pi\rho$ and $\gamma\pi\omega$, which have no bearing on gauge
invariance, should also be included.

\section{Possible Approximation Scheme} \label{sec:approx}

The evaluation of the \emph{full\/} two-pion photoproduction amplitude as
derived in Sec.~\ref{sec:photon} is practically not feasible due to, in
particular, its nonlinear character. This calls for an approximation scheme to
make the problem tractable in practice. While there may be many ways to
approximate the full amplitude given by Eq.~(\ref{eq:2piCurrenta}), we would
like to advocate that --- as alluded to already --- a scheme that preserves the
increasing complexity of the reaction dynamics in terms of dressed loop
structures as presented in the no-loop and one-loop examples of
Figs.~\ref{fig:MNL} and \ref{fig:Loop1}, respectively, is best suited to
reflect the underlying physics. This loop expansion corresponds to an expansion
in powers of the two-body hadronic interaction $X_\gamma$. We know, of course,
that even at the levels of individual loops this is largely an intractable
problem if the loop ingredients are to be calculated completely because of,
again, the nonlinear dynamics of the required four-point interaction currents
for single-meson production~\cite{Haberzettl97} that enter the internal
reaction mechanisms of such loops. However, efficient approximation schemes
have been developed to deal with this complication at the four-point-function
level (see, for example, Refs.~\cite{Haberzettl97,HBMF98a,HNK06,HHN11,HDHH12},
and references therein). Because of its off-shell nature, the requirement of
\emph{local} gauge invariance, in particular, proved to be an invaluable tool
for helping link contributing dynamical mechanisms in a consistent manner (as
described in the Introduction for the example of $NN$
bremsstrahlung~\cite{NH09,HN10}). We can make use of the experience gained
there to treat the present five-point function dynamics of two-meson production
in a similar manner, by demanding that all approximate steps fully preserve
local gauge invariance as an off-shell constraint.

The (dressed) loop structure described in the previous section can be
enumerated in terms of a multiple-scattering series in powers of $X_\gamma$
according to Eqs.~(\ref{eq:MAGS}) and (\ref{eq:Mexpand}) for the underlying
hadronic $N\to \pi\pi N$ vertex $\Fpipi$ of Eq.~(\ref{eq:Fpipi}). Formally, we
may write
\begin{equation}
  \Fpipi  = \sum_{i=0}^\infty \Fpipi_i \ ,
   \label{eq:Mmat-expansion}
  \end{equation}
where the index $i$ enumerates the relevant powers of $X_\gamma$, resulting in
\begin{subequations}
\begin{align}
  \Fpipi_0  & \equiv \sum_\beta f_\beta \ ,  \\
  \Fpipi_1 & \equiv  \sum_\beta \Big( \sum_{\gamma,\alpha} \bar{\delta}_{\beta\gamma} \bar{\delta}_{\gamma\alpha}
  X_\gamma G_0 f_\alpha \Big) \ ,  \\
   \Fpipi_2 & \equiv  \sum_\beta \Big( \sum_{\gamma,\kappa,\alpha} \bar{\delta}_{\beta\gamma}\bar{\delta}_{\gamma\kappa}
   \bar{\delta}_{\kappa\alpha}   X_\gamma G_0 X_\kappa G_0 f_\alpha \nonumber \\
              &\quad\mbox{} \quad\mbox{} \quad\mbox{}
              + \sum_{\alpha} N_{\beta\alpha}G_0f_\alpha \Big) ~,
   \label{eq:Mmat-expansion-1}
  \end{align}
\end{subequations}
etc. The explicit expressions here correspond to the NL, 1L, and 2L
contributions depicted in Fig.~\ref{fig:All2Pi}, of course.

\subsection{Phenomenological hadronic contact vertex}\label{sec:PHCV}

In practice, we suggest to truncate the infinite sum (\ref{eq:Mmat-expansion})
at some order $n$,
\begin{equation}
  \Fpipi \approx \sum_{i=0}^n \Fpipi_i + \Cpipi~,
  \label{eq:Cdefined}
\end{equation}
and account for all higher orders by a remainder term\footnote{%
    For simplicity, we suppress the index $n$ for $\Cpipi$, in particular,
    since the form of the phenomenological ansatz for $\Cpipi$ employed here
    will be independent of $n$; only the fitted values of free parameters will
    depend on $n$.}
$\Cpipi$ that is to be constructed phenomenologically as a \emph{contact term}
(free of singularities) by making an ansatz modeled after the Dirac and isospin
structures of the full vertex $\Fpipi$.

To this end, we note that the most general (Dirac) structure of the full
reaction amplitude $\Fpipi$ for
\begin{equation}
N(p) \to \pi(q_1^{}) + \pi(q_2^{}) + N(p')~,
\label{eq:HadronicReactionProcess}
\end{equation}
where the arguments indicate the corresponding four-momenta, may be written as
\begin{align}
\Fpipi &= a_1^{} + a_2^{} \frac{\slashed{p}}{m} + a_3^{} \frac{\slashed{p}'}{m'}
+ a_4^{} \frac{\slashed{p}'\slashed{p}}{m'm} \nonumber \\
            &+ b_1^{}\frac{\slashed{q}}{m_\pi} + b_2^{} \frac{\slashed{q}\slashed{p}}{m_\pi m}
            + b_3^{} \frac{\slashed{p}'\slashed{q}}{m'm_\pi}
            + b_4^{} \frac{\slashed{p}'\slashed{q}\slashed{p}}{m'm_\pi m}  \ ,
\label{eq:HadrDiracStruc}
\end{align}
where $q \equiv q_1^{} - q_2^{}$ and the coefficients $a_i$ and $b_i$
$(i=1,2,3,4)$ are, in general, complex scalar functions of the momenta. Here,
$m$, $m'$, and $m_\pi$, respectively, stand for
the masses of the initial nucleon, final nucleon, and produced pion.%
  \footnote{These mass parameters are only needed to ensure that all coefficients
  have the same dimensions. Thus having one (average) pion
  mass parameter $m_\pi^{}$ does not preclude treating $\pi^\pm$ and $\pi^0$ as
  distinguishable with different physical masses.}

The most general structure of $\Fpipi$ in isospin space is
\begin{equation}
\Fpipi =  A \left(\hat{\pi}_1^{} \cdot \hat{\pi}_2^{} \right)
+ B \left(\hat{\pi}_1^{} \times \hat{\pi}_2^{} \right) \cdot \vec{\tau} \ ,
\label{eq:HadrIsoStruc}
\end{equation}
where $\hat{\pi}_i\ (i=1,2)$ denotes the outgoing pion fields in isospin space
and $\vec{\tau}$ is the usual Pauli (isospin) operator. The Dirac structures of
coefficients $A$ and $B$ here take the form given by
Eq.~(\ref{eq:HadrDiracStruc}).

Both the Dirac and isospin structures of the full amplitude $\Fpipi$ given by
Eqs.~(\ref{eq:HadrDiracStruc}) and (\ref{eq:HadrIsoStruc}) hold also for any
term $\Fpipi_i$ in Eq.~(\ref{eq:Mmat-expansion}), i.e., at any order in powers
of $X_\gamma$. This means, in particular, that the Dirac structure of $\Fpipi$
also carries over to the remainder term $\Cpipi$ independent of the truncation
order $n$. A natural phenomenological ansatz for $\Cpipi$, therefore, would be
to use the Dirac structure (\ref{eq:HadrDiracStruc}) and replace all eight
coefficients $a_i$, $b_i$ ($i=1,2,3,4$) by individual phenomenological form
factors with parameters that can be fitted to experimental data. The resulting
expressions are presented in Appendix~\ref{app:GenContactCurrent} for future
reference.

In the application given below, in Sec.~\ref{sec:appl}, we have pursued the
simpler approximation of introducing \textit{one\/} overall common form factor.
This approximate ansatz is described in the following.  Ignoring the isospin
structure  for now, we put
\begin{align}
  \Cpipi
     &=  \bigg(a_1^{} + a_2^{} \frac{\slashed{p}}{m} + a_3^{} \frac{\slashed{p}'}{m'}
     + a_4^{} \frac{\slashed{p}'\slashed{p}}{m'm}
            + b_1^{}\frac{\slashed{q}}{m_\pi}
\nonumber \\
     &\qquad\mbox{}
            + b_2^{} \frac{\slashed{q}\slashed{p}}{m_\pi m}
             + b_3^{} \frac{\slashed{p}'\slashed{q}}{m'm_\pi}
            + b_4^{} \frac{\slashed{p}'\slashed{q}\slashed{p}}{m'm_\pi m}\bigg) F~,
\label{eq:CpipiAnsatz}
\end{align}
where the coefficients $a_i$, $b_i$ ($i=1,2,3,4$) are now simple (complex) fit
constants (that may also parametrically depend on the Mandelstam variable $s$
because it is a constant for a given experiment). As described in
Appendix~\ref{appsec:ContCurrentGen}, the form factor
\begin{equation}
F = F(p'^2,q_1^2,q_2^2;p^2)
\label{eq:F4point}
\end{equation}
is a scalar function of the squared external four-momenta. We may take $F$ to
be normalized to unity when all particles are on their respective mass shells,
i.e.,
\begin{equation}
  F(m'^2,m_1^2,m_2^2;m^2)=1~,
  \label{eq:Fnormalization}
\end{equation}
where $m_1$ and $m_2$ are the physical masses of the two pions. The detailed
functional form of $F$ is irrelevant for now, but, in general, $F$ may contain
further fit parameters (see also Sec.~\ref{sec:appl}).

At this point a remark is in order. Although the analyticity and covariance of
the full reaction amplitude $\Fpipi$ is preserved in the contact approximation
for the higher-order loop contribution described above, unitarity is violated.
To maintain unitarity in any type of approximation requires the complex phase
structure of the reaction amplitude to be adjusted consistently as well. This
is a highly non-trivial issue and beyond the scope of the present work.

\begin{figure*}[t]\centering
  \includegraphics[width=0.8\textwidth]{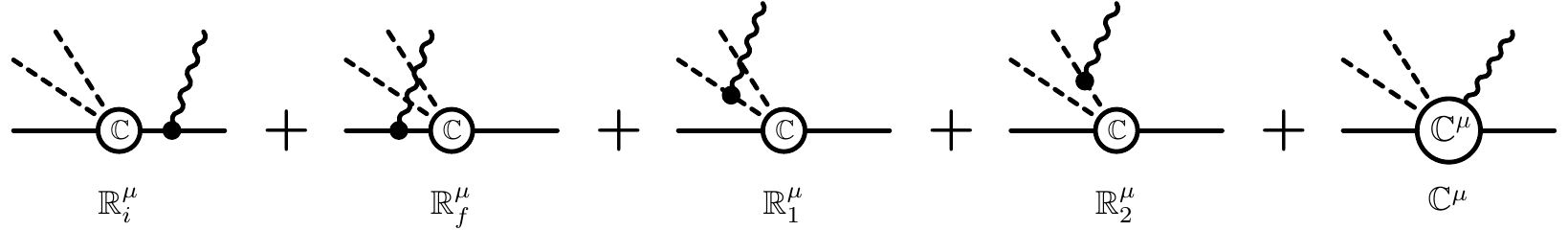}
  \caption{\label{fig:ContCurrent}%
Phenomenological current $\Rpipi^\mu$ given by Eq.~(\ref{eq:Rmu}) subsuming
higher-order loop contributions. The hadronic four-point vertex labeled
$\Cpipi$ is the contact term introduced by Eq.~(\ref{eq:Cdefined}), with its
phenomenological form given by (\ref{eq:CpipiAnsatz}), and the last diagram
depicts the five-point contact current $\Cpipi^\mu$, whose on-shell form is
given in Eq.~(\ref{eq:CmuAllonshell}).}
\end{figure*}

\subsection{Phenomenological current for higher-order loops}%
\label{sec:PhenContactCurrent}

The next step is to construct a two-pion production current $\Rpipi^\mu$ that
results from the mechanisms subsumed in $\Cpipi$. Using the loop expansion, the
full photoproduction amplitude of Eq.~(\ref{eq:2piCurrent-0}) may be written as
\begin{align}
M^\mu_{\pi\pi} &= \sum_{i=0}^\infty \left( - G_0^{-1}\{ G_0 \Fpipi_i S\}^\mu S^{-1} \right)
\nonumber\\
&\approx  \sum_{i=0}^n \left( - G_0^{-1}\{ G_0 \Fpipi_i S\}^\mu S^{-1} \right)+ \Rpipi^\mu
\nonumber\\
&=\sum_{i=0}^n \Mpipi^\mu_i + \Rpipi^\mu~,
\label{eq:Mpp-expansion}
\end{align}
where the sum over $\Mpipi^\mu_i$  subsumes two-meson production processes that
are to be treated explicitly, with two-pion production loops up to order $n$.
Lowest-order examples are the no-loop processes $\Mpipi^\mu_0$ of
Fig.~\ref{fig:MNL} and the one-loop processes $\Mpipi^\mu_1$ depicted in
Fig.~\ref{fig:Loop1}.

The approximate treatment of higher-order loops is provided by the remainder
current $\Rpipi^\mu$, which arises from coupling the photon to the
phenomenological hadronic contact term $\Cpipi$. In detail, one has
\begin{align}
\Rpipi^\mu &= -G_0^{-1}\{ G_0 \Cpipi S\}^\mu S^{-1}~,
\nonumber\\
 &=
   \Rpipi^\mu_i+\Rpipi^\mu_f
  + \Rpipi^\mu_1+ \Rpipi^\mu_2+\Cpipi^\mu  ~,
  \label{eq:Rmu}
\end{align}
as depicted in Fig.~\ref{fig:ContCurrent}. This is \emph{not} a contact current
since the first four contributions contain the polar contributions due to the
photon coupling to the initial ($i$) and final ($f$) baryons and the two
outgoing pions (1, 2) given by
\begin{subequations}
\begin{align}
\Rpipi^\mu_i   &= \Cpipi S J_{N_i}^\mu~,
\\
 \Rpipi^\mu_f+\Rpipi^\mu_1+ \Rpipi^\mu_2  &= D^\mu G_0 \Cpipi~,
\end{align}
\end{subequations}
where $D^\mu$ given by Eq.~(\ref{eq:G0DmuG0}) subsumes the currents for all
three outgoing hadrons.

The contact current $\Cpipi^\mu$ corresponding to the last diagram in
Fig.~\ref{fig:ContCurrent} is derived following the general procedure outlined
in Appendix~\ref{app:GenContactCurrent}. The corresponding phenomenological
current is given in Eq.~(\ref{eq:CmuAll}) in full detail for off-shell hadrons.
For the application of Sec.~\ref{sec:appl}, however, we only need the version
where all external hadrons are on shell; it reads
\begin{align}
  \Cpipi^\mu &= - e_i F_i \bigg[ \Big( a_2 + a_4 \Big)
  + \Big(b_2  + b_4^{}\Big)\frac{\slashed{q} }{m_\pi}\bigg]\frac{\gamma^\mu}{m}
    \nonumber\\
    &\quad\mbox{}
  - e_f^{} F_f \frac{\gamma^\mu}{m'}\bigg[ \Big(a_3  + a_4  \Big)
  +\Big(b_3 +b_4^{}\Big)\frac{ \slashed{q}}{m_\pi }\bigg]
  \nonumber\\
  &\quad\mbox{}
   -\big(e_1 F_1 -e_2 F_2\big) \big( b_1^{}+ b_2^{}
    + b_3^{}   + b_4^{} \big) \frac{\gamma^\mu}{m_\pi}
    \nonumber\\
    &\quad\mbox{}
  +\bigg[\Big(a_1^{}  +a_2^{}   + a_3^{}  + a_4^{} \Big)
            +  \frac{\slashed{q}}{m_\pi } \Big(b_1^{}  + b_2^{}  + b_3^{} + b_4^{}  \Big)\bigg]C^\mu_F~,
            \label{eq:CmuAllonshell}
\end{align}
where
\begin{subequations}
\begin{align}
  F_i & = F\big(m'^2,m_1^2,m_2^2;(p+k)^2\big)~,
\\
  F_f & = F\big((p'-k)^2,m_1^2,m_2^2;m^2\big)~,
\\
  F_1 & = F\big(m'^2,(q_1-k)^2,m_2^2;m^2\big)~,
\\
  F_2 & = F\big(m'^2,m_1^2,(q_2-k)^2;m^2\big)
\end{align}
\end{subequations}
accounts for kinematical situations with an intermediate off-shell hadron in
the first four diagrams of Fig.~\ref{fig:ContCurrent}. [Note that within the
present on-shell context, $F$ effectively is \textit{separable\/} in all four
squared-momentum contributions; cf.\ Eq.~(\ref{eq:Fseparable}).] As explained
in Appendix~\ref{app:GenContactCurrent}, the four Kroll-Ruderman-type terms
with $\gamma^\mu$ couplings --- one for each hadron leg --- arise from applying
the gauge derivative to the Dirac structure of $\Cpipi$. The auxiliary scalar
current $C^\mu_F$ is obtained by coupling the photon to the internal vertex
structure described by the form factor. Assuming $F$ to be normalized to unity,
according to (\ref{eq:Fnormalization}), the on-shell expression for $C^\mu_F$,
according to (\ref{eq:S+T}), is given as the manifestly nonsingular contact
current
\begin{align}
  C^\mu_F &=-e_1^{} (2q_1^{} - k)^\mu \frac{F_1-1}{(q_1^{}-k)^2-m_1^2}\, H_1
  \nonumber\\
  &\quad\mbox{}
  -e_2^{} (2q_2^{} - k)^\mu \frac{F_2-1}{(q_2^{}-k)^2-m_2^2} \; H_2
    \nonumber\\
  &\quad\mbox{}
  -e_f^{} (2p'-k)^\mu \frac{F_f-1}{(p'-k)^2-m'^2}\; H_f
  \nonumber\\
  &\quad\mbox{}
  -e_i^{} (2p+k)^\mu \frac{F_i-1}{(p+k)^2-m^2}\; H_i
  ~,
\label{eq:CauxFonshell}
\end{align}
where
\begin{align}
H_1 &=1-\left(1-\delta_2 F_2\right)\left(1-\delta_f F_f\right)\left(1-\delta_i F_i\right)~.
\label{eq:HfunctionOnshell}
\end{align}
The functions $H_2$, $H_f$, and $H_i$ are obtained from this expression by
cyclic permutation of indices $\{12fi\}$. The factors $\delta_x$ for
$x=1,2,f,i$ are unity if the corresponding particle carries charge; they are
zero otherwise.

The four-divergence of the contact current (\ref{eq:CmuAllonshell}) satisfies
\begin{align}
  k_\mu \Cpipi^\mu
  &= e_1^{} \Cpipi(p',q_1^{}-k,q_2^{};p)+e_2^{} \Cpipi(p',q_1^{},q_2^{}-k;p)
\nonumber\\
  &\quad\mbox{}
    +e_f^{} \Cpipi(p'-k,q_1^{},q_2^{};p)-e_i \Cpipi(p',q_1^{},q_2^{};p+k)~,
\label{eq:gWTIcontact}
\end{align}
which is the explicit version of the generalized  WTI (\ref{eq:gWTIintM}) for
the present case.

The contact current $\Cpipi^\mu$ thus provides a separate, independent
generalized  WTI for the entire remainder current $\Rpipi^\mu$, just like each
of the $i$-th order loop currents $\Mpipi_i^\mu$ in (\ref{eq:Mpp-expansion}),
as was shown in the preceding Sec.~\ref{sec:photon}. The present treatment,
therefore, remains fully \emph{locally\/} gauge invariant across all orders. Note
that by construction, the generic form of the hadronic contact term
$\Cpipi^\mu$ underlying the approximate current $\Rpipi^\mu$ remains the same
at all orders, however, the values of the corresponding free fit parameters
modeled after Eq.~(\ref{eq:HadrDiracStruc}) will change depending on how many
loop orders $\Mpipi_i^\mu$ are taken into account explicitly.

It should be emphasized in this context that the sole purpose of incorporating
the phenomenological remainder current $\Rpipi^\mu$ would be to provide an
approximate account of otherwise neglected higher-loop contributions. As such,
therefore, this current is not necessary for preserving gauge invariance and
could be omitted entirely (which presumably would be justified when the order
$n$ of explicit loop contributions is sufficiently high). However, if it is
incorporated, it must be made locally gauge invariant as described here.

\subsection{Lowest-order approximation}\label{sec:lowestorder}

\begin{figure}[t]\centering
  \includegraphics[width=0.99\columnwidth]{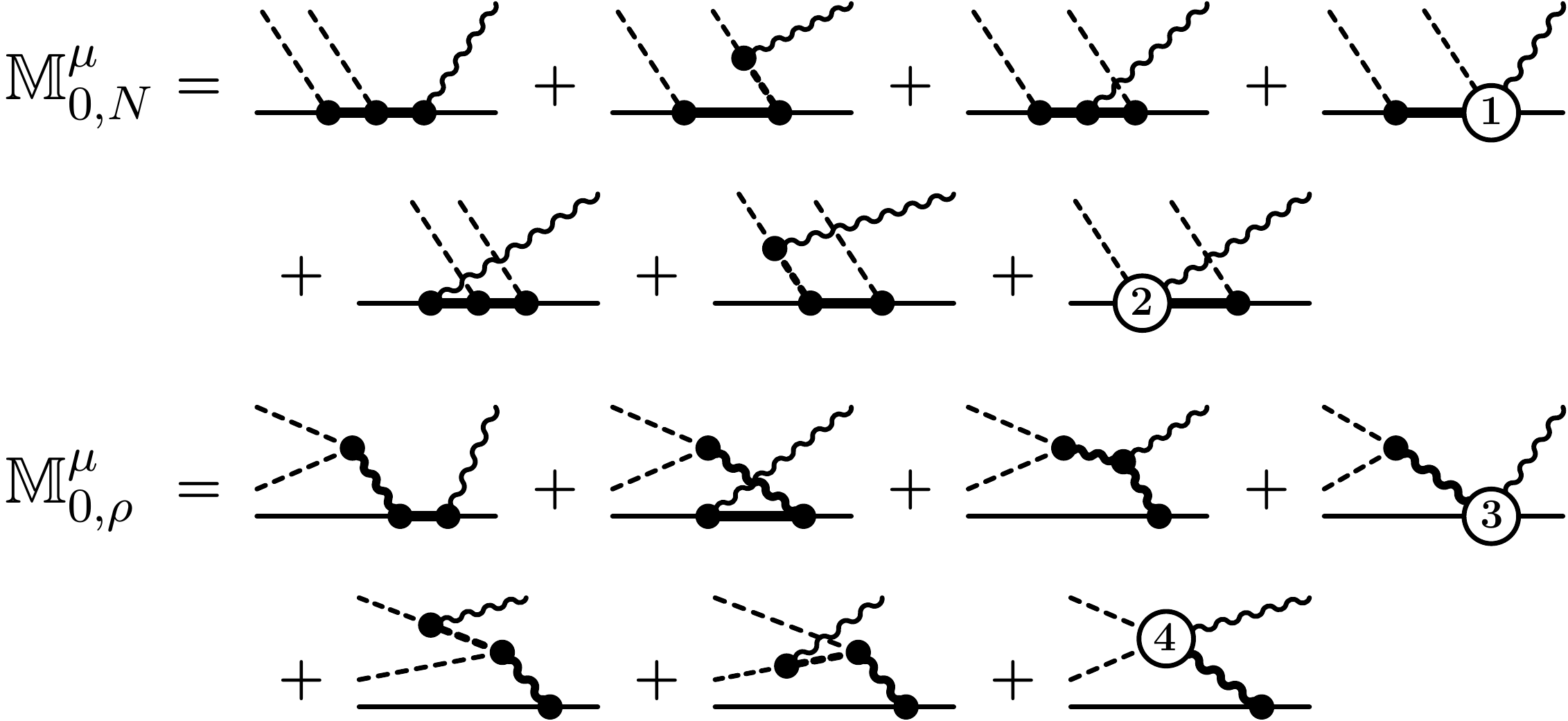}
  \caption{\label{fig:NoLoopExplicit}%
Explicit diagrams for two-pion photoproduction at the no-loop level,
corresponding to Eq.~(\ref{eq:Mnoloop}), providing a full account of the
topology inherent in the diagrams of Fig.~\ref{fig:MNL}. Internal thick lines
subsume hadrons compatible with the reaction. Labels 1--4 indicate contact-type
four-point currents. Depending on the level of sophistication, these diagrams
indicate microscopic interaction currents incorporating two-body final-state
interactions~\cite{HNK06} or simple phenomenological contact currents. Details
for the latter case are given in Appendix~\ref{appsec:4point}.  (For more
discussion, see Sec.~\ref{sec:lowestorder}.)}
\end{figure}

The lowest-order approximation of $\pi\pi$ photoproduction is given by
\begin{equation}
  M^\mu_{\pi\pi} \approx \Mpipi_0^\mu + \Rpipi^\mu~,
\label{eq:MnoloopwR}
\end{equation}
where
\begin{equation}
\Mpipi_0^\mu=\Mpipi_{0,N}^\mu+\Mpipi_{0,\rho}^\mu
\label{eq:Mnoloop}
\end{equation}
corresponds to the no-loop currents depicted in Fig.~\ref{fig:MNL} that
separates into two separately gauge-invariant contributions, depending on
whether the two pions are produced sequentially off the nucleon
($\Mpipi_{0,N}^\mu$) or the $\rho$ meson ($\Mpipi_{0,\rho}^\mu$), with each
mechanism breaking down into seven topologically distinct graphs as shown in
Fig.~\ref{fig:NoLoopExplicit}. Each group of seven diagrams respectively
corresponds to explicit renderings of the $\text{NL}_1$ and $\text{NL}_2$
diagrams of Fig.~\ref{fig:MNL}.

%
%

Hence, if all mechanisms depicted in these lowest-level diagrams are
implemented fully, this requires dressing all vertices and propagators
according to the description given in Sec.~\ref{sec:II.B} and, in particular,
it requires accounting for all two-body final-state interactions in terms of
contact-type four-point interaction currents (labeled by 1--4 in
Fig.~\ref{fig:NoLoopExplicit}) such that \textit{local\/} gauge invariance is
preserved fully. This corresponds to \emph{full\/} solutions of the underlying
$\gamma N \to \pi N$ and $\gamma \rho \to \pi\pi$ problems at a level of
sophistication that so far has never been undertaken because of the inherent
nonlinearities of these problems. At their most sophisticated, such two-body
subsystem dynamics are treated in linearized coupled-channel approaches that
account for dressing and final-state effects. The two-pion production
calculation reported in Ref.~\cite{KJLMS09}, for example, corresponds to such
an approximate treatment of the no-loop diagrams of
Fig.~\ref{fig:NoLoopExplicit} however, without properly accounting for gauge
invariance. Moreover, no attempt was made to account for higher-order loops,
thus effectively setting $\Rpipi^\mu=0$ in Eq. (\ref{eq:MnoloopwR}).

\subsubsection{Tree-level approximation}\label{sec:treelevel}

At its most elementary, one may interpret the diagrams in
Fig.~\ref{fig:NoLoopExplicit} as tree-level diagrams, with Feynman propagators
with physical masses, and vertices with physical coupling constants and
phenomenological cutoff functions, based on effective Lagrangians. This is
straightforward for the usual $s$-, $u$-, and $t$-channel diagrams of
single-meson-production dynamics corresponding to diagrams like the
correspondingly labeled ones from Figs.~\ref{fig:Msutc} and \ref{fig:Mrho}, for
example. In fact, this is an approximation widely used in the literature for
single-meson production. The preservation of local gauge invariance, however,
demands that the corresponding contact-type interaction currents (labeled 1--4
in Fig.~\ref{fig:NoLoopExplicit}) be constructed in a manner that preserves
local gauge invariance in terms of an \emph{off-shell\/} generalized  WTI. The
advantages of proceeding in this way are threefold. First, the underlying
single-meson production processes will of course be gauge invariant by
construction. Second, and crucial for the present application, the two-meson
production will be gauge invariant as well, \textit{without\/} any additional
work and the two contributing mechanisms $\Mpipi_{0,N}^\mu$ and
$\Mpipi_{0,\rho}^\mu$ will be gauge invariant \textit{separately\/}. Third, if
one ever wishes to undertake the calculation of \textit{three or more\/}
meson-production processes based on the same elementary interaction mechanisms,
the corresponding amplitudes \textit{will be gauge invariant as well\/}. In other
words, implementing local gauge invariance correctly at the lowest level will
carry through to all levels of more complex dynamical situations.

Approximate treatments of interaction currents in terms of contact currents
that preserve local gauge invariance have been suggested in
Ref.~\cite{HNK06,HHN11} and its variations have been used by a number of
authors (including the present ones) in the study of one-meson photoproduction
reactions~\cite{HDHH12,NOH08,HHN12}. Explicit forms for the present application
are given in Appendix~\ref{appsec:4point}.

We mention that the majority of existing two-meson photo- and electroproduction
models correspond to tree-level approximations of $\Mpipi^\mu_0$ of
Eq.~(\ref{eq:Mnoloop}) with some variations. None of them includes the
remainder current $\Rpipi^\mu$ and none preserve local gauge invariance, except
Refs.~\cite{NOH06,MON11}.

Before leaving this subsection, it should also be mentioned that while gauge
invariance, analyticity, and covariance of the two-meson photo- and
electroproduction are preserved in a tree-level approximation, unitarity is
violated. Note that the origin of this kind of unitarity violation is different
from that introduced by approximating the higher-order loop contributions of
the $N \to \pi \pi N$ hadronic amplitude by a contact interaction as described
in Sec.~\ref{sec:PHCV}.

\section{\boldmath Application to $\gamma N \to K K \Xi$} \label{sec:appl}

As a first application of the present formalism that will also allow us to
assess the effect of accounting for higher-order loop contributions in terms of
a phenomenological five-point current $\Rpipi^\mu$, we will calculate the
$\gamma N \to K K \Xi$ reaction in the no-loop approximation of
Eq.~(\ref{eq:MnoloopwR}), with tree-level approximations for $\Mpipi_0^\mu$ as
described in the preceding section. Here, we replace the produced two pions by
two kaons and the nucleon in the final state by the cascade particle $\Xi$.
For this particular reaction, the term equivalent to $\Mpipi_{0,\rho}^\mu$ in
Eq.~(\ref{eq:Mnoloop}) is absent since the exchanged meson (the analog of the
intermediate $\rho$ meson in Fig.~\ref{fig:NoLoopExplicit}) would need to have
strangeness quantum number $|S|=2$ and no such meson has been observed so far.
In summary, therefore, we employ the (approximate) description
\begin{equation}
M^\mu_{KK} = \Mpipi^\mu_Y + \Rpipi^\mu
\label{eq:KKphoto}
\end{equation}
for this process, where the topological structure of $\Mpipi^\mu_Y$ used here
is given by the $\Mpipi^\mu_{0,N}$ group of diagrams in
Fig.~\ref{fig:NoLoopExplicit}, with outgoing $\Xi$ baryon and intermediate
hyperons $Y$.

To model the tree-level approximation to $\Mpipi^\mu_Y$, we basically follow
Refs.~\cite{NOH06,MON11}. The contributing intermediate states of the diagrams
displayed in Fig.~\ref{fig:NoLoopExplicit}, in addition to kaons and
ground-state baryons, also subsume other relevant mesons and baryon resonances,
respectively. The four-point contact currents indicated by labels 1 and 2 in
Fig.~\ref{fig:NoLoopExplicit} used here are described in
Appendix~\ref{appsec:4point}. They differ from those employed in
Refs.~\cite{NOH06,MON11} by manifestly transverse contributions that do not
affect gauge invariance, constructed along the lines of
Eq.~(\ref{eq:addionalTmu}); in particular, see remarks below
Eq.~(\ref{eq:KRtadjust}).
For further details of the model for $\Mpipi^\mu_Y$, we refer to
Refs.~\cite{NOH06,MON11}. The main difference to those works is the inclusion
here of an overall five-point remainder current $\Rpipi^\mu$ to approximately
account for the effect of higher-order loops, as described in
Sec.~\ref{sec:PhenContactCurrent}. As a consequence, the free parameters of
$\Mpipi^\mu_Y$ of Refs.~\cite{NOH06,MON11} --- especially, the coupling
constants of the above-threshold resonances --- are readjusted here to
reproduce the $\Xi$ photoproduction data. The corresponding values are given in
Table~\ref{tbl:fmu0-tree}. All other parameter values are kept the same as
given in Ref.~\cite{MON11}.

Regarding the details for the remainder current $\Rpipi^\mu$, we note that the
isospin structure (\ref{eq:HadrIsoStruc}) for the underlying $N \to K K \Xi$
vertex has separate contributions for isospins $T=0,\,1$ of the $K\Xi$
subsystem in the final state. This isospin structure carries over  to the
contact current $\Cpipi^\mu$ of Eq.~(\ref{eq:CmuAllonshell}), resulting in
eight parameters $a^T_i$, $b^T_i$ ($i=1,\ldots,4$) for each isospin channel.
Allowing for a parametric dependence on $s$, we make the ansatz
\begin{subequations}\label{eq:s-dependent}
\begin{align}
a^T_i & = \tilde{a}^T_i \exp\left[-\alpha_a \left(\frac{s - s_0^{}}{2 m_N^{} \Lambda_S}\right)^2\right]
   \left(\frac{s - s_0^{}}{2m_N^{} \Lambda_S}\right)^2~,
\\
b^T_i & = \tilde{b}^T_i \exp\left[-\alpha_b \left(\frac{s - s_0^{}}{2m_N^{} \Lambda_S}\right)^2 \right]
   \left(\frac{s - s_0^{}}{2m_N^{} \Lambda_S}\right)^2~,
\end{align}
\end{subequations}
where $s \equiv (k + p)^2$ is the invariant mass of the reaction and $s_0^{}
\equiv (2m_K + m_\Xi)^2$. The scale parameter $\Lambda_S$ is fixed at
$\Lambda_S = 1$\,GeV. The constants $\tilde{a}^T_i$ and $\tilde{b}^T_i$, as
well as $\alpha_j$ ($j=a, b$), are fit constants.  For simplicity, we set
$\alpha_a^{} = \alpha_b^{}$, and take $\tilde{a}_i$ and $\tilde{b}_i$ to be real in
the present work. Furthermore, since at present there are data available only
for the single charged channel corresponding to $\gamma p \to K^+ K^+ \Xi^-$,
we set $a_i^1 = b_i^1 = 0$ in the present application.  These simplifications
seem to be sufficient to reproduce the presently available data. The values
obtained in our fits are given in Table~\ref{tbl:fcontactampl}.

%
\begin{table}[t]\centering
\caption{\label{tbl:fmu0-tree} Adjusted parameter values entering
$\Mpipi_Y^\mu$ of Eq.~(\ref{eq:KKphoto}). All other parameter values are
kept the same as given in Ref.~\cite{MON11}.} \bgroup
\renewcommand{\arraystretch}{1.3}
\begin{tabular}{ld{2.4}}
\hline\hline
\multicolumn{1}{l}{Product of coupling constants\qquad\qquad{}}    & \multicolumn{1}{l}{Values}     \\
\hline
\multicolumn{1}{l}{$g_{N\Lambda K}g_{\Xi\Lambda K}$ for $\Lambda(1890)3/2^+$} &   1.4  \\
\multicolumn{1}{l}{$g_{N\Sigma K}g_{\Xi\Sigma K}$ for $\Sigma(2250)5/2^-$}    &   0.06 \\
\multicolumn{1}{l}{$g_{N\Sigma K}g_{\Xi\Sigma K}$ for $\Sigma(2030)7/2^+$}    &   1.9 \\
\hline\hline
\end{tabular}
\egroup
\end{table}
%
%

%
\begin{table}[b!]\centering
\caption{\label{tbl:fcontactampl}    Fitted values of parameters
$\tilde{a}_i^T$ and $\tilde{b}_i^T$ appearing in
Eq.~(\ref{eq:s-dependent})  for isospin channels $T=0,1$ of the outgoing $K\Xi$
subsystem. In this work, we set $\tilde{a}_i^1 = \tilde{b}_i^1 = 0$ since only a
single charged channel ($\gamma p \to K^+ K^+ \Xi^-$) is considered.  Also,
$\alpha_a^{} = \alpha_b^{} = 2.4462$.}
 \bgroup
\renewcommand{\arraystretch}{1.2}
\begin{tabular}{@{\qquad}c@{\qquad}d{2.4}@{\qquad}d{2.4}@{\qquad}}
\hline\hline
\multicolumn{1}{c}{\raisebox{2.7ex}{}$i$} & \multicolumn{1}{c}{$\tilde{a}^0_i$~[fm]\qquad\qquad{}} &  \multicolumn{1}{c}{$\tilde{b}^0_i$~[fm]\qquad\qquad{}}
\\[.5ex]
 \hline
1  &    8.4268 &   1.8314   \\
2  &    1.5403 &  -1.6502   \\
3  &   -2.3021 &   4.0460   \\
4  &   -0.0014 &   1.4192   \\
\hline\hline
\end{tabular}
\egroup
\end{table}
%
%

For the present application, the single form factor $F$ we choose here for the
hadronic $NKK\Xi$ vertex according to Eq.~(\ref{eq:CpipiAnsatz}) is only needed
in the context of having all external hadrons for the diagrams in
Fig.~\ref{fig:ContCurrent} on their respective mass shells. Since for any of
the corresponding kinematic situations only one of the intermediate hadrons is
off-shell, effectively, without lack of generality, we may write the form
factor as a separable product of functions in the form
\begin{equation}
  F= f_\Xi(p'^2) f_K(q_1^2)  f_K(q_2^2) f_N(p^2)~.
  \label{eq:Fseparable}
\end{equation}
Any form factor $F(p'^2,q_1^2,q_2^2;p^2)$ can be reduced to this effective form
within the present on-shell context. Following Refs.~\cite{NOH06,MON11}, we
employ
\begin{subequations}\label{eq:expoFF}
\begin{align}
  f_K(q^2)& ={\frac{\Lambda_K^2 - m_K^2}{\Lambda_K^2 - q^2 }} ~,
  \\
  f_x(p^2)& =\frac{ \Lambda_x^4}{ \Lambda_x^4 + (p^2 - m_x^2)^2}
\end{align}
\end{subequations}
for the form factors entering the current $\Rpipi^\mu$, where $x=\Xi,\,N$, with
associated masses $m_x$. As cutoff parameters, we choose
$\Lambda_\Xi=\Lambda_N=900$\,MeV and $\Lambda_K=1500$\,MeV.

\begin{figure}[t!]\centering
  \includegraphics[width=.95\columnwidth]{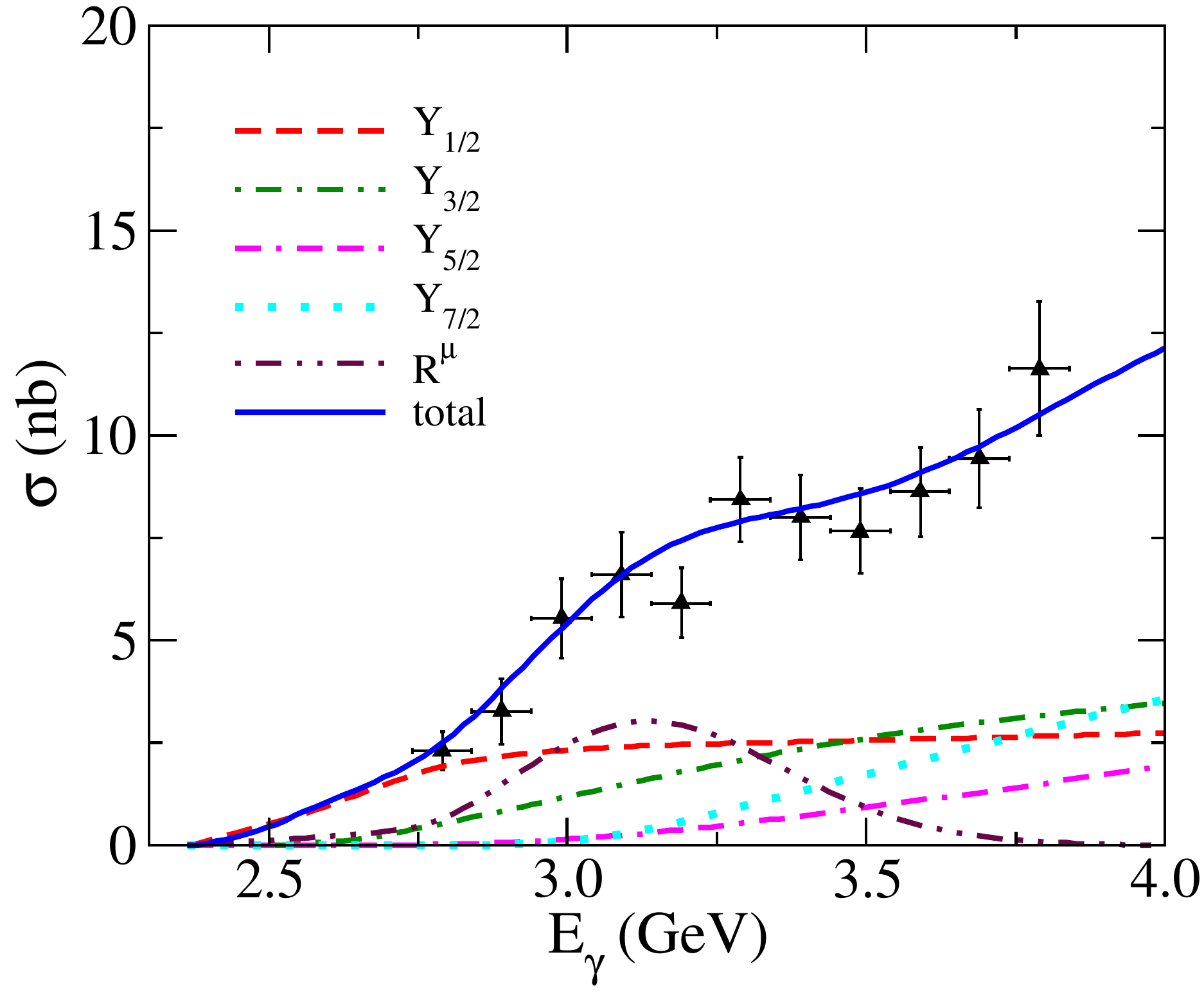}
  \caption{\label{fig:txsc}%
Total cross section for $\gamma p \to K^+ K^+ \Xi^-$ as a function of incident
photon energy. The dashed (red) line corresponds to the spin-1/2 hyperons
contribution; the dash-dotted (green) line to the spin-3/2 hyperons; the
long-dashed (magenta) to spin-5/2 and dotted (cyan) to spin-7/2 resonance
contributions. The long-dash-double-dotted (maroon) line corresponds to the
phenomenological five-point current ($\Rpipi^\mu$) contribution. Data are from Ref.~\cite{CLAS07b}.}
\end{figure}

\begin{figure}[t]\centering
  \includegraphics[width=.95\columnwidth]{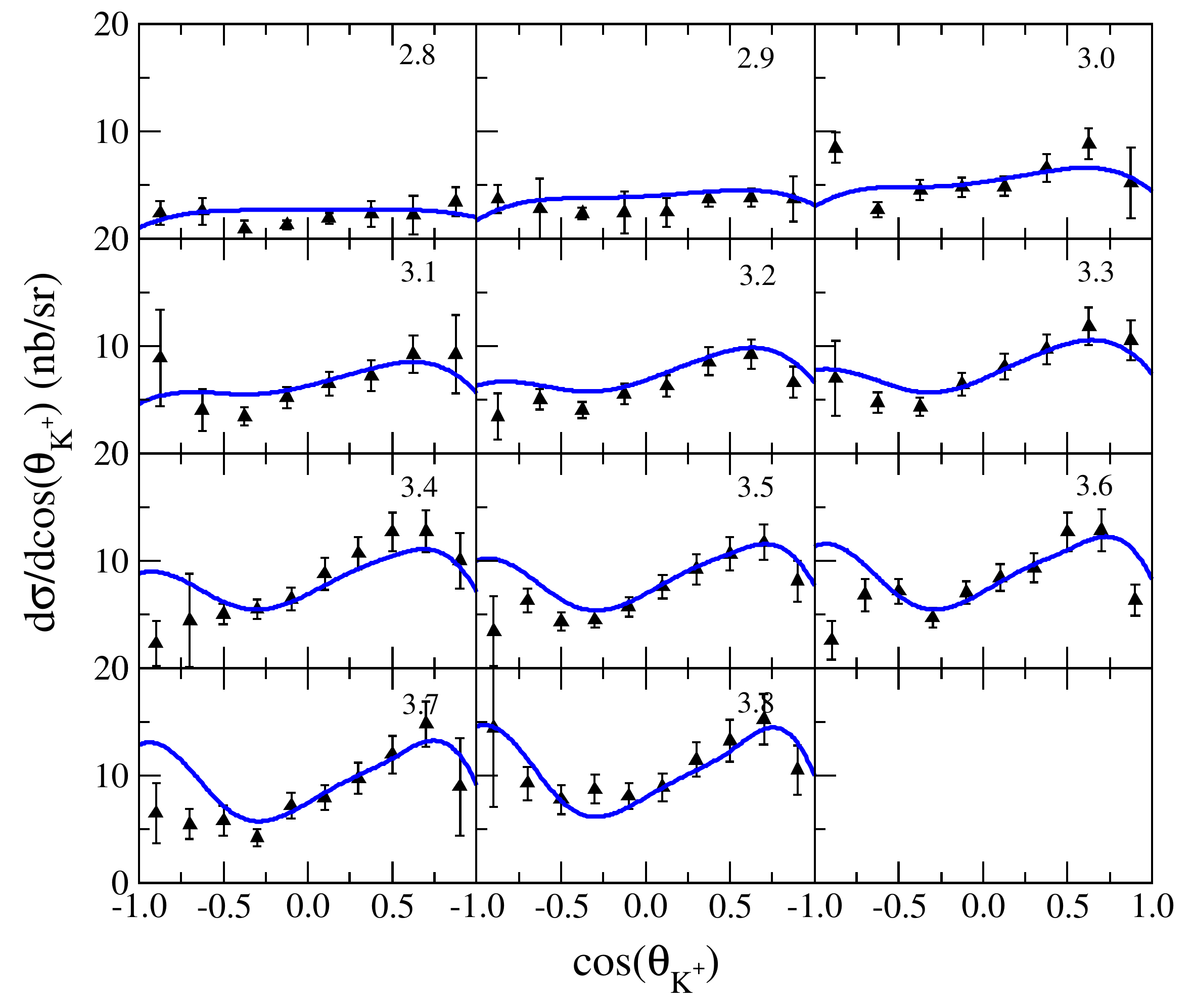}
  \caption{\label{fig:dxsc}%
Differential cross section for $\gamma p \to K^+ K^+ \Xi^-$ as a function of
$K^+$ emission angle in the center-of-mass system. Data are from
Ref.~\cite{CLAS07b}.}
\end{figure}

\begin{figure}[!t]\centering
  \includegraphics[width=.95\columnwidth]{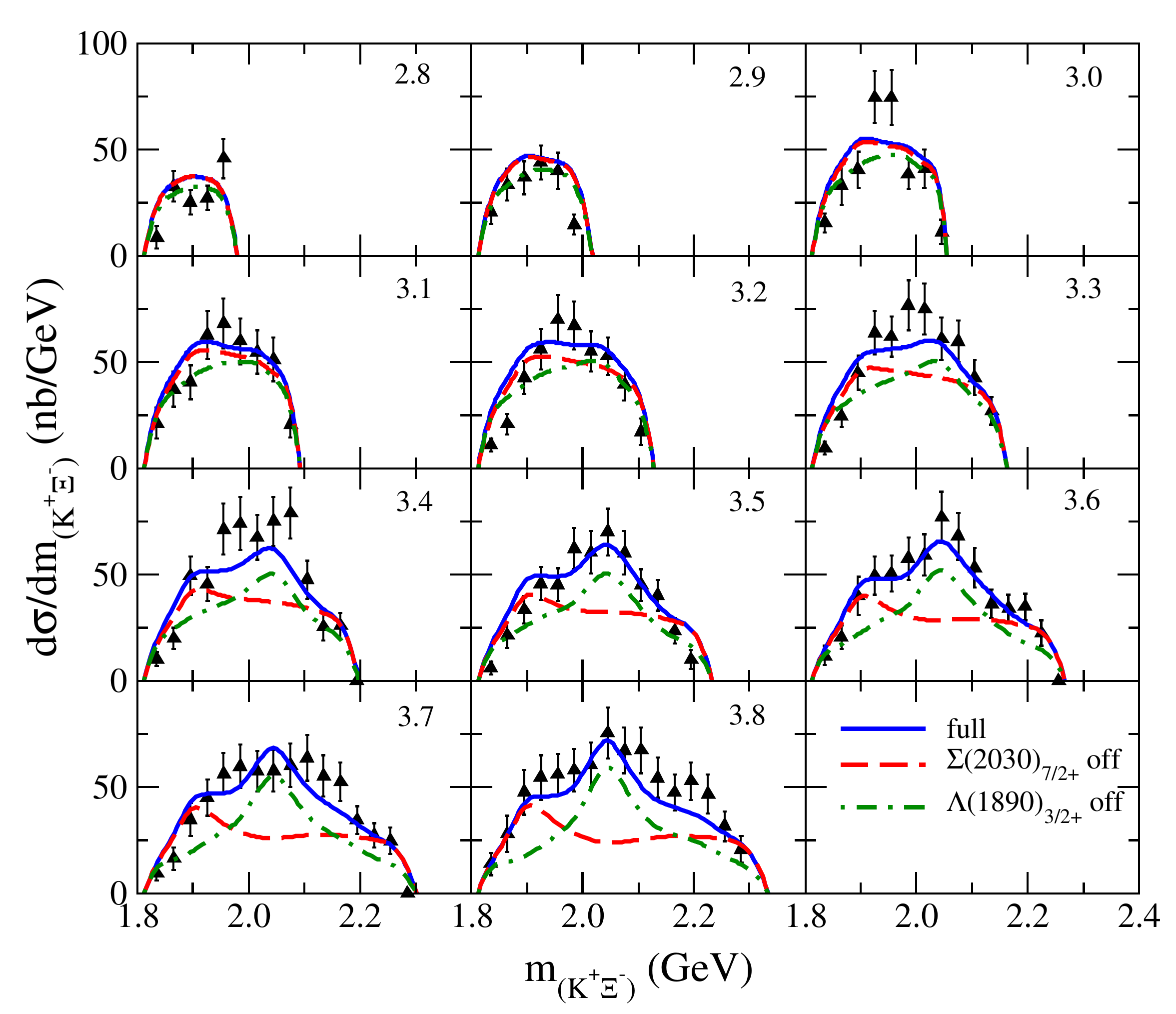}
  \caption{\label{fig:invmass}%
$K^+\Xi^-$ invariant-mass distribution in $\gamma p \to K^+ K^+ \Xi^-$, with
full results shown as solid (blue) lines. The dashed (red) and dash-dotted
(green) curves are obtained when the resonances $\Sigma(2030)7/2^+$ and
$\Lambda(1890)3/2^+$, respectively, are switched off. Data are from
Ref.~\cite{CLAS07b}.}
\end{figure}

\begin{figure}[!t]\centering
  \includegraphics[width=.95\columnwidth]{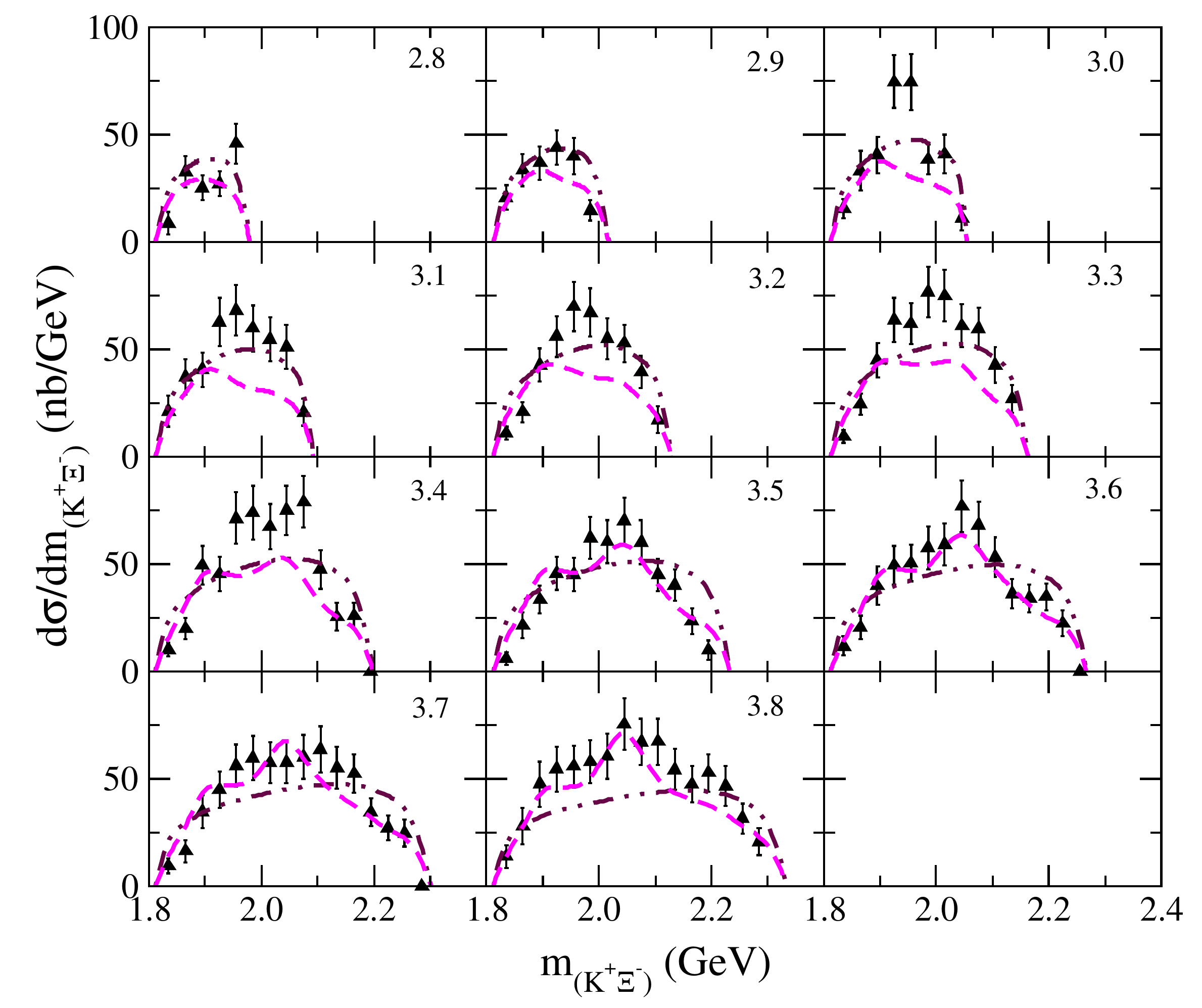}
  \caption{\label{fig:background}%
Effect of remainder current $\Rpipi^\mu$ on the $K^+\Xi^-$ invariant-mass
distribution in $\gamma p \to K^+ K^+ \Xi^-$.  The dashed (magenta) curves
result from switching off the phenomenological remainder
current $\Rpipi^\mu$ compared to the respective solid (blue) lines in
Fig.~\ref{fig:invmass}. The dash-double-dotted (maroon) curves show the results
of fitting the data using only the current $\Rpipi^\mu$ by itself. Data are from
Ref.~\cite{CLAS07b}.}
\end{figure}

It should be noted that the parameterization~(\ref{eq:s-dependent}) is minimal
as far as reproducing the data is concerned, but  by restricting $\tilde{a}_i$
and $\tilde{b}_i$ to be real, we manifestly violate unitarity. Similarly, in
the hyperon resonance contributions entering in $\Mpipi^\mu_Y$ in the
tree-level approximation, the associated coupling constants are also chosen to
be real and the resonance widths are taken as constants ignoring their energy
dependences. The presently available data (cross sections and invariant masses)
are rather insensitive to these features of the parameters, unlike some of the
spin observables which tend to be more sensitive to such details of the model.
Given this situation and the fact that the detailed analysis of  $\Xi$
photoproduction reaction is not the main objective of the present work, while
in principle analytic properties from $S$-matrix theory should be imposed to
improve upon the approximations employed in the present work, we defer such
improvements to future studies dedicated to more detailed analyses once the
corresponding database becomes more complete.

Figure~\ref{fig:txsc} shows the total cross section results for the reaction
$\gamma p \to K^+ K^+ \Xi^-$. The dynamical content of the present model is
also displayed. We find that the spin-1/2 hyperons dominate at lower energies.
The contribution of the remainder current $\Rpipi^\mu$, especially around
$E_\gamma \approx 3.2$\,GeV, is seen to be considerable. However, at this
stage, it is not clear whether, as intended, the effect of $\Rpipi^\mu$ points
to missing explicit higher-order contributions to provide a better resolution
of detailed dynamics that produce the bump in the cross section (see also
discussion below regarding Fig.~\ref{fig:background}) or whether it simply
mimics possible hyperon resonance contributions not included in the present
lowest-order model for $\Mpipi_Y^\mu$. In any case, from the values of the
coupling constants given in Table~\ref{tbl:fmu0-tree}, it is clear that the
resonance content of the reaction is affected by the presence of $\Rpipi^\mu$
because the magnitudes of the strengths of the intermediate hyperons are now
reduced compared to what was found in the previous model
calculation~\cite{MON11} (and the sign of the $\Sigma(2030)7/2^+$ coupling is
changed as well). These are issues that remain to be investigated in future
analyses when a more complete database becomes available.

The $K^+$ angular distribution in the center-of-mass frame of the system,
displayed in Fig.~\ref{fig:dxsc}, is reasonably well reproduced by the present
model calculation. The effects of the above-threshold hyperon resonances on the
$K^+\Xi^-$ invariant-mass distribution are shown in Fig.~\ref{fig:invmass}. We
find that the $\Lambda(1890)3/2^+$ resonance contributes considerably,
especially in the lower invariant-mass region, while the $\Sigma(2030)7/2^+$
resonance affects very much the higher invariant-mass region. Both resonances
are crucial for reproducing the data, especially the $\Sigma(2030)7/2^+$. We
note that the $\Sigma(2250)5/2^-$ resonance needed for reproducing a bump
structure observed in the total cross section of the hadronic reaction $K^- p
\to K^+ \Xi^-$, as was shown in Ref.~\cite{JOHN15}, is negligible for the
present photoreaction $\gamma N \to K K \Xi$.

The effect of the remainder current $\Rpipi^\mu$ on the $K\Xi$ invariant mass
is illustrated in Fig.~\ref{fig:background}. Switching off its contribution
produces the dashed (magenta) curves. Comparing this with the full results
shown in Fig.~\ref{fig:invmass} reveals that the effect of $\Rpipi^\mu$ is
substantial for energies in the range of $3.1 < E_\gamma < 3.4$ GeV, as
discussed already in the context of the total cross section of
Fig.~\ref{fig:txsc}.  It should be pointed out that the remainder current
$\Rpipi^\mu$ is important not just for improving the overall description of the
total cross section, but for the differential cross section as well (the
corresponding effect is not shown explicitly in Fig.~\ref{fig:dxsc}). The
dash-double-dotted (maroon) curves in the same figure show the results of
fitting the $K^+\Xi$ invariant-mass data considering only the remainder current
$\Rpipi^\mu$ by itself. (The fit was constrained by including total and
differential cross-section data as well; the corresponding results are not
shown here.) This clearly indicates the necessity to include some resonance
contributions to reproduce the data.

We emphasize here that the present calculation does not exclude the possibility
of other resonances contributing to these reaction processes. It simply reveals
the most relevant resonances to describe the existing data. Clearly, to better
constrain the choice of resonances, automatized penalty-based selection
techniques, like the one recently described in Ref.~\cite{LASSO18}, need to be
implemented for future analyses. Moreover, for a more detailed analysis, it is
strongly desirable to have more accurate data, especially, in the hadronic
$\bar{K} N \to K \Xi$ process, where the currently existing data are of poor
quality.

\section{Summary and Discussion}\label{sec:summary}

Maintaining gauge invariance is trivial in any photo-process if \emph{all\/}
currents that contribute to the reaction are constructed in a manner that
preserves their individual (generalized) Ward-Takahashi identities. The present
considerations show that then putting together these currents in groups where
each member can be related to the same topologically distinct hadronic process
will not only imply  gauge invariance for the entire group, but it will also
ensure that this group as a whole will provide the correct four-divergence
contribution if it appears as a subprocess within a larger and more complicated
process. Consistency of the construction of the microscopic dynamics in terms
of currents that satisfy \emph{off-shell\/} WTIs is the key here. Mere current
conservation alone does not help, because then one must start all over again
when going over to a new problem. As a simple illustration of this point, let
us consider the no-loop current given in Fig.~\ref{fig:MNL}. The fact that the
four-point and two-point currents appearing in the graphs satisfy their
respective off-shell  WTIs \emph{automatically\/} ensures gauge invariance of
the NL currents. If, on the other hand, the construction of the single-pion
photoproduction amplitude $M^\mu$ had been restricted to providing a conserved
current only, and not the full off-shell  WTI, the NL graphs of
Fig.~\ref{fig:MNL} would \emph{not\/} be gauge invariant, and would not even
provide a conserved current. In other words, one would need an additional
\emph{unphysical} mechanism to even construct a conserved current. Not
considering at the outset the full set of diagrams that are needed for gauge
invariance would make matters even worse.

With this microscopic consistency in mind, we have presented here a
gauge-invariant theory for the production of two pions off the nucleon that
applies equally well to real and virtual photons. The formalism is based on an
extension of the field-theory approach of Ref.~\cite{Haberzettl97} originally
developed by one of the authors for single-pion photoproduction off the
nucleon. In analogy to the single-pion case, we first constructed a complete
description of the hadronic production process $N\to \pi\pi N$ by accounting
for the multiple-scattering series of the interacting final $\pi\pi N$ system
to all orders in terms of the Faddeev-type three-body AGS
amplitudes~\cite{AGS67,Sandhas72}. Coupling then the electromagnetic field to
this hadronic system by employing the gauge derivative~\cite{Haberzettl97}
produced the closed-form expression of Eq.~(\ref{eq:2piCurrent}) for the full
double-pion production current $\Mpp^\mu$ that is gauge invariant as a matter
of course. We emphasize in this respect the efficacy of the gauge-derivative
procedure to identify and link \emph{all\/} relevant reaction mechanisms in a
microscopically consistent manner.

Most importantly for practical purposes, we  have provided here a consistent
expansion scheme for the full current in terms of groups of contributing
currents that are easily identifiable by the topological complexity of the
underlying hadronic processes and that are separately gauge invariant as a
group. We  have explicitly discussed in this manner the no-loop currents of
Fig.~\ref{fig:MNL} and the one-loop currents of Fig.~\ref{fig:Loop1}.

Existing theoretical models based on baryon and meson degrees of freedom
can all be subsumed under the no-loop scenario of Fig.~\ref{fig:MNL}, more
explicitly depicted in Fig.~\ref{fig:NoLoopExplicit}. However, none of the
models actually incorporates the full subsystem-process information in terms
of, for example, the $\gamma N \to \pi N$ or the $\gamma\rho\to\pi\pi$
production currents which, as Fig.~\ref{fig:NoLoopExplicit} clearly shows,
would be necessary for a consistent description and which would be fairly
straightforward to do given the technology available for treating such
subprocesses in a gauge-invariant manner~\cite{HBMF98a,HNK06,HHN11}. As the
details of Fig.~\ref{fig:NoLoopExplicit} show, at the no-loop level practically
all the theoretical effort needs to be expended on the adequate modeling of
these subprocesses.

The situation is quite a bit more complicated at the two-loop level depicted in
Fig.~\ref{fig:Loop1}. Apart from the additional complication of the loop
integrations itself and the fact that the meson-baryon and meson-meson
amplitudes $X$ must be available, the most complicated ingredient in each of
the three gauge-invariant groups of graphs is the occurrence of the interaction
current $X^\mu$ for $\gamma\pi N\to \pi N$ in the two groups labeled 1L$_1$ and
1L$_2$, and for $\gamma \pi \pi \to\pi\pi$ in the 1L$_3$ group.%
    \footnote{Recall here that $\pi$ and $N$ on the initial sides of the reactions
    are just generic placeholders for any allowed meson and baryon, respectively,
    since they occur as intermediate states of the diagrams.}
Equation~(\ref{eq:Xmu}) shows that the interaction current $X^\mu$ itself is
fairly complicated, requiring another double-loop integration for its full
calculation, and it may not be possible to evaluate Eq.~(\ref{eq:Xmu}) in an
actual application. However, in Appendix~\ref{appsec:ContCurrentGen} we provide
a detailed, general description how any interaction current can be incorporated
in a locally gauge-invariant manner by a phenomenological contact-type current.
Local gauge invariance, therefore, need never be an issue even if other
approximations may be necessary to render the problem manageable.

\begin{figure*}\centering
  \includegraphics[width=0.95\textwidth,clip=]{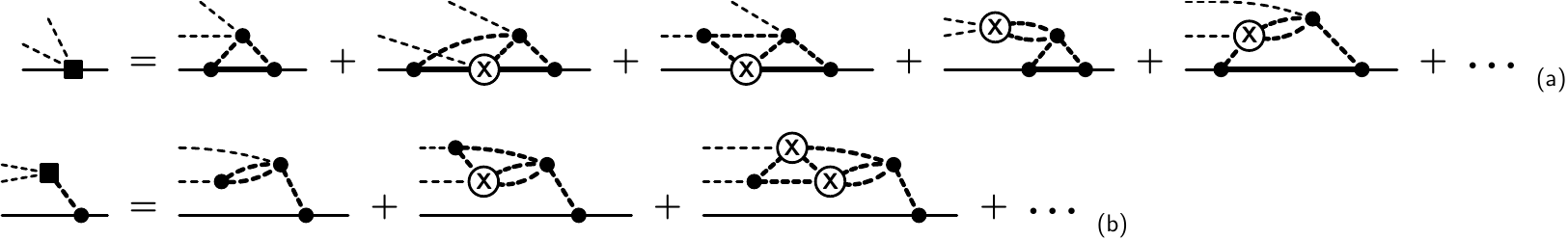}
  \caption{\label{fig:Contact28}%
Hadronic two-pion production mechanism resulting from a three-pion vertex that
start out as a four-body process until one of the mesons gets absorbed. The
thick dashed and solid internal lines stand for any meson or baryon,
respectively, compatible with the process. (a) Contact-type mechanisms summing
up processes where at least one connection is made with the baryon line after
the initial production. Part (b) sums up the three-body multiple-scattering
series within the three-meson system, without intermediate reconnection with
the baryon line. Once the final $\pi\pi N$ system occurs, subsequent
interactions for both cases are topologically equivalent to that of the
$\rho\pi\pi$ production vertex in Fig.~\ref{fig:basicprocess}(b) resulting in
the graphs shown in Fig.~\ref{fig:All2PiPrime}. The contact-term contributions
(a) are chosen to avoid double-counting with the mechanisms shown in
Fig.~\ref{fig:All2PiPrime}.}
\end{figure*}

\begin{figure}[t!]\centering
  \includegraphics[width=1.0\columnwidth,clip=]{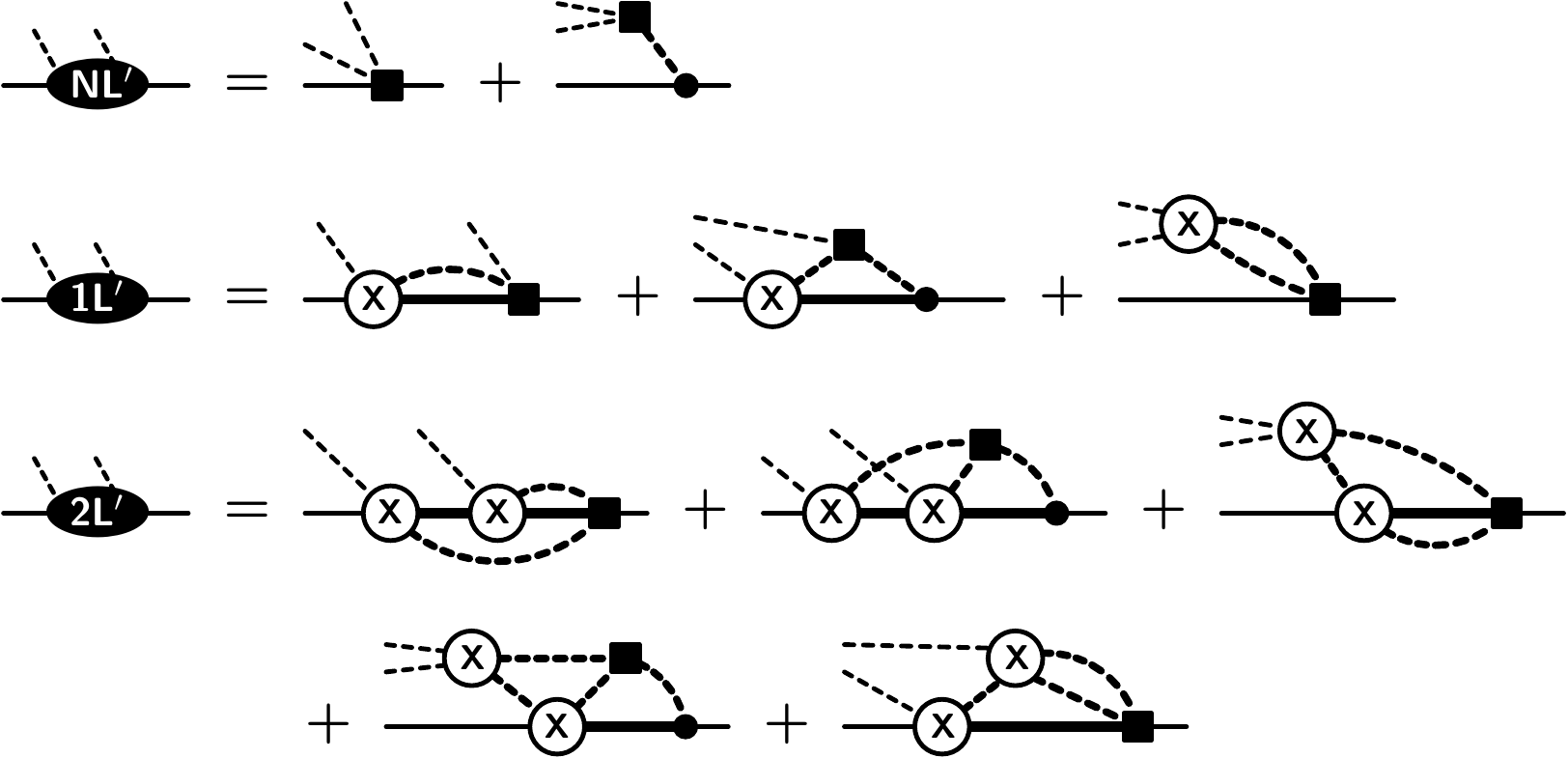}
  \caption{\label{fig:All2PiPrime}%
Higher-order double-pion-production contributions using the initial mechanisms
depicted in Fig.~\ref{fig:Contact28}. The no-loop (NL$^\prime$), one-loop
(1L$^\prime$), and two-loop  (2L$^\prime$) diagrams given here are exactly
analogous to those of Fig.~\ref{fig:All2Pi}.}
\end{figure}

\begin{figure}[t!]\centering
  \includegraphics[width=1.0\columnwidth,clip=]{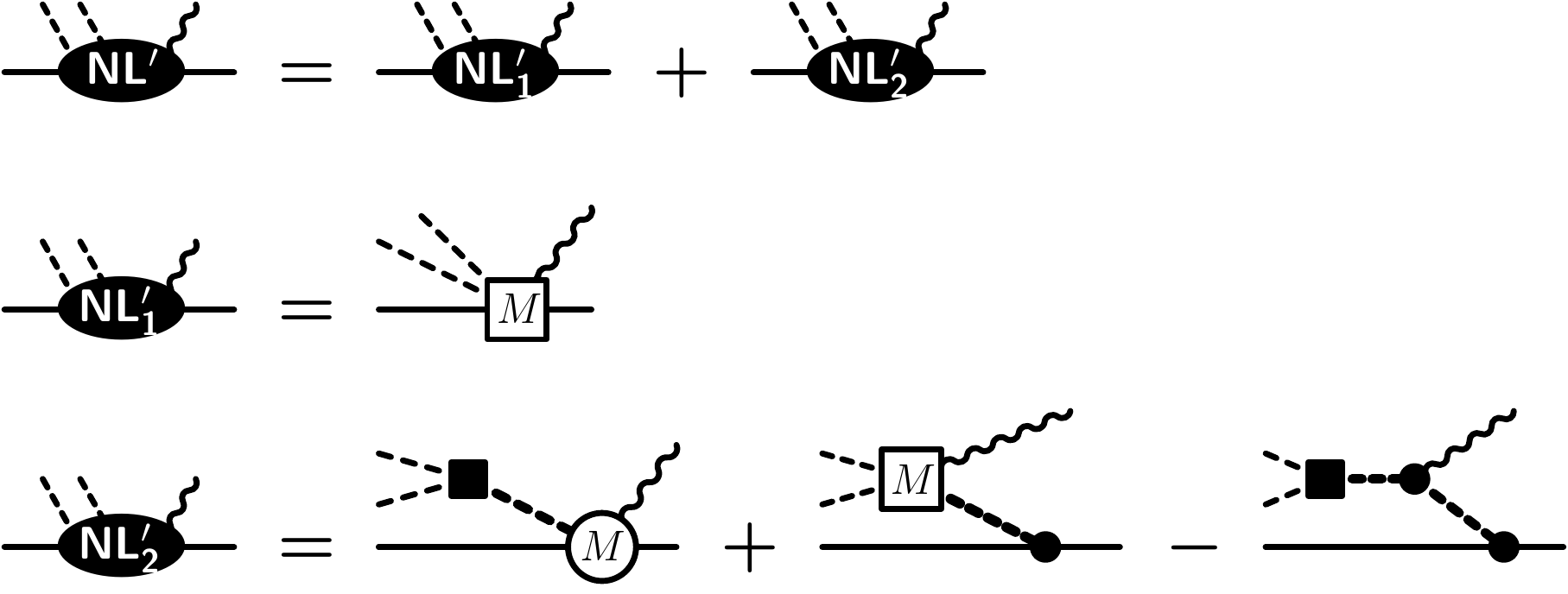}
  \caption{\label{fig:PhotoPrimeNL}%
Current contributions resulting from the no-loop (NL$'$) diagrams of
Fig.~\ref{fig:All2PiPrime}. The four- and five-point currents indicated by
square boxes labeled $M$ are given in Fig.~\ref{fig:MboxCurrents}.}
\end{figure}

The usefulness of an approximate treatment of neglected current contributions
was demonstrated here for the reaction $\gamma N \to K K \Xi$ by describing the
neglected higher-order loop contributions beyond the tree level in terms of a
phenomenological remainder current $\Rpipi^\mu$. The results reported in
Sec.~\ref{sec:appl} show that despite the relative simplicity of the ansatz,
the remainder current is indeed capable of contributing substantially to a good
quantitative description of the reaction. In addition, the corresponding
results point to where additional efforts need to be expended for a better
description of the experimental data.

In deriving the present formalism, we have relied heavily on the dynamically
correct formulation in terms of local gauge invariance because it provides a
very convenient framework that allowed us to identify and consistently link in
a straightforward manner all topologically relevant microscopic mechanisms. We
emphasize, however, that as a consistent field-theory-based formulation, the
resulting amplitudes satisfy as well the usual properties of Lorentz covariance
and analyticity. With respect to unitarity, however, care must be exercised
when approximating or truncating the full formalism, as mentioned in
Secs.~\ref{sec:PHCV} and \ref{sec:treelevel}.

In summary, we have presented here a formulation of the two-pion production
process off the nucleon based on field theory that is of the same level of
rigor as the one-pion production described in Ref.~\cite{Haberzettl97}. We hope
that the present formulation of two-pion photoproduction will be of similar
usefulness. Moreover, we emphasize that since the present formulation is based
solely on the topological properties of the underlying hadronic production
processes (cf.\ Fig.~\ref{fig:basicprocess}), it applies equally well to the
photoproduction of any two mesons off any baryon resulting from topologically
similar basic hadronic mechanisms.

\begin{figure}[t!]\centering
  \includegraphics[width=1.0\columnwidth,clip=]{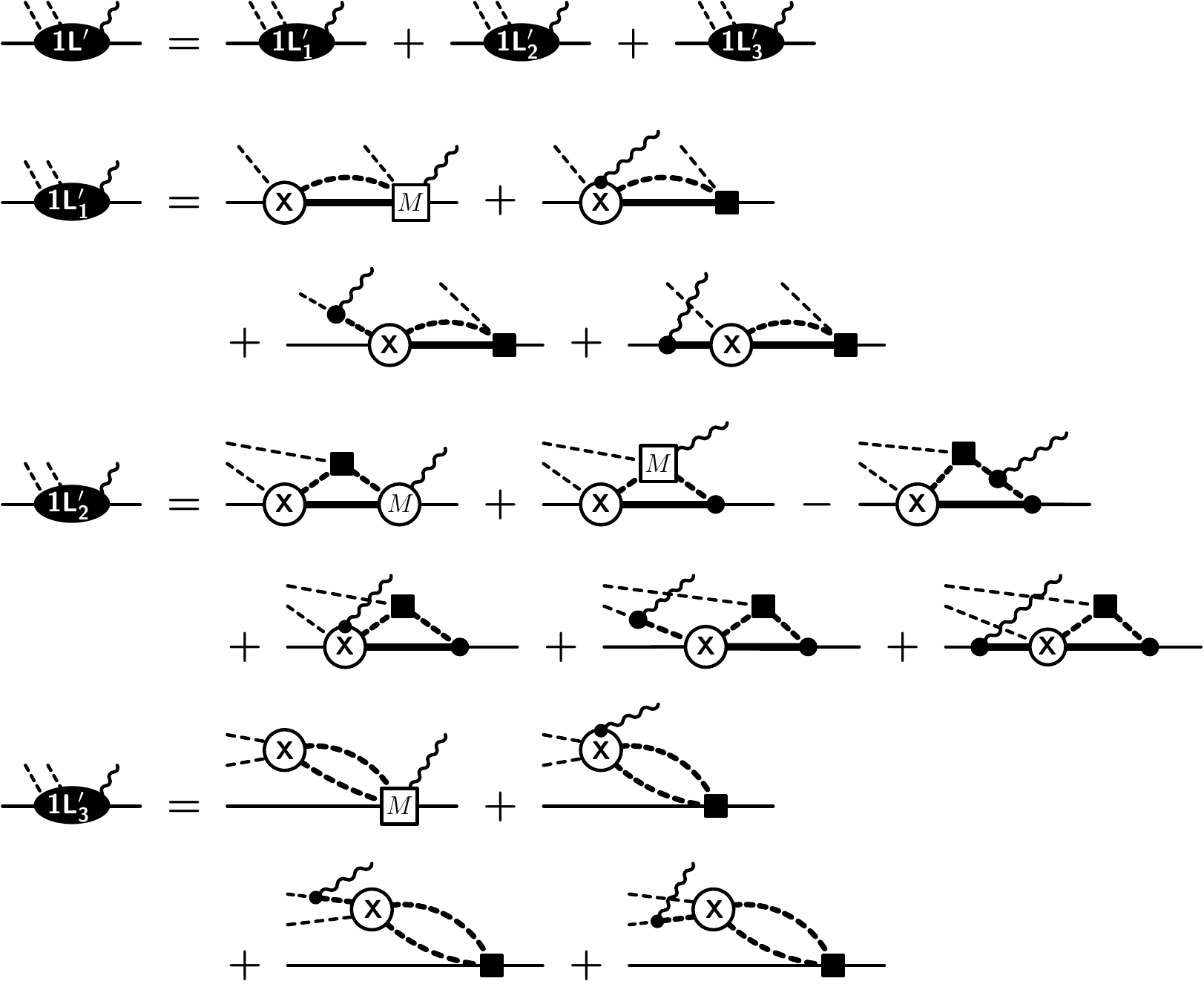}
  \caption{\label{fig:PhotoPrime1L}%
  Current contributions resulting from the one-loop (1L$'$) diagrams of
  Fig.~\ref{fig:All2PiPrime}. See Fig.~\ref{fig:PhotoPrimeNL} for details.}
\end{figure}

\begin{figure*}[t!]\centering
  \includegraphics[width=.75\textwidth,clip=]{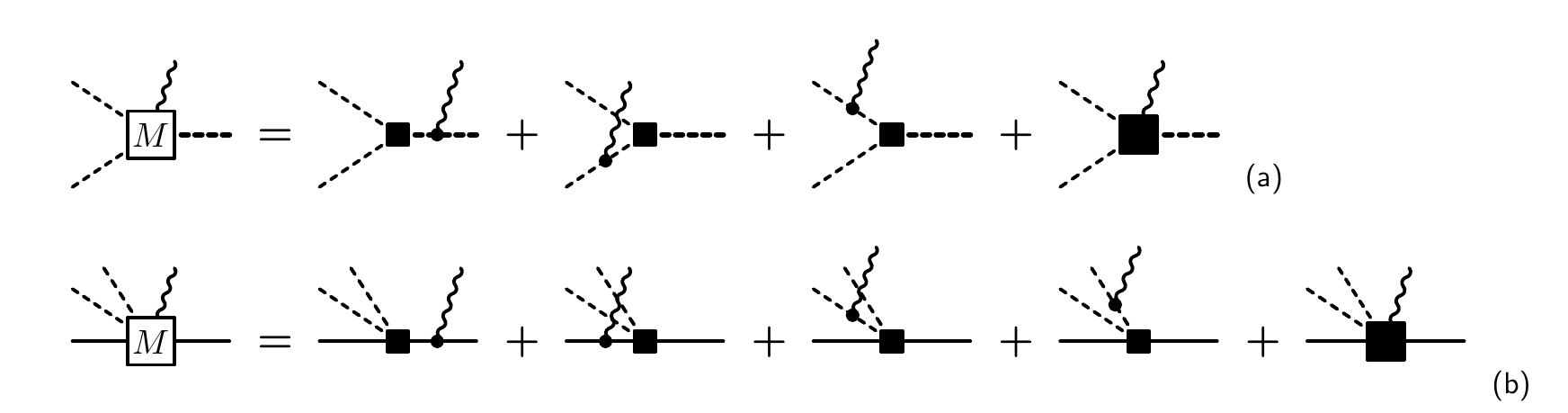}
  \caption{\label{fig:MboxCurrents}%
Two-pion photoproduction currents associated with (a) the effective two-pion vertex of Fig.~\ref{fig:Contact28}(b) and (b) the contact vertex
of Fig.~\ref{fig:Contact28}(a).
The respective last diagrams here (solid square boxes with photon attached) depict the four-point and five-point interaction currents
of the respective processes.}
\end{figure*}

\acknowledgments H.H. acknowledges partial support by the U.S. Department
Energy, Office of Science, Office of Nuclear Physics, under Award Number
DE-SC0016582. The work of K.N. was partially supported by the FFE Grant No.\
41788390 (COSY-58). The work of Y.O. was supported by the National Research
Foundation of Korea under Grant Nos.\ NRF-2018R1D1A1B07048183 and
NRF-2018R1A6A1A06024970.

\appendix

\section{\boldmath Incorporating four-meson vertices like $\omega \to \pi\pi\pi$}
\label{app:4meson}

In this Appendix, we address the question how processes based on $n$-pion
vertices for $n\ge 3$ can be incorporated into the formalism. As mentioned
earlier, the dynamics resulting from such vertices requires at least a
treatment in terms of an $(n+1)$-body problem. For the simplest possible case,
e.g., the $\pi\pi\pi\omega$ vertex depicted in Fig.~\ref{fig:basicprocess}(c),
this means that we need at least a four-body treatment. It is possible to do
that in principle, i.e., the formalism for doing so
exists~\cite{GS67c,Sandhas74,Sandhas75,AGS70}, but presumably there is little
practical value to do so in full because of the enormous complexity of the
relativistic version of that problem. Instead, we will take our cues from the
full four-body problem of the $\pi\pi\pi N$ system and then reduce its
complexity to a three-body problem by reabsorbing one of the mesons into either
another meson or the nucleon, similar to the simple  example depicted in
Fig.~\ref{fig:basicprocess}(c).

Using the AGS four-body theory~\cite{GS67c,Sandhas74,Sandhas75,AGS70}, one
finds two classes of graphs where the initial four-body system eventually is
reduced to a three-body system because one of the three intermediate mesons is
absorbed either in the baryon or another meson. If somewhere along the line
before the final absorption at least one interaction with the baryon takes
place, we find the processes depicted in Fig.~\ref{fig:Contact28}(a), and if
all scattering processes happen exclusively within the three-meson system, we
obtain Fig.~\ref{fig:Contact28}(b). We thus obtain a contact-type $\pi\pi NN$
vertex for Fig.~\ref{fig:Contact28}(a) that behaves topologically like the
sequential two-pion process of Fig.~\ref{fig:basicprocess}(a) where the
intermediate nucleon propagation has shrunk to a point and an intermediate
effective three-meson vertex for Fig.~\ref{fig:Contact28}(b) that is
topologically equivalent to the intermediate $\pi\pi\rho$ vertex of
Fig.~\ref{fig:basicprocess}(b). Taken as effective \enquote{elementary}
production processes, their resulting three-body dynamics is \emph{exactly\/}
like that of the basic processes depicted in Fig.~\ref{fig:Fvertices}, and we
may apply the full formalism developed for them to the new effective two-pion
production mechanisms of Fig.~\ref{fig:Contact28}. Up to the two-loop level,
therefore, we obtain the processes depicted in Fig.~\ref{fig:All2PiPrime}
which, apart from the fact that certain intermediate baryon lines have shrunk
to a point, is completely analogous to Fig.~\ref{fig:All2Pi}.

Attaching the photon is now equally straightforward, producing the diagrams of
Fig.~\ref{fig:PhotoPrimeNL} at the no-loop (NL$'$) level and of
Fig.~\ref{fig:PhotoPrime1L} for the one-loop (1L$'$) currents. Both sets of
graphs are analogous to those of Figs.~\ref{fig:MNL} and \ref{fig:Loop1},
respectively. The four- and five-point currents resulting from the mechanisms
of Fig.~\ref{fig:Contact28} are given in Fig.~\ref{fig:MboxCurrents}. The
crucial parts of these currents are the respective interaction currents
(depicted as solid square boxes with an incoming photon attached) because
formally they follow from coupling the photon to the respective interiors of
the mechanisms depicted in Fig.~\ref{fig:Contact28} which in practice cannot be
calculated such that the usual gauge-invariance constraints of an interaction
current can be expected to hold true. However, as shown in the subsequent
Appendix~\ref{app:GenContactCurrent}, it is straightforward to find
phenomenological approximations of these interaction currents that allow one to
maintain the full off-shell gauge invariance of both currents in
Fig.~\ref{fig:MboxCurrents}. The corresponding prescriptions for doing so
consistently with whatever description one chooses for the underlying hadronic
processes allowing for iterative refinements of approximations have been given
in Ref.~\cite{HNK06}, and there is no need here to repeat that discussion. In
practical applications, therefore, one may employ phenomenological models for
the two-pion production mechanisms in Fig.~\ref{fig:Contact28} without
sacrificing gauge invariance. For the process in
Fig.~\ref{fig:MboxCurrents}(b), in particular, we note that it is topologically
equivalent to the remainder current $\Rpipi^\mu$ of Fig.~\ref{fig:ContCurrent}.
Therefore, if a meaningful approximate treatment of the contact-type hadronic
process depicted in Fig.~\ref{fig:Contact28}(a) can be extracted, one may use
this as the basis for an approximate treatment along the lines outlined in
Sec.~{\ref{sec:PhenContactCurrent}}.

\begin{figure}[t!]\centering
  \includegraphics[width=.6\columnwidth,clip=]{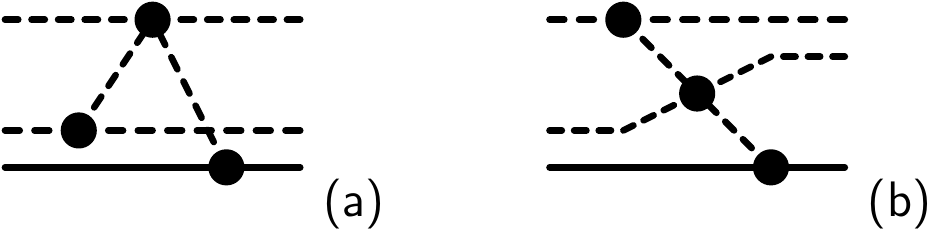}
  \caption{\label{fig:3BF}%
Examples of four-meson vertices producing three-body forces.
There are several more examples of this kind.}
\end{figure}

The procedure just described takes care of two-pion systems evolving out of the
initial four-body system created via a nucleon and a three-pion vertex. All
subsequent interactions, however, follow the three-body dynamics described in
Sec.~\ref{sec:Faddeev}. To bring back the possibility of intermediate
four-meson interactions, one may add any number of mechanisms involving
four-meson vertices to the driving term (\ref{eq:AGSdrive}) of the AGS
equations. Some of the simplest examples are the three-body-force graphs shown
in Fig.~\ref{fig:3BF}. Denoting such processes by  $B_{\beta\gamma}$, one finds
that the expansion of $M_\beta$ given in Eq.~(\ref{eq:Mexpand}) then needs to
be modified in lowest order as
\begin{equation}
M_\beta \to   M'_\beta = M_\beta +\sum_{\gamma,\alpha} B_{\beta\gamma} G_0
X_\gamma G_0 \bar{\delta}_{\gamma\alpha} F_\alpha +\cdots~,
\end{equation}
i.e., the additional terms are linear in $X_\gamma$, whereas the nonlinear
mechanisms of Fig.~\ref{fig:NLexample} are of third order in $X_\gamma$.
However, either one of such effects requires three-loop integrations at their
respective lowest orders. For the graphs shown in Fig.~\ref{fig:3BF}, the extra
term here corresponds to subjecting the final $\pi\pi N$ system of the one-loop
graphs in Fig.~\ref{fig:All2Pi} to the corresponding three-body forces.

We will not pursue this point any further since we suspect that it may be of
limited practical value in view of the complexity of such mechanism. Also, for
the same reason, we will not consider even more complex mechanisms with an even
higher number of mesons produced at initial or intermediate stages. In any
case, we would like to emphasize once more that, should the inclusion of more
complicated mechanisms be deemed necessary for practical applications, the
corresponding currents may simply be added without affecting gauge invariance
of the existing approach because the currents of topologically independent
hadronic graphs satisfy their own independent gauge-invariance constraints and
this can be treated independently.

\section{\boldmath Generalized contact currents} \label{app:GenContactCurrent}

In this Appendix, we provide a generic expression for the phenomenological
contact current describing the coupling of the photon to the interior
interaction region of a hadronic transition process. Based on this general
result, we will provide the full off-shell version of the five-point contact
current for $\gamma N \to \pi\pi N$ whose on-shell form was provided in
Sec.~\ref{sec:PhenContactCurrent} and the four-point contact currents used in
the subprocesses of $\gamma N\to KK\Xi$ in Sec.~\ref{sec:appl}.

The results follow by applying the gauge derivative introduced in
Ref.~\cite{Haberzettl97} as a formal way of applying minimal substitution to
interacting systems. The resulting current will satisfy the appropriate
generalized  WTI (\ref{eq:gWTIintM}) mandated by local gauge
invariance~\cite{Haberzettl97}. However, it may lack additional transverse
current contributions that are required by additional physical constraints.
Such transverse currents do not contribute to gauge invariance and thus are
inaccessible to the gauge-derivative procedure. For contact currents, in
particular, the gauge-derivative result needs to be amended by a manifestly
transverse current to address the `violation of scaling problem' at high
energies~\cite{DL72}. (The expressions presented here generalize the generic
four-point-function results given in the appendix of Ref.~\cite{HWH15}.)

%
\begin{figure*}[t!]\centering
\includegraphics[width=.7\textwidth,clip=]{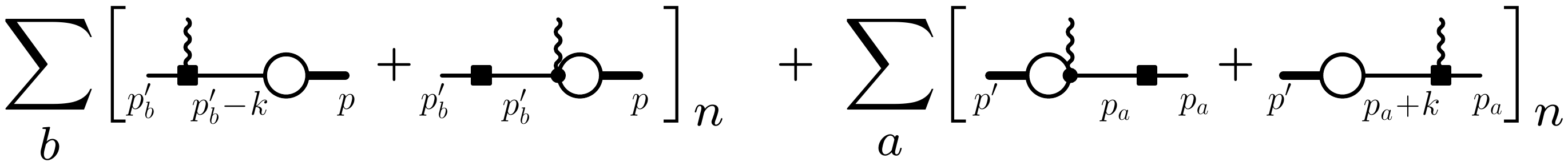}
  \caption{\label{fig:contact}%
  Schematic diagram corresponding to Eq.~(\ref{eq:Cpipimu}) for applying the
  gauge derivative to the $n$-th linearly independent contribution to the
  hadronic process (\ref{eq:expandhadron}). Sums over $b$ and $a$ enumerate
  contributions from the respective final and initial hadron legs to which a
  photon (wavy line) is attached. Solid squares depict coupling operators $G_n$
  and open circles correspond to form factors $F_n$. Thick lines with momentum
  labels $p$ or $p'$ generically stand for all incoming or outgoing legs,
  respectively. Lines labeled with momenta $p^{}_a$ or $p'_b$ indicate the
  particle to which the photon couples; all other particle lines (incoming or
  outgoing) not participating in the electromagnetic interaction are omitted
  for clarity. The photon carries a momentum $k$ into the process, i.e., all
  downstream momenta are accordingly increased. Overall, the initial and final
  momentum sets $p$ and $p'$, respectively, are fixed and satisfy
  (\ref{eq:PhotoConservedMomentum}); only the internal momenta can increase
  ($p_a+k$) or decrease ($p'_b-k$) accordingly.}
\end{figure*}
%

\subsection{Contact current for arbitrary hadronic transition}%
\label{appsec:ContCurrentGen}

An arbitrary hadronic transition process for $N_i$ incoming hadrons going over
to $N_f$ outgoing hadrons may generically be written as
\begin{equation}
\Fpipi = \sum_n G_n F_n
\label{eq:expandhadron}
\end{equation}
where the sum extends over all linearly independent  operator structures $G_n$
and (complex) scalar functions $F_n$. An example of this would be the structure
given in Eq.~(\ref{eq:HadrDiracStruc}) for $N\to \pi\pi N$, but generically
this applies also to any other transition like $N\to \pi N$, $\rho\to\pi\pi$,
$NN\to NN$, etc. (The latter contact currents are needed for
bremsstrahlung~\cite{NH09,HN10}.)

In general, assuming four-momentum conservation, the form factors $F_n$ depend
on the $(N_i+N_f)(N_i+N_f-1)/2$ scalar invariants one can form from the
$N_i+N_f-1$ independent particle momenta. Without lack of generality, they may
all be written as squares of individual momenta, or as squares of sums or
differences of two momenta, and we may always choose the set of all $N_i+N_f$
squares of external momenta to be among the squares necessary for a complete
description of the specific dynamical situation described by $F_n$. As we shall
see by construction in the following, limiting our discussion in the following
to this subset of squared momenta will be sufficient for the present
phenomenological purposes.

In the following, incoming and outgoing hadron momenta will be, respectively,
unprimed, $p_a$, and enumerated by index $a=1,\ldots,N_i$, and primed, $p'_b$,
and enumerated by index $b=1,\dots,N_f$.

Hence, for the present phenomenological purpose, we assume that the form
factors $F_n$ depend only\footnote{For a three-point vertex, the squares of all
    three external momenta allow for a \emph{complete} description of the
    invariant vertex structure, but for four or more external legs, this is no
    longer true.}
on the squares of all incoming and outgoing hadronic momenta $p_a^2$ and
$p'^2_b$, respectively, i.e.,
\begin{equation}
  F_n = F_n(p'^2_1,\ldots,p'^2_{N_f};p^2_1,\ldots,p^2_{N_i})~,
\end{equation}
ignoring other possible squares like $(p'_b-p_a)^2$, etc. In general,  we allow
for off-shell momenta, but four-momentum conservation,
\begin{equation}
  \sum_a p_a = \sum_b p'_b~,
  \label{eq:HadronConservedMomentum}
\end{equation}
is always assumed. Note that we suppress here the isospin structure in
(\ref{eq:expandhadron}) because it has no bearing on the basic approach.
(However, isospin dependence will give rise to the charge parameters $e_x$
below associated with legs of external particles $x$.)

When applying the gauge derivative~\cite{Haberzettl97}, it will act on
four-momenta in both coupling $G_n$ operators and form factors $F_n$. In
general, if depending on momenta, $G_n$ may depend on arbitrary powers of the
momentum of a particular particle leg and it may depend on products of momenta
from different legs. When applying the gauge derivative to such products, they
need to be treated according to the usual product rule for derivatives.
However, regarding the photon coupling to a particular leg, the momentum
dependence arising from other legs is irrelevant, in other words, the gauge
derivative effectively acts like a partial derivative that affects only the
momentum of the active leg. Furthermore, we need to account for whether that
momentum pertains to an incoming leg or an outgoing leg.

The contact current,
\begin{align}
\Cpipi^\mu &= -\{\Fpipi\}^\mu +\Tpipi^{\,\mu}
\nonumber\\
 & = \sum_n\left[ -\{G_n F_n\}^\mu +G_n \Tpipi^{\,\mu}_n\right],
 \label{eq:CTpipimuGeneral}
\end{align}
associated with coupling a photon (with Lorentz index $\mu$) to the interior of
the process $\Fpipi$ is obtained by applying the gauge derivative
$\{\cdots\}^\mu$ here~\cite{Haberzettl97} and adding a judiciously constructed
manifestly transverse current $\Tpipi^\mu=\sum_n G_n \Tpipi_n$ to avoid the
high-energy scaling violation discussed by Drell and Lee~\cite{DL72}. The gauge
derivative results in
\begin{align}
\{G_n F_n\}^\mu  &=\left\{G_n\,F_n\right\}^\mu_{\text{out}}+\left\{F_n\,G_n\right\}^\mu_{\text{in}}
\nonumber\\[1.5ex]
&=  \sum_b \left[\big\{G^{(b)}_n\big\}^\mu_b F_n+ G^{(b)}_n \big\{F_n\big\}^\mu_b \right]
\nonumber\\
&\quad\mbox{}
   + \sum_a \left[ \big\{F_n\big\}^\mu_a G_n^{(a)}+F_n \big\{G_n^{(a)}\big\}^\mu_a \right]~,
\end{align}
where the terms are ordered here according to where couplings occur during the
reaction, with the couplings containing outgoing momenta (sum over $b$),
denoted by $G_n^{(b)}$, placed on the left and couplings containing incoming
momenta (sum over $a$), denoted by $G_n^{(a)}$, placed on the right. The
ordering is necessary  to properly account for the fact that the momentum $k$
carried into the reaction by the photon is available only for particles
downstream of the photon coupling. Diagrammatically, this is depicted in
Fig.~\ref{fig:contact}. With four-momenta of all external particles taking part
in the photoprocess at their fixed values, with four-momentum conservation,
\begin{equation}
  \sum_a p_a +k = \sum_b p'_b~,
  \label{eq:PhotoConservedMomentum}
\end{equation}
 we obtain
\begin{align}
  \{G_n F_n\}^\mu &=
  \sum_b \bigg[\left\{G_n^{(b)}(p'_b-k)\right\}^\mu F_n^{(b)}\big((p'_b-k)^2\big)
  \nonumber\\
  &\qquad\qquad\mbox{}
     +G_n^{(b)}(p'_b)\left\{F_n^{(b)}\big((p'_b-k)^2\big)\right\}^\mu\bigg]
     \nonumber\\
     &\mbox{}\quad
     +\sum_a \bigg[\left\{F_n^{(a)}\big(p_a^2\big)\right\}^\mu G_n^{(a)}(p_a)
  \nonumber\\
  &\qquad\qquad\mbox{}
     +F_n^{(a)}\big((p_a+k)^2\big) \left\{G_n^{(b)}(p_a)\right\}^\mu
     \bigg]~,
     \label{eq:Cpipimu}
\end{align}
where the four terms under the sums and their momenta are shown in
Fig.~\ref{fig:contact}. The notation $F_n^{(b)}\big((p'_b-k)^2\big)$ means that
all hadron legs carry the momentum appropriate for the photo process, except
the outgoing particle $b$, which has $p'_b-k$ to ensure four-momentum
conservation across the hadronic process. Similarly for
$F_n^{(a)}\big((p_a+k)^2\big)$, all momenta are fixed at their external values,
except for the particle $a$ coming into the hadronic process after it has
interacted with the photon and thus has $p_a+k$.

The gauge derivatives of the coupling operators,
\begin{subequations}\label{eq:KRtypeGamma}
\begin{align}
\Gamma^\mu_{n,b} &\equiv -\left\{G_n^{(b)}(p'_b-k)\right\}^\mu~,
\\
\Gamma^\mu_{n,a} &\equiv -\left\{G_n^{(a)}(p_a)\right\}^\mu~,
\end{align}
\end{subequations}
provide Kroll-Ruderman-type (KRt) photon couplings arising from the hadronic
coupling operators  independent of any form-factor dependence. When evaluating
these gauge derivatives, only the momenta of respective outgoing and incoming
particles $b$ and $a$ are contributing.  An example of such KRt couplings is
furnished by the $\gamma^\mu$ couplings in Eq.~(\ref{eq:KRforNNpipi}) for
$\gamma N N \pi\pi$. The details of such couplings depend on a particular
application, but, in general, their four-divergences are given by
\begin{subequations}\label{eq:WardIdentity}
\begin{align}
  k_\mu \Gamma_{n,b}^\mu &=  e_b^{} \left[G_n^{(b)}(p'_b-k)-G_n^{(b)}(p'_b)\right]~,
  \\
  k_\mu \Gamma_{n,a}^\mu &=  e_a^{} \left[G_n^{(a)}(p_a)-G_n^{(a)}(p_a+k)\right]~,
\end{align}
\end{subequations}
where the parameters $e_b^{}$ and $e_a^{}$ describe the charges of particles $b$ and
$a$, respectively.\footnote{To obtain these explicit charge factors, the
   charge operators that result from applying the gauge derivative must be
   combined with the appropriate isospin dependence that
   was suppressed here~\cite{Haberzettl97}.}
These results are essential for deriving the four-divergence of the contact
current $\Cpipi^\mu$ given in Eq.~(\ref{eq:genWTIcontactF}) below. Note that
their structure at the elementary level of a single particle leg is the same as
the general result (\ref{eq:gWTIintM}) for interaction currents. The subsequent
Sec.~\ref{secsubsub:proof} will provide further discussion of elementary gauge
derivatives needed for evaluating the KRt currents (\ref{eq:KRtypeGamma}) and a
proof of Eqs.~(\ref{eq:WardIdentity}).

To evaluate the action of coupling the photon to the form factors in
(\ref{eq:Cpipimu}), we note that using the elementary gauge derivative
$\{q^2\}^\mu=Q(2q+k)^\mu$ for a particle of momentum $q$ with charge $Q$, one
can easily show that the gauge derivative of a scalar function $f(q^2)$
associated with that particle is given by
\begin{equation}
  \left\{f(q^2)\right\}^\mu = Q(q'+q)^\mu \frac{f(q'^2)-f(q^2)}{q'^2-q^2}~,
    \label{eq:f(q^2)coupling}
\end{equation}
where $q'=q+k$ is the particle's momentum after the electromagnetic
interaction. This has the structure of a manifestly nonsingular
finite-difference derivative. (This is an exact result, not a phenomenological
ansatz.) For the following, it will be useful to introduce abbreviations
\begin{equation}
  s_a=(p_a+k)^2
  \qtext{and}
  u_b=(p'_b-k)^2
\end{equation}
for Mandelstam-like squares associated with initial and final momenta,
respectively, to express some of the squared momenta appearing here.

Straightforward algebra shows now that the full contact current
(\ref{eq:CTpipimuGeneral}) may be written as
\begin{equation}
\Cpipi^\mu=\sum_n\Big[\Kpipi^\mu_n + G_n(\Spipi^\mu_n+\Tpipi^\mu_n)\Big] ~,
\label{eq:CmuTcomplete}
\end{equation}
where
\begin{equation}
\Kpipi^\mu_n =
  \sum_b \Gamma^\mu_{n,b}\,F_n^{(b)}(u_b)
  +\sum_a  \Gamma^\mu_{n,a}\,F_n^{(a)}(s_a)
  \label{eq:KRtAll}
\end{equation}
subsumes the contributions from the KRt electromagnetic couplings
(\ref{eq:KRtypeGamma}) to the initial and final four-momenta in the hadronic
coupling operator $G_n$. The scalar finite-difference coupling
(\ref{eq:f(q^2)coupling}) to the form factors then produces the current
$\Spipi^{\,\mu}_n$. Adding a transverse current $\Tpipi^{\,\mu}_n$, to be discussed
presently, we may write
\begin{align}
\Spipi^{\,\mu}_n+\Tpipi^{\,\mu}_n
&=-\sum_b e_b^{} (2p'_b-k)^\mu \frac{F_n^{(b)}(u_b)-\hat{F}_n+\hat{F}_n H_n}{u_b-p'^2_b}
\nonumber\\
&\quad\mbox{}
    -\sum_a e_a^{} (2p_a+k)^\mu \frac{F_n^{(a)}(s_a)-\hat{F}_n+\hat{F}_n H_n}{s_a-p^2_a}
\nonumber\\
&=-\sum_b e_b^{} (2p'_b-k)^\mu \frac{F_n^{(b)}(u_b)-\hat{F}_n}{u_b-p'^2_b}\,H_n^{(b)}
\nonumber\\
&\quad\mbox{}
    -\sum_a e_a^{} (2p_a+k)^\mu \frac{F_n^{(a)}(s_a)-\hat{F}_n}{s_a-p^2_a}\,H_n^{(a)}~,
    \label{eq:S+T}
\end{align}
where $H_n$ is the dimensionless function:
\begin{equation}
H_n=\prod_{b}\left[1- \delta_b \frac{F_n^{(b)}(u_b)}{\hat{F}_n}\right]
\prod_{a}\left[1- \delta_{a} \frac{F_n^{(a)}(s_a)}{\hat{F}_n}\right]
~,
\label{eq:Hprod}
\end{equation}
with $\delta_x=1$ (for $x=b,a$) if particle $x$ carries charge; otherwise it is
zero. The function $\hat{F}_n$ is given by
\begin{equation}
  \hat{F}_n = F_n(p'^2_1,\ldots,p'^2_{N_f};p^2_1,\ldots,p^2_{N_i}) ~,
  \label{eq:hatFgeneric}
\end{equation}
which is an unphysical value of $F_n$ because its momenta here are related by
Eq.~(\ref{eq:PhotoConservedMomentum}), and not by Eq.~(\ref{eq:HadronConservedMomentum}).
The function $H_n$ thus vanishes whenever any of the denominators in
Eq.~(\ref{eq:S+T}) vanishes, thus rendering Eq.~(\ref{eq:S+T}) manifestly
nonsingular. The functions $H^{(b)}_n$  and $H^{(a)}_n$ are obtained by
removing one of the factors from $H_n$ resulting in
\begin{subequations}\label{eq:Hn(ba)falloff}
\begin{align}
  H_n^{(b)} &= 1-\frac{\hat{F}_n H_n}{\hat{F}_n- \delta_b F_n^{(b)}(u_b)}
  =\sum_{b'\neq b}\delta_{b'}F_n^{(b')}(u_{b'})+\cdots~,
  \\
  H_n^{(a)} &= 1-\frac{\hat{F}_n H_n}{\hat{F}_n- \delta_a F_n^{(a)}(s_a)}
  =\sum_{a'\neq a}\delta_{a'}F_n^{(a')}(s_{a'})+\cdots~.
\end{align}
\end{subequations}
The right-most expressions here show that the leading fall-off behavior at
large energies is that of the form factors; explicit expressions for higher
orders are omitted here.

The transverse current $\Tpipi_n^{\mu}$ in Eq.~(\ref{eq:S+T}) arises solely
from the sum of terms proportional to $H_n$. In other words, the
four-divergence of
\begin{equation}
\Tpipi^\mu_n
=-\bigg[\sum_b e_b^{} \frac{(2p'_b-k)^\mu }{u_b-p'^2_b}
    +\sum_a e_a^{} \frac{(2p_a+k)^\mu }{s_a-p^2_a}\bigg] \hat{F}_n H_n
    \label{eq:Talone}
\end{equation}
vanishes identically,
\begin{equation}
  k_\mu \Tpipi^\mu \equiv 0~,
\end{equation}
because of charge conservation,
\begin{equation}
\sum_b e_b^{}  = \sum_a e_a^{}~.
\label{eq:hadronChargeConserved}
\end{equation}
However, it is this transverse current that leads to the factors $H_n^{(b)}$
and $H_n^{(b)}$ in Eq.~(\ref{eq:S+T}) providing the necessary high-energy
fall-off behavior to prevent the violation of scaling discussed in
Ref.~\cite{DL72}. Without this transverse current (i.e., for $H_n\equiv 0$),
these factors would all be unity. This limit defines $\Spipi^{\mu}_n$, which
therefore by itself provides the full four-divergence of Eq.~(\ref{eq:S+T}),
\begin{equation}
  k_\mu \Spipi^\mu_n
  =\sum_b e_b^{} F_n^{(b)}(u_b) - \sum_a e_a^{} F_n^{(a)}(s_a)~.
  \label{eq:gWTIscalarGeneral}
\end{equation}
Note here that this result would correspond to the complete generalized  WTI of
Eq.~(\ref{eq:gWTIintM}) for scalar coupling, where $G_n$ is a constant and
$\Kpipi^\mu_n\equiv 0$. For bare particles, with a constant value for the form
factors, this four-divergence vanishes because of charge conservation
(\ref{eq:hadronChargeConserved}).

It is now a simple exercise to verify that the full contact current
$\Cpipi^\mu$ given by the general expression (\ref{eq:CmuTcomplete}) satisfies
\begin{align}
  k_\mu \Cpipi^\mu &= \sum_b e_b^{} \left(\sum_n G^{(b)}_n(p'_b-k)F^{(b)}_n(u_n)\right)
  \nonumber\\
  &\qquad\mbox{}
  -\sum_a e_a^{} \left(\sum_n G^{(a)}_n(p_a+k) F_n^{(a)}(s_a)\right)~,
  \label{eq:genWTIcontactF}
\end{align}
where the sums in the parentheses provide different kinematical situations for
$\Fpipi$ of Eq.~(\ref{eq:expandhadron}), with all external four-momenta fixed
at the values appropriate for the photoprocess except for particles $b$ and $a$
in the sums which respectively have four-momenta $p'_b-k$ and $p_a+k$ so that
the photoprocess four-momentum conservation (\ref{eq:PhotoConservedMomentum})
holds true. The four-divergence (\ref{eq:genWTIcontactF}) thus corresponds to
Eq.~(\ref{eq:gWTIintM}) proving that $\Cpipi^\mu$ is indeed a locally
gauge-invariant contact current.

The construction presented here for $\Cpipi^\mu$ of Eq.~(\ref{eq:CmuTcomplete})
is the \textit{minimal\/} result that satisfies all constraints. However,
regarding the full dynamics of a contact-type interaction current, in general
there may be additional \textit{transverse\/} contributions that cannot be
determined uniquely by the gauge-derivative procedure and thus require a more
detailed microscopic calculation.  We emphasize, however, that the gauge
derivative will identify and consistently link all contributing topologically
distinct reactions mechanisms even if the explicit electromagnetic coupling
operator may be subject to transversality ambiguities (cf.\ subsequent
Sec.~\ref{secsubsub:proof}). As a case in point, we mention the transverse
treatment of final-state interactions for the $\gamma N\to \pi N$ reaction in
Ref.~\cite{HNK06}.

In principle, from a phenomenological point of view, one may add terms to the
right-hand side of Eq.~(\ref{eq:CmuTcomplete}), as long as they are transverse,
nonsingular, fall off fast enough at high energies as to not violate
scaling~\cite{DL72}, and  are independent of individual particle indices. The
manifestly transverse expression
\begin{align}
\textbf{T}^{\mu}_n=
  \left[\BGamma^{\mu}_n \left(\Spipi^{\,\nu}_n+\Tpipi^{\,\nu}_n\right)
  -\BGamma^{\,\nu}_n \left(\Spipi^{\mu}_n+\Tpipi^{\mu}_n\right)\right] k_\nu
  \label{eq:addionalTmu}
\end{align}
satisfies all of these constraints for any well-behaved current operator
$\BGamma^{\mu}_n$. To be consistent within the present approach this operator
should be based on the KRt currents (\ref{eq:KRtypeGamma}), leading to the
ansatz
\begin{equation}
  \BGamma^{\mu}_n = \sum_a \hat{\Gamma}^{\mu}_{n,a}-\sum_b \hat{\Gamma}^{\mu}_{n,b}~,
\label{eq:KRtadjust}
\end{equation}
where $\hat{\Gamma}^{\mu}_{n,x}$ is equal to $\Gamma^{\mu}_{n,x}$ with charge
$e_x$ replaced by $\delta_x$ (for $x=b,a$). With this choice,
Eq.~(\ref{eq:addionalTmu}) provides the simplest additional transverse current
that treats contributions from incoming and outgoing particle legs on an equal
footing and utilizes only dynamical elements that are already part of the
current $\Cpipi^\mu$. Thus, in phenomenological approaches one may exploit
these features and add linear combinations $\textbf{T}^{\mu}=\sum_n c_n
\textbf{T}^{\mu}_n$, with (dimensionless) fit parameters $c_n$, without
affecting gauge invariance. This basically readjusts the form-factor weights
for each KRt coupling in Eq.~(\ref{eq:KRtAll}) and the overall hadronic
coupling operator $G_n$ for the auxiliary current
$\Spipi^{\nu}_n+\Tpipi^{\nu}_n$ in a gauge-invariant manner. We mention in this
context that the four-point contact currents given in Refs.~\cite{NOH06,MON11}
utilize this freedom, corresponding to coefficients $c_n=1$ (with $n=1$).

\subsubsection{Proof of Eqs.~(\ref{eq:WardIdentity})}\label{secsubsub:proof}

For the proof of Eqs.~(\ref{eq:WardIdentity}), we will only discuss the
currents $\Gamma^{\mu}_{n,a}$ for incoming particles $a$; the proof for
outgoing particles $b$ can then easily be found along the same lines. Thus,
suppressing all extraneous indices, we shall prove that
\begin{equation}
  \Gamma^\mu = -\{ G(p) \}^\mu
  \label{eq:Gamma0}
\end{equation}
implies
\begin{equation}
  k_\mu \Gamma^\mu= Q\left[G(p)-G(p+k)\right]~,
  \label{eq:kGamma0}
\end{equation}
where $G(p)$ is the relevant hadronic coupling operator of the (incoming)
particle with charge $Q$ and momentum  $p$.

If $G(p)$ is a constant or linear in the momentum, the result follows trivially
because the gauge derivative of a constant vanishes, and if it is linear in the
momentum, there are only two possibilities~\cite{Haberzettl97},
\begin{equation}
\GD{p^\nu} = Qg^{\mu\nu}\qqtext{and} \GD{\fs{p}} = Q\gamma^\mu~,
\end{equation}
which both satisfy (\ref{eq:kGamma0}). The KRt currents in
Eq.~(\ref{eq:KRforNNpipi}) of the subsequent Sec.~\ref{appsec:5point} are of
this simple kind. The situation is more complicated if the coupling is of a
higher order because the gauge derivative is not unique for such cases; it
depends on the order of the (commuting) factors since factors downstream (i.e.,
to the left) of the gauge-derivative action acquire the extra photon momentum
$k$. For example, the expressions
\begin{subequations}
\begin{align}
  \GD{p^\lambda p^\nu} &= Q\left[g^{\mu\lambda} p^\nu + (p+k)^\lambda g^{\mu\nu}\right]~,
  \\
  \GD{p^\nu p^\lambda } &= Q\left[g^{\mu\nu} p^\lambda + (p+k)^\nu g^{\mu\lambda}\right]
\end{align}
\end{subequations}
differ by a manifestly transverse term. Such transverse terms have no bearing
on gauge invariance and may be ignored for the present discussion. (Of course,
they may be quite relevant for the physics of a given problem.) In general,
choosing any particular order for higher-order momentum dependences will be
subject to such transversality ambiguities. One may choose to deal with this
ambiguity in a democratic fashion by symmetrizing the expression, i.e.,
\begin{equation}
\frac{1}{2}\GD{p^\lambda p^\nu+p^\nu p^\lambda }
=\frac{Q}{2}\left[g^{\mu\lambda} (2p+k)^\nu + g^{\mu\nu} (2p+k)^\lambda \right]~,
\end{equation}
because this treats all factors on an equal footing. (Moreover, it will provide
current expressions that are symmetric in $p$ and $p'=p+k$, which clearly is
desirable.) This is easy to do for two factors like $p^\lambda p^\nu$ here or
$\fs{p}p^\nu$, etc.\footnote{Note
    that the gauge derivatives of $\fs{p}^2$ and $p^2$ differ
    by a manifestly transverse term, i.e., $\GD{\fs{p}^2}=\GD{p^2} + Q
    i\sigma^{\mu\nu}k_\nu$.}
For either one of such cases one may easily show explicitly that
Eq.~(\ref{eq:kGamma0}) is true irrespective of the chosen order, whether one
symmetrizes or not.

For higher orders of coupling that may occur for higher-spin particles, there
are $m!$ ordering possibilities for $m$ momentum factors. However, for the
purpose of the proof, there is no need to consider symmetrization explicitly.
We may just take one arbitrary order, number the factors from right to left,
and write
\begin{equation}
  G(p) = f_m(p)f_{m-1}(p)\ldots f_{n+1}(p) f_n(p)\ldots f_1p)~,
\end{equation}
where each factor here is linear in $p$. Since we know that
Eq.~(\ref{eq:kGamma0}) is true for linear and quadratic couplings, we may now
prove that it is true for $G(p)$ by induction assuming that it is true for
\begin{equation}
  G_n(p) = f_n(p)\ldots f_1p)
\end{equation}
and show that it remains true for
\begin{equation}
  G_{n+1}(p) = f_{n+1}(p)\,G_n(p)~.
\end{equation}
Using the product rule, the  gauge derivative for $n+1$ coupling factors is
given as
\begin{align}
\Gamma^{\mu}_{n+1} &\equiv-\GD{G_{n+1}(p)}
\nonumber\\
&= -\GD{ f_{n+1}(p) } G_n(p)- f_{n+1}(p+k) \GD{G_n(p)}~.
\label{eq:GDn+1}
\end{align}
With
\begin{equation}
  -k_\mu \GD{ f_{n+1}(p)}= Q\left[f_{n+1}(p)-f_{n+1}(p+k)\right]
\end{equation}
and
\begin{equation}
 -k_\mu \GD{G_n(p)} = Q\left[G_{n}(p)-G_{n}(p+k)\right]
\end{equation}
as stipulated, we then immediately find upon taking the four-divergence of
(\ref{eq:GDn+1}) and inserting these expressions that indeed
\begin{align}
  k_\mu \Gamma^{\mu}_{n+1}
  &=Q\left[G_{n+1}(p) -G_{n+1}(p+k)\right]~,
\end{align}
which proves (\ref{eq:kGamma0}) for $n+1$ factors. Thus, letting $n+1$ go to
$m$  proves the assertion.

It should be obvious now that this will remain true for a fully symmetrized KRt
current since each of the $m!$ current terms in the sum will yield the same
expression (\ref{eq:kGamma0}). Dividing then by $m!$ to properly normalize the
current shows that Eq.~(\ref{eq:kGamma0}) also holds true for the symmetrized
version.

\subsection{\boldmath Five-point contact current for $\gamma N \to \pi\pi N$}%
\label{appsec:5point}

The generic expressions of the previous section will now be applied to
generalize the  phenomenological five-point contact current for $\gamma N \to
\pi\pi N$ provided in Sec.~\ref{sec:PhenContactCurrent}. The most general
phenomenological ansatz for the hadronic $\pi \pi NN$ vertex
(\ref{eq:HadrDiracStruc}) is given by
\begin{align}
  \Cpipi(p',q_1,q_2;p)
     &=  a_1^{}F_1 + a_2^{} \frac{\fs{p}}{m}F_2 + a_3^{} \frac{\fs{p}'}{m'}F_3 + a_4^{} \frac{\fs{p}'\fs{p}}{m'm}F_4
\nonumber \\
     &\quad\mbox{}
            + b_1^{}\frac{\fs{q}}{m_\pi}F_5 + b_2^{} \frac{\fs{q}\fs{p}}{m_\pi m}F_6 + b_3^{} \frac{\fs{p}'\fs{q}}{m'm_\pi}F_7
\nonumber \\
     &\quad\mbox{}
            + b_4^{} \frac{\fs{p}'\fs{q}\fs{p}}{m'm_\pi m}F_8
\label{eq:Cpipifullpi}
\end{align}
in terms of eight phenomenological form factors,
\begin{equation}
  F_i = F_i(p'^2,q_1^2,q_2^2;p^2)~,\qtext{for} i=1,\ldots,8~,
\end{equation}
instead of the one of Eq.~(\ref{eq:CpipiAnsatz}), with fit constants $a_j$,
$b_j$ ($j=1,\ldots,4$) for the eight independent coupling operators.  [We
suppress here the isospin dependence (\ref{eq:HadrIsoStruc}) of the vertex
which will double the number of fit parameters; see Sec.~\ref{sec:appl}.] The
momenta here are defined as in Eq.~(\ref{eq:HadronicReactionProcess}).

Following the previous section, the associated contact current may be written
as
\begin{align}
  \Cpipi^\mu &= \Kpipi^\mu
  +a_1 C^\mu_{1} +a_2^{} \frac{\fs{p}}{m} C^\mu_{2} + a_3^{} \frac{\fs{p}'}{m'} C^\mu_{3}
   \nonumber \\
            &\quad\mbox{}
             + a_4^{} \frac{\fs{p}'\fs{p}}{m'm} C^\mu_{4} + b_1^{}\frac{\fs{q}}{m_\pi} C^\mu_{5} + b_2^{} \frac{\fs{q}\fs{p}}{m_\pi m}C^\mu_{6}
 \nonumber\\
&\quad\mbox{}
           + b_3^{} \frac{\fs{p}'\fs{q}}{m'm_\pi} C^\mu_{7} + b_4^{} \frac{\fs{p}'\fs{q}\fs{p}}{m'm_\pi m} C^\mu_{8}~,
            \label{eq:CmuAll}
\end{align}
where the Dirac structure of the coupling operators in (\ref{eq:Cpipifullpi})
gives rise to 16 KRt contributions
\begin{align}
  \Kpipi^\mu &= - e_i^{} \Big[ a_2 \frac{\gamma^\mu}{m} F_{2,i}+ a_4 \frac{\fs{p}'\gamma^\mu }{m'm} F_{4,i}
\nonumber\\
&\qquad\qquad\mbox{}
  + b_2 \frac{\fs{q}\gamma^\mu }{m_\pi m} F_{6,i} + b_4^{} \frac{\fs{p}'\fs{q}\gamma^\mu }{m'm_\pi m}F_{8,i}\Big]
  \nonumber\\
  &\quad\mbox{}
  - e_f^{} \Big[ a_3 \frac{\gamma^\mu}{m'} F_{3,f} + a_4 \frac{\gamma^\mu \fs{p}}{m'm} F_{4,f}
\nonumber\\
&\qquad\qquad\mbox{}
  +b_3 \frac{\gamma^\mu \fs{q}}{m' m_\pi} F_{7,f}+b_4^{} \frac{\gamma^\mu \fs{q}\fs{p}}{m'm_\pi m}F_{8,f}\Big]
  \nonumber\\
  &\quad\mbox{}
   -e_1^{} \Big[ b_1\frac{\gamma^\mu}{m_\pi} F_{5,1}+ b_2 \frac{\gamma^\mu \fs{p}}{m_\pi m} F_{6,1}
\nonumber\\
&\qquad\qquad\mbox{}
    + b_3 \frac{\fs{p}'\gamma^\mu}{m' m_\pi } F_{7,1}
   + b_4^{} \frac{\fs{p}'\gamma^\mu \fs{p}}{m'm_\pi m} F_{8,1}\Big]
  \nonumber\\
  &\quad\mbox{}
   +e_2^{} \Big[ b_1\frac{\gamma^\mu}{m_\pi} F_{5,2}+ b_2 \frac{\gamma^\mu \fs{p}}{m_\pi m} F_{6,2}
\nonumber\\
&\qquad\qquad\mbox{}
    + b_3 \frac{\fs{p}'\gamma^\mu}{m' m_\pi } F_{7,2}
   + b_4^{} \frac{\fs{p}'\gamma^\mu \fs{p}}{m'm_\pi m} F_{8,2}\Big]~.
   \label{eq:KRforNNpipi}
\end{align}
The parameters $e_i^{}$, $e_f^{}$, $e_1^{}$, and $e_2^{}$ are the individual charges of the
incoming and outgoing nucleons and the two pions related by charge
conservation,
\begin{equation}
  e_1^{}+e_2^{}+e_f^{}=e_i^{}~.
\end{equation}
The functions $F_{j,x}$, for $j=1,\ldots, 8$, are defined by
\begin{subequations}\label{eq:Ckinematics}
\begin{align}
F_{j,i}&=F_j(p'^2,q^2_1,q^2_2;s)~,
  \\
F_{j,f}&=F_j(u,q^2_1,q^2_2;p^2)~,
  \\
F_{j,1}&=F_j(p'^2,t_1,q^2_2;p^2)~,
  \label{eq:kinpion1}
  \\
F_{j,2}&=F_j(p'^2,q^2_1,t_2;p^2)~,
  \label{eq:kinpion2}
\end{align}
\end{subequations}
with Mandelstam-type variables
\begin{equation}
  s=(p+k)^2~,\qquad u=(p'-k)^2
\end{equation}
for the nucleons and
\begin{equation}
  t_1^{}=(q_1^{}-k)^2~,\qquad t_2^{} = (q_2^{}-k)^2
\end{equation}
for the pions. The indices $x=i,f,1,2$, therefore, serve as reminders of the
kinematic situations of intermediate off-shell hadrons in the first four
diagrams labeled $\Rpipi^\mu_x$ in Fig.~\ref{fig:ContCurrent}.

The eight auxiliary scalar contact currents $C^\mu_j$, for $j=1,\dots,8$,
follow from Eq.~(\ref{eq:S+T}) as
\begin{align}
  C^\mu_j &=
  -e_1^{} (2q_1^{}-k)^\mu \frac{F_{j,1}-\hat{F}_{j,0}}{t_1^{}-q_1^2}\; H^{(1)}_j
  \nonumber\\
  &\quad\mbox{}
  -e_2^{} (2q_2^{}-k)^\mu \frac{F_{j,2}-\hat{F}_{j,0}}{t_2^{}-q_2^2}\; H^{(2)}_j
    \nonumber\\
  &\quad\mbox{}
  -e_f^{} (2p'-k)^\mu \frac{F_{j,f}-\hat{F}_{j,0}}{u-p'^2}\; H^{(f)}_j
  \nonumber\\
  &\quad\mbox{}
  -e_i^{} (2p+k)^\mu \frac{F_{j,i}-\hat{F}_{j,0}}{s-p^2}\; H^{(i)}_j
  ~,
\label{eq:CauxAll}
\end{align}
where
\begin{equation}
  \hat{F}_{j,0} = F_j(p'^2,q_1^2,q_2^2;p^2)
\end{equation}
is the form factor at the unphysical squared hadronic four-momenta defined by
the photoreaction. The multiplicative (dimensionless) functions in Eq.~(\ref{eq:CauxAll}) are given by
\begin{align}
H^{(1)}_{j} &=
\delta_2 \frac{F_{j,2}}{\hat{F}_{j,0}} + \delta_f \frac{F_{j,f}}{\hat{F}_{j,0}}+ \delta_i \frac{F_{j,i}}{\hat{F}_{j,0}}
  -\delta_2\delta_f \frac{F_{j,2} F_{j,f}}{\hat{F}_{j,0}^2}
  \nonumber\\
  &\quad\mbox{}
  -\delta_f\delta_i \frac{F_{j,f} F_{j,i}}{\hat{F}_{j,0}^2}
  -\delta_i\delta_2 \frac{F_{j,i} F_{j,2}}{\hat{F}_{j,0}^2}
  +\delta_2\delta_f\delta_i\frac{F_{j,2} F_{j,f} F_{j,i}}{\hat{F}_{j,0}^3}~,
\label{eq:Hfunction}
\end{align}
and cyclic permutation of indices $\{12fi\}$. The factors $\delta_x$ indicate
whether the corresponding particle is charged or not in the usual manner.

With external hadrons taken on-shell and one common phenomenological form
factor $F$, instead of eight, the off-shell current $\Cpipi^\mu$ of
Eq.~(\ref{eq:CmuAll}) reduces to the on-shell result (\ref{eq:CmuAllonshell})
given in Sec.~\ref{sec:PhenContactCurrent}. Structurally, the off-shell current
(\ref{eq:CmuAll}) here satisfies the same generalized  WTI
(\ref{eq:gWTIcontact}) as the on-shell current, as required by the general
result (\ref{eq:gWTIintM}).

%
\begin{figure*}[t!]\centering
  \includegraphics[width=.9\textwidth,clip=]{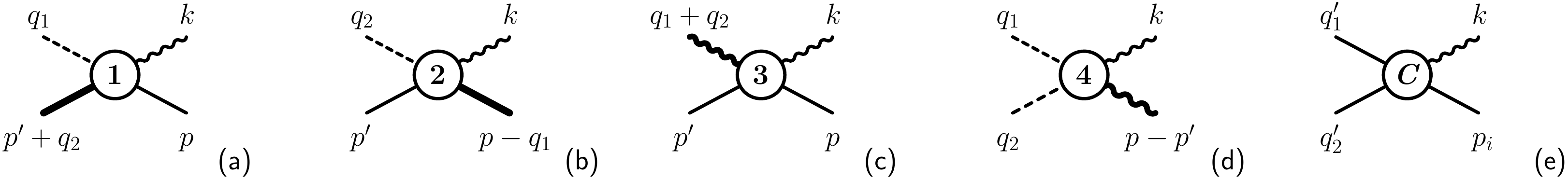}
  \caption{\label{fig:CCkinematics}%
  Contact-current diagrams with associated four-momenta. Processes with labels
  1--4 in the first four diagrams correspond to contact currents labeled
  similarly in the two-meson production processes depicted in
  Fig.~\ref{fig:NoLoopExplicit}, with explicit four-momenta as in
  Eq.~(\ref{eq:gammaN-Npipi}). The last diagram (e) depicts a generic rendering
  of the first four diagrams, with corresponding hadrons shown as solid lines
  here and generic momenta used in Eq.~(\ref{eq:Aux4Current}).}
\end{figure*}
%

\subsection{Four-point contact currents}%
\label{appsec:4point}

The kinematics of the four contact currents labeled 1--4 in
Fig.~\ref{fig:NoLoopExplicit} for the no-loop calculation of the process
\begin{equation}
  \gamma(k) + N(p) \to \pi_1^{}(q_1^{})+\pi_2^{}(q_2^{})+N'(p')
  \label{eq:gammaN-Npipi}
\end{equation}
are indicated in Fig.~\ref{fig:CCkinematics}. Their momentum dependences as
used here thus are given by
\begin{subequations}
\begin{align}
 \gamma N\to \pi B:&& \Cpipi_1^\mu &= \Cpipi_1^\mu(p'+q_2,q_1;k,p)~,
 \\
 \gamma B\to \pi N':&& \Cpipi_2^\mu &=\Cpipi_2^\mu(p',q_2;k,p-q_1)~,
\\
 \gamma N\to \rho N':&& \Cpipi_3^{\lambda\mu} &=\Cpipi_3^{\lambda\mu}(p',q_1+q_2;k,p)~,
\\
 \gamma \rho \to \pi\pi:&& \Cpipi_4^{\mu\lambda} &=\Cpipi_4^{\mu\lambda}(q_1,q_2;k,p-p')~,
\end{align}
\end{subequations}
where $B$ indicates any intermediate baryon compatible with the process. Also,
$N'$, $\pi$, and $\rho$ are placeholders for any outgoing baryon,
(pseudo)scalar meson, and intermediate vector meson, respectively. The Lorentz
index $\lambda$ is that of the vector meson.

It follows from the generic treatment in Appendix \ref{appsec:ContCurrentGen}
that the currents may be written as
\begin{subequations}\label{eq:4current}
\begin{align}
  \Cpipi^\mu_1 &= e_1^{} F^{(1)}_1  \,\hat{\Gamma}^\mu_\pi+ G_1 C^\mu_1~,
  \\
  \Cpipi^\mu_2 &= e_2^{} F^{(2)}_2  \,\hat{\Gamma}_\pi + G_2 C^\mu_2~,
  \\
  \Cpipi^{\lambda\mu}_3 &= e_\rho F^{(\rho)}_\rho\, \hat{\Gamma}^{\lambda\mu}_\rho+ G^\lambda_\rho C^\mu_\rho~,
  \\
  \Cpipi^{\mu\lambda}_4 &=  \big(e_1^{} F^{(1)}_{\pi\pi}-e_2^{} F^{(2)}_{\pi\pi}\big)\,
    \hat{\Gamma}^{\mu\lambda}_{\pi\pi} + G^\lambda_{\pi\pi} C^\mu_{\pi\pi}~,
\end{align}
\end{subequations}
where the hadronic coupling operators for the first two currents are
\begin{equation}
  G_k = g G^{\pm} \frac{\fs{q}_k}{M+m}~,\qqtext{for} k=1,2~,
\end{equation}
where $m$ and $M$, respectively, are the masses of the nucleon and the baryon,
and $G^\pm$ is either $\gamma_5$ or $(M+m)/(M-m)$ depending on whether initial
and final baryon parities are the same or not (see Ref.~\cite{NOH06} for details;
modifications of the present vector coupling structure by allowing scalar
contributions as well are also discussed there). The couplings of the vector
meson to the baryon and the meson, respectively, are given as\footnote{All
coupling operators here assume that propagators are
  written with factors of $i$ stemming from the usual Feynman rules, i.e., $S(p)=i/(\fs{p}-m)$ etc.}
\begin{equation}
  G^\lambda_\rho=-ig_N \left[\gamma^\lambda -i\kappa \frac{\sigma^{\lambda\tau}}{2m}(q_1+q_2)_\tau\right]
\end{equation}
and
\begin{equation}
  G^\lambda_{\pi\pi} = -ig_M(q_1 -q_2)^\lambda~.
\end{equation}
The constants $g$, $g_N$, and $g_M$ are strength parameters, and $\kappa$ is
the anomalous moment of the vector meson. The form factors $F_1$, $F_2$,
$F_\rho$, and $F_{\pi\pi}$ are phenomenological scalar functions depending on
the squared four-momenta of all associated hadron legs.
For the specific kinematics of Figs.~\ref{fig:CCkinematics}(a)-(d), indicated
by an upper index, their dependence reads
\begin{subequations}
\begin{align}
  F^{(1)}_1&=F_1\big(t_1^{},(p’+q_2)^2;p^2\big)~,
  \\[1ex]
  F^{(2)}_2&=F_2\big(q_1^2,t_2^{};(p-q_1)^2\big)~,
  \\[1ex]
  F^{(\rho)}_\rho&=F_\rho\big(t_\rho,p'^2;p^2\big)~,
  \\[1ex]
  F^{(1)}_{\pi\pi}&=F_{\pi\pi}\big(t_1^{},q_2^2;(p-p')^2\big)~,
  \\[1ex]
  F^{(2)}_{\pi\pi}&=F_{\pi\pi}\big(q_1^2,t_2^{};(p-p’)^2\big)~,
\end{align}
\end{subequations}
where
\begin{equation}
  t_1^{}=(q_1^{}-k)^2~, \quad t_2^{}=(q_2^{}-k)^2~, \quad t_\rho = (q_1^{}+q_2^{}-k)^2~.
\end{equation}
Note that
\begin{equation}
  F_1(q^2,p'^2;p^2) = F_2(q^2,p^2;p'^2)
\end{equation}
by symmetry.

Following (\ref{eq:KRtypeGamma}), the KRt current operators are found as
\begin{subequations}
\begin{align}
\hat{\Gamma}^\mu_\pi & = -gG^\pm \frac{\gamma^\mu}{M+m}~,
\\[1ex]
\hat{\Gamma}^{\lambda\mu}_\rho &=g_N  \frac{\sigma^{\lambda\mu}}{2m}\kappa~,
\\[1ex]
\hat{\Gamma}^{\mu\lambda}_{\pi\pi} &=ig_M g^{\mu\lambda}~,
\end{align}
\end{subequations}
where, compared to (\ref{eq:KRtypeGamma}), the charges have been pulled out
explicitly in (\ref{eq:4current}).

The final pieces for the complete description of the
currents~(\ref{eq:4current}) are the auxiliary scalar currents corresponding to
the generic results of Eq.~(\ref{eq:S+T}). Utilizing the generic contact
current depicted in Fig~\ref{fig:CCkinematics}(e), they may be written as
\begin{subequations}\label{eq:Cmulist}
\begin{align}
  C^{\mu}_1 &= C^{\mu}_{1BN}[F_1](q_1,p'+q_2;p)~,
  \\[1ex]
  C^{\mu}_2 &= C^{\mu}_{2N'B}[F_2](q_2,p';p-q_1)~,
  \\[1ex]
  C^{\mu}_\rho &= C^{\mu}_{\rho N'N}[F_\rho](q_1+q_2,p';p)~,
  \\
  C^{\mu}_{\pi\pi} &= C^{\mu}_{12\rho}[F_{\pi\pi}](q_1,q_2;p-p')~,
\end{align}
\end{subequations}
where the right-hand sides here are specific applications of the corresponding
generic result. The auxiliary current for Fig~\ref{fig:CCkinematics}(e) is
given as
\begin{align}
  C^{\mu}_{12i}[h](q'_1,q'_2;p_i)
  &=
  -e_1^{} \frac{h_1(t_1)-\hat{h}}{t_1- q'^2_1}\,H^{(1)}_{2i}[\hat{h}]
  \nonumber\\
  &\quad\mbox{}
  -e_2^{} \frac{h_2(t_2)-\hat{h}}{t_2- q'^2_2}\,H^{(2)}_{i1}[\hat{h}]
  \nonumber\\
  &\quad\mbox{}
  -e_i^{} \frac{h_i(s_i)-\hat{h}}{s_i- p^2_i}\,H^{(i)}_{12}[\hat{h}]~,
  \label{eq:Aux4Current}
\end{align}
written here as a functional of a generic hadronic form factor $h$, with
four-momenta of incoming and outgoing hadrons labeled by $i$ and $1,2$,
respectively, as in Fig.~\ref{fig:CCkinematics}(e). Using Mandelstam-type
squared four-momenta
\begin{equation}
  s_i=(p_i+k)^2~,\quad t_1=(q'_1-k)^2~,\quad t_2=(q'_2-k)^2~,
\end{equation}
the relevant kinematic situations for the form factor read
\begin{equation}
 h_i=h(q'^2_1,q'^2_2;s_i)~,\quad h_1=h(t_1,q'^2_2;p_i^2)~,\quad h_2=h(q'^2_1,t_2;p_i^2)
\end{equation}
and the analog of Eq.~(\ref{eq:Hn(ba)falloff}) for the present case is
\begin{equation}
  H^{(1)}_{2i}[h]=\delta_2 \frac{h_2}{\hat{h}}+\delta_i \frac{h_i}{\hat{h}}-\delta_2 \delta_i \frac{h_2 h_i}{\hat{h}^2}~,
\end{equation}
where $\delta_k=1$ if particle $k$ is charged and zero otherwise. The other
functions, $H^{(2)}_{i1}[h]$ and $H^{(i)}_{12}[h]$, are obtained by cyclic
permutation of indices $\{12i\}$. As usual, the expression
\begin{equation}
  \hat{h} = h(q'^2_1,q'^2_2;p_i^2)
\end{equation}
provides the unphysical extrapolation to the squared external hadronic
four-momenta to render Eq.~(\ref{eq:Aux4Current}) manifestly nonsingular, similar
to Eq.~(\ref{eq:hatFgeneric}).

Finally, note that the corresponding contact currents given in
Refs.~\cite{NOH06,MON11} differ from the ones in (\ref{eq:4current}) here by
manifestly transverse terms obtained by making use of the freedom of adding
specific terms constructed along the lines of the generic form
(\ref{eq:addionalTmu}); see, in particular, remarks after
Eq.~(\ref{eq:KRtadjust}).


\end{document}